\documentclass{aa}  

\usepackage{graphicx}
\usepackage{txfonts}
\usepackage{orcidlink}
\begin{document}

\title{Characterising the stellar differential rotation based on largest-spot statistics from ground-based photometry
\thanks{The Johnson B- and V-band differential photometry and the numerical time series analysis results are available in electronic form at the CDS via anonymous ftp to \texttt{cdsarc.u-strasbg.fr} (130.79.128.5) or via \texttt{http://cdsweb.u-st rasbg.fr/cgi-bin/qcat?J/A+A/nnn/Annn}}}

\author{Mikko Tuomi\inst{1,2}\fnmsep\thanks{\email{mikko.tuomi@helsinki.fi}}\orcidlink{0000-0002-4164-4414} \and
Jyri J. Lehtinen\inst{3,1} \and
Gregory W. Henry\inst{4} \and
Thomas Hackman\inst{1}\orcidlink{0000-0002-1412-2610}
}

\institute{University of Helsinki, Department of Physics, PO Box 64, 00014, Finland
\and
University of Hertfordshire, Center for Astrophysics, College Lane Campus, Hatfield, Hertfordshire, UK, AL10 9AB
\and Finnish Centre for Astronomy with ESO (FINCA), University of Turku, Vesilinnantie 5, 20014 Turku, Finland
\and
Center of Excellence in Information Systems, Tennessee State University, Nashville, TN 37209, USA (Retired)
}

\date{Received May 2024; accepted later}

\abstract
{}
{Stellar spot distribution has consequences on the observable periodic signals in long-time baseline ground-based photometry. We model the statistics of the dominating spots of two young and active Solar-type stars, V889 Her and LQ Hya, in order to obtain information on the underlying spot distribution, rotation of the star, as well as the orientation of the stellar axis of spin.}
{By calculating estimates for spot-induced periodicities in independent subsets of photometric data, we obtain statistics based on the dominating spots in each subset, giving rise to largest-spot statistics accounting for stellar geometry and rotation, including differential rotation.}
{Our simple statistical models are able to reproduce the observed distribution of photometric signals rather well. This also enables us to estimate the dependence of angular velocity of the spots as a function of latitude. Our results indicate that V889 Her has a non-monotonic differential rotation curve with a maximum angular velocity between latitudes of 37-40 deg and lower angular velocity at the pole than the equator. Our results for LQ Hya indicate that the star rotates much like a rigid body. Furthermore, the results imply that the monotonic Solar differential rotation curve may not be a universal model for other solar-type stars.}
{The non-monotonicity of the differential rotation of V889 Her is commonly produced in magnetohydrodynamic simulations, which indicates that our results are realistic from a theoretical perspective.}


\maketitle

\section{Introduction}

The simplest possible means of obtaining information on stellar activity-induced phenomena such as starspots is photometric monitoring of active stars with ground-based telescopes. Such monitoring campaigns have traditionally been applied to determine stellar rotation periods and spot activity cycles. The ground-based surveys are still highly valuable for the simple reason that they provide long-baseline windows into the brightness evolution and variability of nearby stars.

Typically, such long-baseline differential photometry is used to determine variations in the star's apparent rotation period, as the co-rotating starspots appear and disappear on the visible disk of the stellar surface and leave their periodic mark on the light curve. This enables characterisation of the variations in the rotation period, indicative of starspots at different altitudes affected by the surface differential rotation of the star. Estimates of stellar differential rotation have been obtained in this manner for a number of stars \citep[e.g.][]{kajatkari2014,lehtinen2016,siltala2017}. The method has also been applied to chromospheric activity time series \citep{mittag2024}.

Interpreting  variations in photometric periods directly as a measure of differential rotation is rather problematic. Another effect that can cause variations in the observed period in the light curves of even rigidly rotating active stars is the evolution of active regions. Spot evolution is the more influential the smaller the ratio of spot lifetime to the rotation period is. However, generally differentiating between the two has been challenging \citep{basri2020}.

But with photometric data, it is only possible to assume that an apparent spot-induced photometric rotation period observed at a given time interval is probably representative of the behaviour of the largest non-axisymmetric spot structure over that interval. Such an assumption has been made (implicitly or explicitly) throughout the literature when estimating the stellar differential rotation.

Moreover, the amplitude of the photometric period variations depends on the latitude range of the spots on the stellar surface. The spot configurations of rapidly rotating active stars are generally dominated by very strong polar spots \citep[e.g.][]{willamo2019}, but there is also evidence of low latitude spots extending all the way to the stellar equator \citep{lehtinen2022}. Crucially, the polar spots are also largely axisymmetric structures with respect to the stellar rotation, which greatly diminishes their dominance over the light curve shape, effectively ensuring that the photometric spot modulation probably samples the differential rotation across a broad latitude range.

Here we demonstrate a new approach for interpreting the period variations of photometric signals together with simultaneously determined signal amplitudes. This enables characterisation of rotation period variations, indicative of starspots at different latitudes and consequently qualitative and quantitative estimates for the stellar differential rotation.

\section{Target stars}

We have applied our technique of largest-spot statistics to the photometry of two nearby, young, solar-type stars: V889 Her and LQ Hya. These two stars have been selected because of their comparably young age, rapid rotation, and availability of long-baseline photometric data.

\object{V889 Her} (HD 171488, HIP 91043) is a young and active Solar analog in the sense that it has the same spectral type of G2 V \citep{montes2001} except for age of only $\sim 50$ Myr \citep{strassmeier2003}. Its mass of 1.06$\pm$0.02 $M_{\odot}$, radius of 1.09$\pm$0.05 $R_{\odot}$ and effective temperature of 5830$\pm$50 K \citep{strassmeier2003} indicate that the star is very similar to the Sun in all respects except for the age. The young age also explains the rapid rotation of the star with a period of 1.3454$\pm$0.0013 days \citep{lehtinen2016} and manifests itself as large, persistent spot structures at the polar regions \citep{willamo2019}.

Due to an abundance of spots on its surface and their distinctive effect on the photometric variability of V889 Her, it has been possible to determine the differential rotation of the star. \citet{marsden2006} reported an equatorial rotation period $P_{\rm eq} =$ 1.313$\pm$0.004 days, and estimated $\Delta \Omega$, or the rotational shear to be $0.402\pm0.044$ rad day$^{-1}$. To obtain this estimate \citet{marsden2006} assumed the star to have a monotonic solar-type differential rotation curve. A consistent estimate of $\Delta\Omega = 0.319\pm0.114$ rad day$^{-1}$ was recently reported by \citet{brown2024}.

\object{LQ Hya} (HD 82558, HIP 46816) is a K2 V dwarf with likewise an estimated age of 50 Myr \citep{tetzlaff2011} but a lower mass of 0.8$\pm$0.1 $M_\odot$ \citep{tetzlaff2011}, a radius of $R = 0.58$ R$_{\odot}$ \citep{lehtinen2020} and an effective temperature of 4934$\pm$70 K \citep{tsantaki2014}. Its rotation period has been estimated to be 1.60405$\pm$0.00040 d \citep{lehtinen2022}. A distinct feature of the rotational profile is that the star does not demonstrate much evidence for differential rotation, but rather appears to rotate much like a rigid body based on photometric data. The differential rotation obtained by \citep{lehtinen2022} corresponds to a rotational shear of 0.067 rad day$^{-1}$. As is the case with V889 Her, LQ Hya also has a persistent polar spot structure \citep{cole-kodikara2019}.

The lifetimes of spots on the surfaces of the two target stars are unknown. However, similar young and active Solar-type stars have been estimated to have spot lifetimes of 10-300 days for the large spot structures that can be studied photometrically \citep{namekata2019,ioannidis2020}. Other authors have reported similar results based on Kepler photometry and different statistical techniques \citep{basri2022}. In particular, \citet{namekata2019} estimated the lifetimes of 56 spots on the surfaces of such stars. They did not report shorter spot lifetimes than about 10 days, and found that such short lifetimes were rather rare for large spots. For rapidly rotating stars with $P_{\rm rot} < 10$ days, the spot lifetimes were rather uniformly between 10 and 80 days, which suggests that similar lifetimes can be expected for V889 Her and LQ Hya.

\section{Material and methods}

\subsection{Observations}

We have analysed the long-baseline differential Johnson B- and V-band photometry of the target stars, obtained with the T3 0.4 m Automatic Photoelectric Telescope (APT) at the Fairborn Observatory, Arizona \citep{fekel2005}. The photometric observations of V889 Her were obtained over a baseline of 29.14 years from May 1994 to July 2023 and consist of 2780 individual B- and 2755 V-band observations. For LQ Hya, there are 3730 B- and 3653 V-band observations over a baseline of 35.41 years from December 1987 to May 2023.

Typical photometric errors for this telescope range between 0.003 and 0.004 mag \citep{henry1995}. The comparison stars used for obtaining the differential photometry are \object{HD 171286} for V889 Her and \object{HD 82477} for LQ Hya.

To maximize the number of observations available for the light curve analysis, we followed \cite{lehtinen2016} and combined the B- and V-band data into single unified time-series scaled according to the V-band observations. We did this by fitting a linear model of the form $B = c_1V - c_0$ into the sets of coeval B- and V-band observations and used these models to transform the B-band observations onto the V-band observations.

We calculated estimates for the periodicities and the corresponding amplitudes based on these data sets for independent data subsets with baselines $T_{\rm s}$ of 15 and 20 days. For V889 Her, these subsets enabled obtaining estimates for 192 and 166 independent periods and amplitudes with the Continuous Period Search (CPS) code \citep{lehtinen2011,lehtinen2016}. The CPS algorithm provides a flexible fit of a truncated Fourier series determining the optimal order of the harmonic light curve model according to the Bayesian information criterion. For LQ Hya we obtained 250 and 213 such pairs of periods and amplitudes. Although these baselines are somewhat greater than the lowest reported lifetimes of large spots on the surfaces of similar stars \citep{namekata2019,ioannidis2020}, we consider them a rather safe choice as larger spots also appear to live longer than smaller ones \citep{giles2017}.

\begin{figure*}
\center
\includegraphics[angle=270,width=0.48\textwidth,clip]{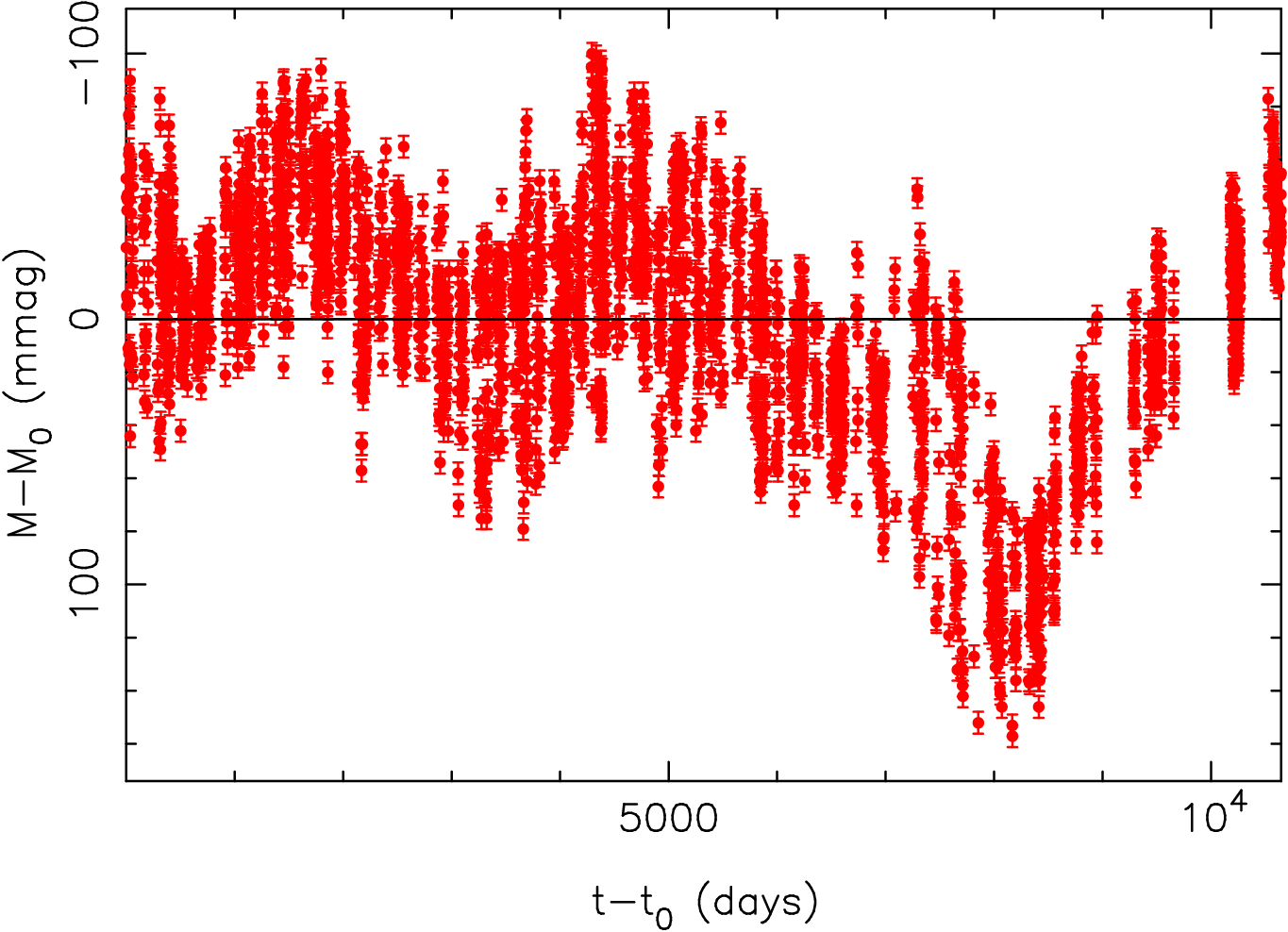}
\includegraphics[angle=270,width=0.48\textwidth,clip]{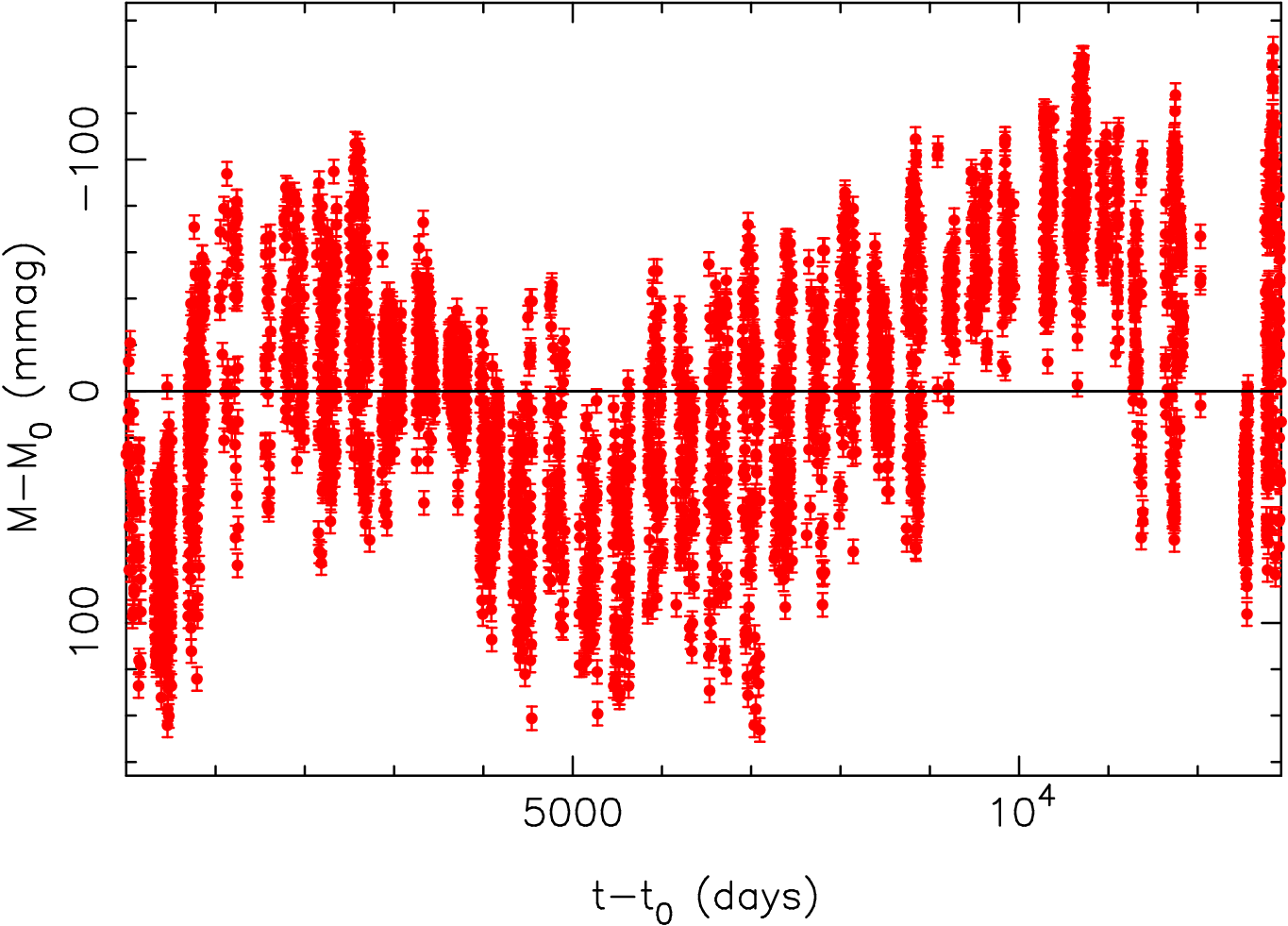}

\includegraphics[angle=270,width=0.48\textwidth,clip]{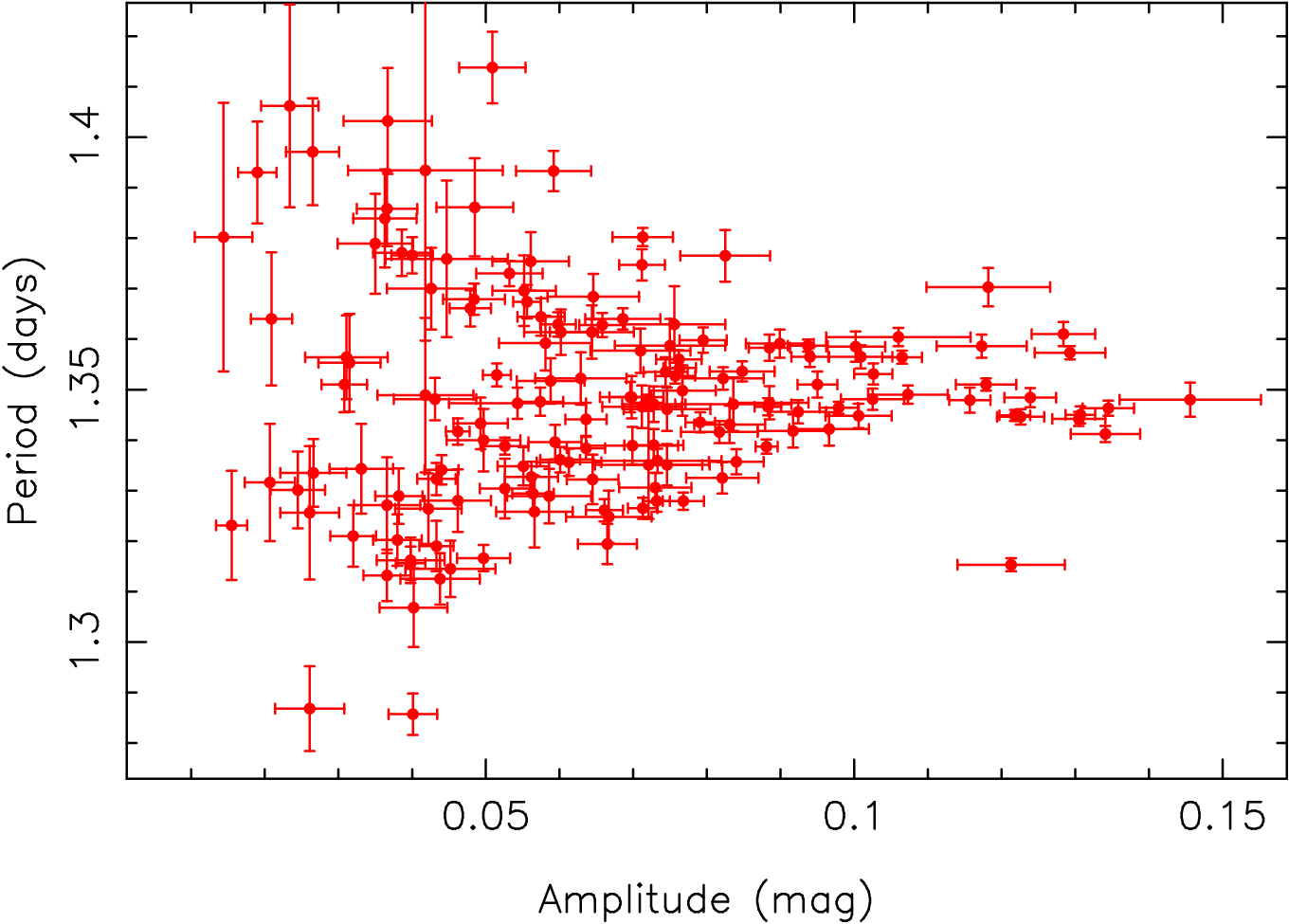}
\includegraphics[angle=270,width=0.48\textwidth,clip]{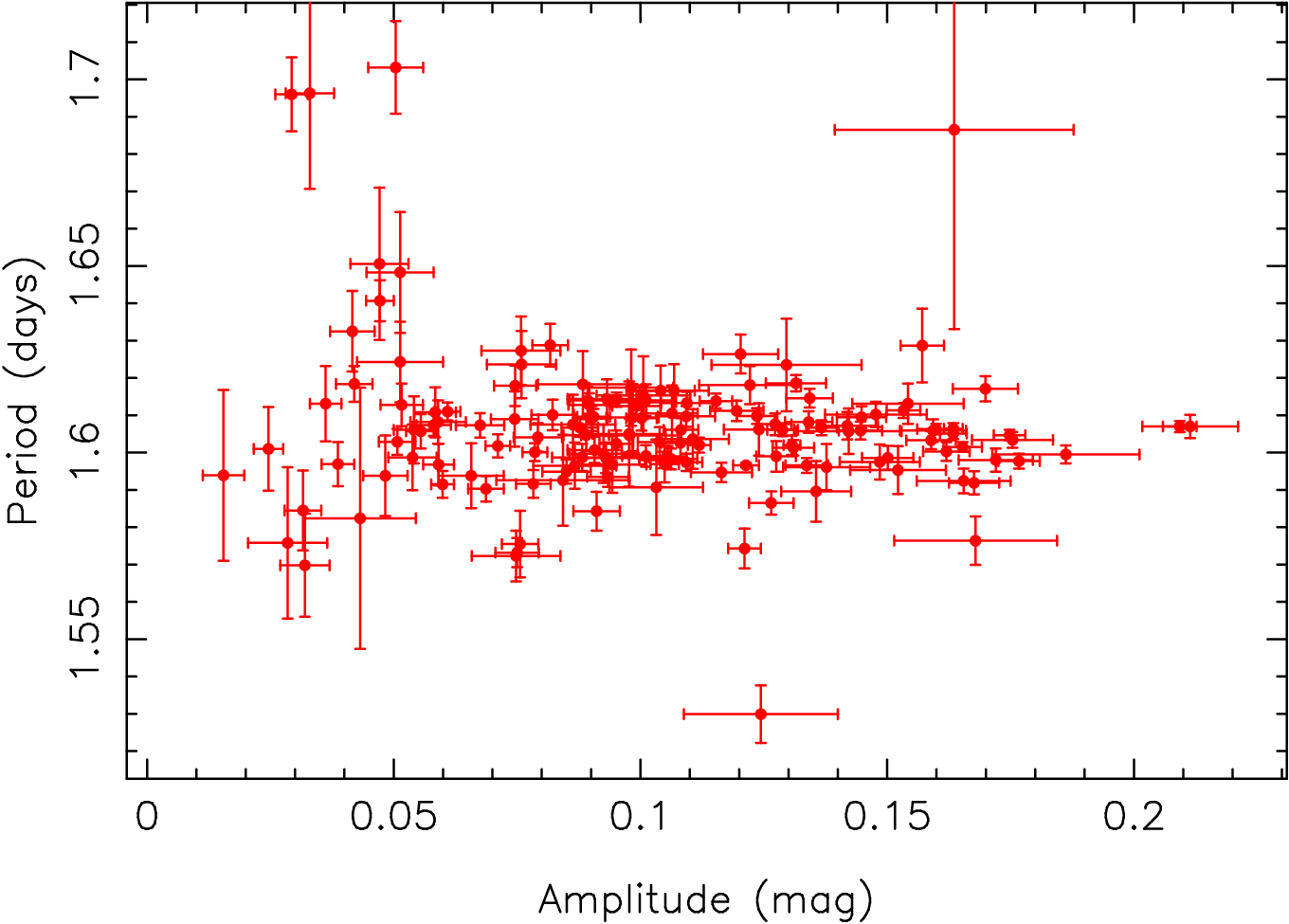}
\caption{Full V-band light curves (top panels) and amplitude and period data (bottom panels) of largest spots for V889 Her (left panels) and LQ Hya (right panels). The reference times $t_{0}$ in the light curves are 2449481 and 2447141 JD for V889 Her and LQ Hya, respectively. Mean magnitudes $M_{0}$ have been subtracted from both data sets. The periods and amplitudes are shown for $T_{\rm s} = 15$ days.}\label{fig:data}
\end{figure*}

We neglected all such subsets for which there were indications of the superposition of two or more periodicities of comparable amplitude. This corresponds to removing subsets during which the dominating spots were too short-lived or were replaced by other emerging spot structures. Such situations generally produce light curves with double minima, corresponding to two or more simultaneous and independent spotted regions on the star. By choosing the subsets with only single light curve minima, we could limit the samples to those subsets that were considered more probable to be dominated by one spotted region. This enabled us to obtain more reliable estimates for the largest-spot statistics\footnote{Obviously, the largest spot does not refer to the spot or a group of spots that is actually the largest one but the one whose photometric effect is the largest given its geometry.}. As a result, we neglected some of the subsets and base our results on 154 and 134 signals for V889 Her and 146 and 120 signals for LQ Hya, respectively. These restrictions yield more reliable statistics for the rotation of the target stars because the largest spot dominates all remaining subsets and its parameters are thus more reliable. We present these period-amplitude pairs for both stars and the 15-day baseline in Fig. \ref{fig:data}. We note that the overall distributions of period-amplitude pairs appear similar for other baselines albeit with fewer data points.

Our choices for subset baselines are less arbitrary than it may seem. The baseline of 15 days was selected because lower baselines, say 10 days, would contain at most only ten epochs for the target albeit with two wavelength bands. This is because the stars have only been observed once a night. We thus considered a baseline of 15 days to correspond to the bare minimum out of which periodicities could be detected reliably. Moreover, increasing the baseline decreases the number of independent detections of periodicities, and risks them being contaminated by the superposition of two or more spots with a higher probability.

\begin{figure*}
\center
\includegraphics[angle=270,width=0.32\textwidth,clip]{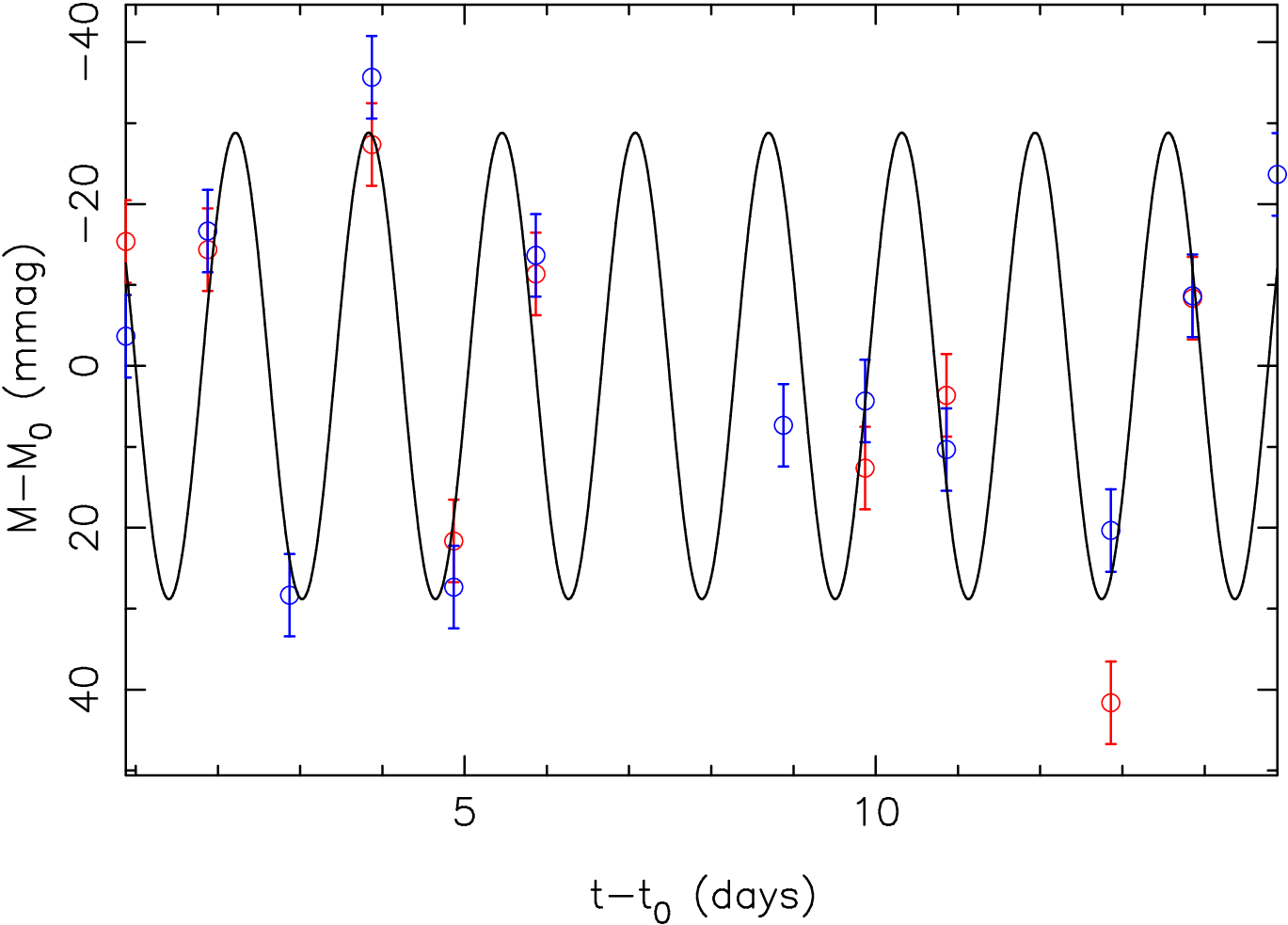}
\includegraphics[angle=270,width=0.32\textwidth,clip]{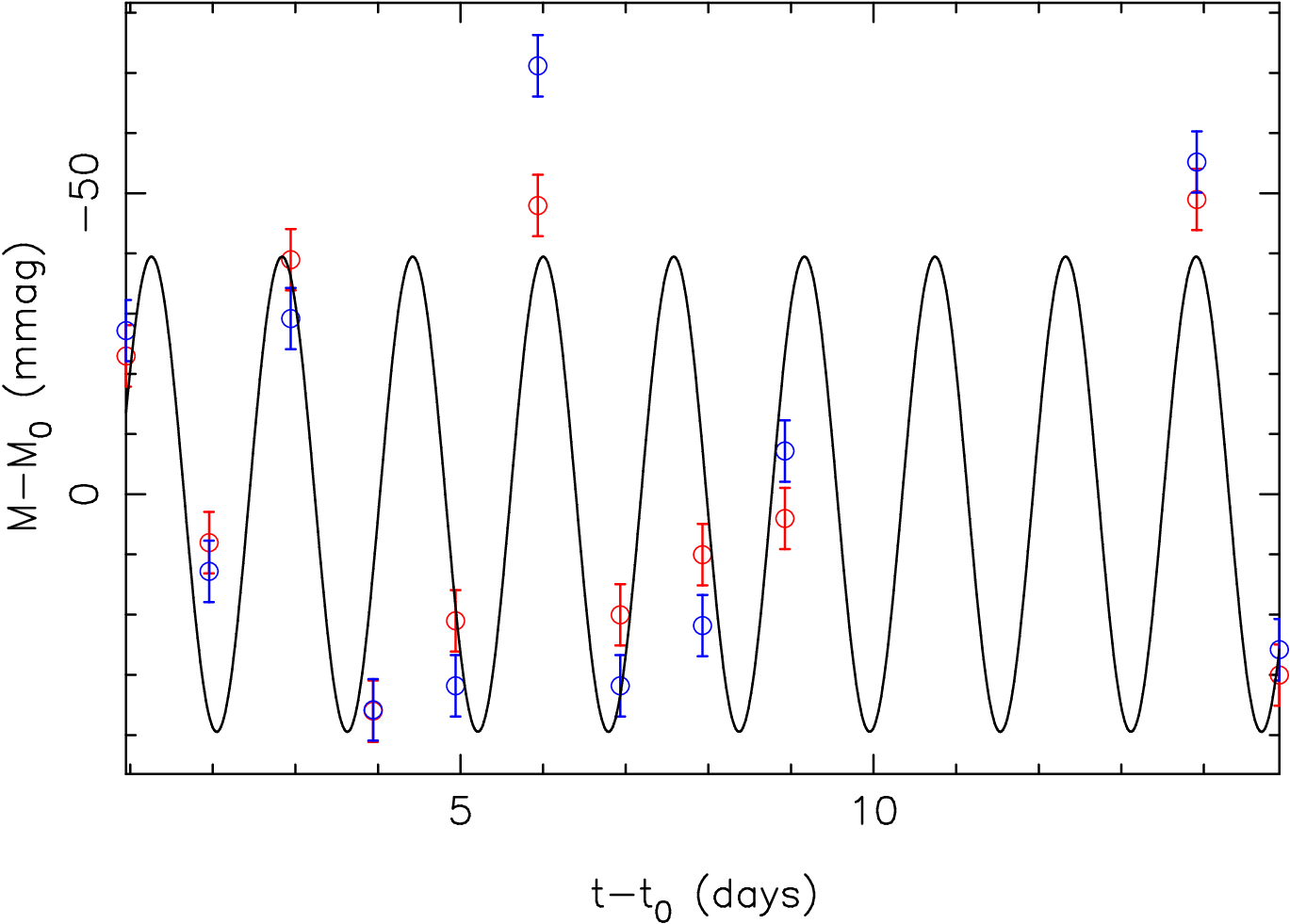}
\includegraphics[angle=270,width=0.32\textwidth,clip]{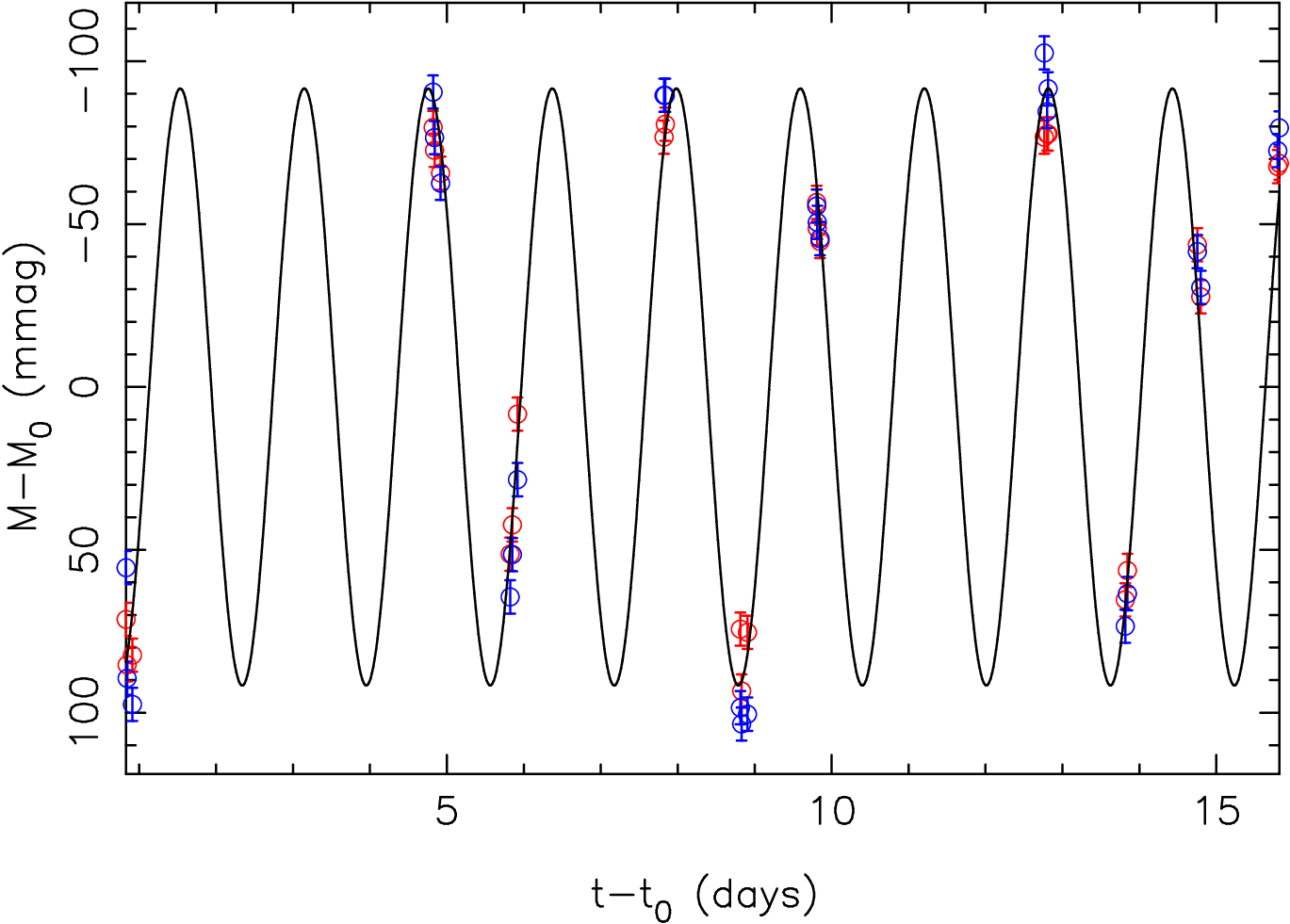}

\includegraphics[angle=270,width=0.32\textwidth,clip]{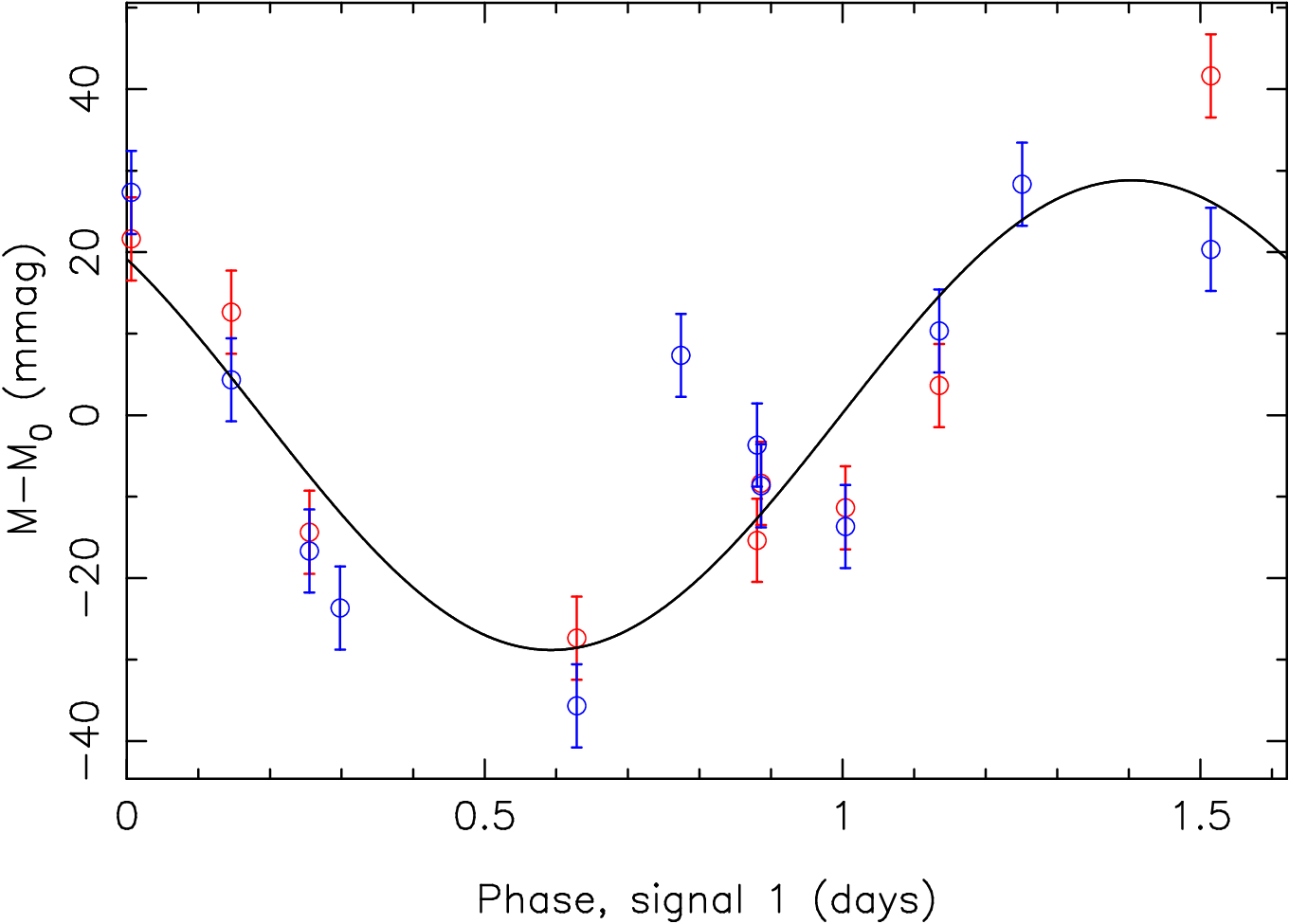}
\includegraphics[angle=270,width=0.32\textwidth,clip]{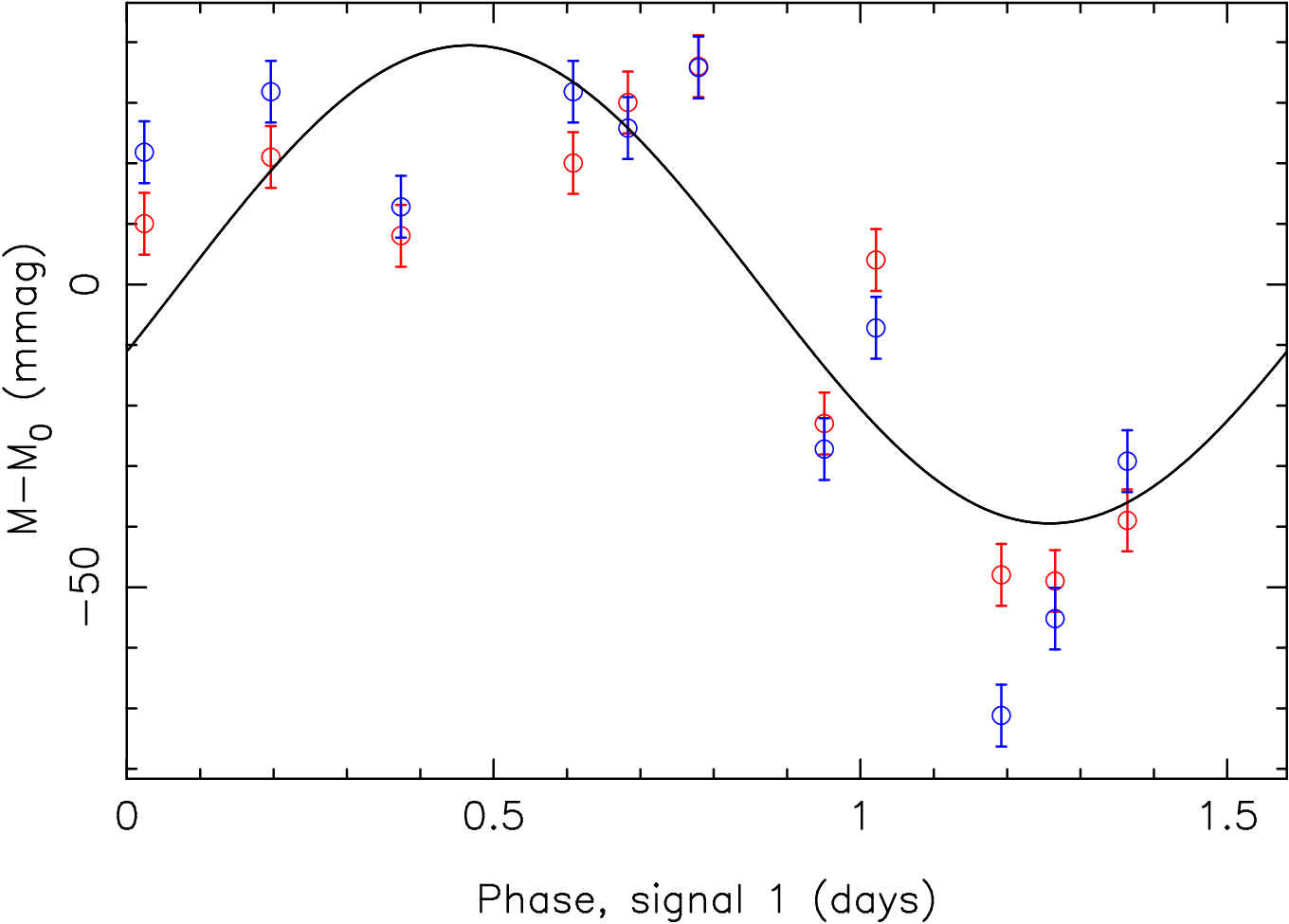}
\includegraphics[angle=270,width=0.32\textwidth,clip]{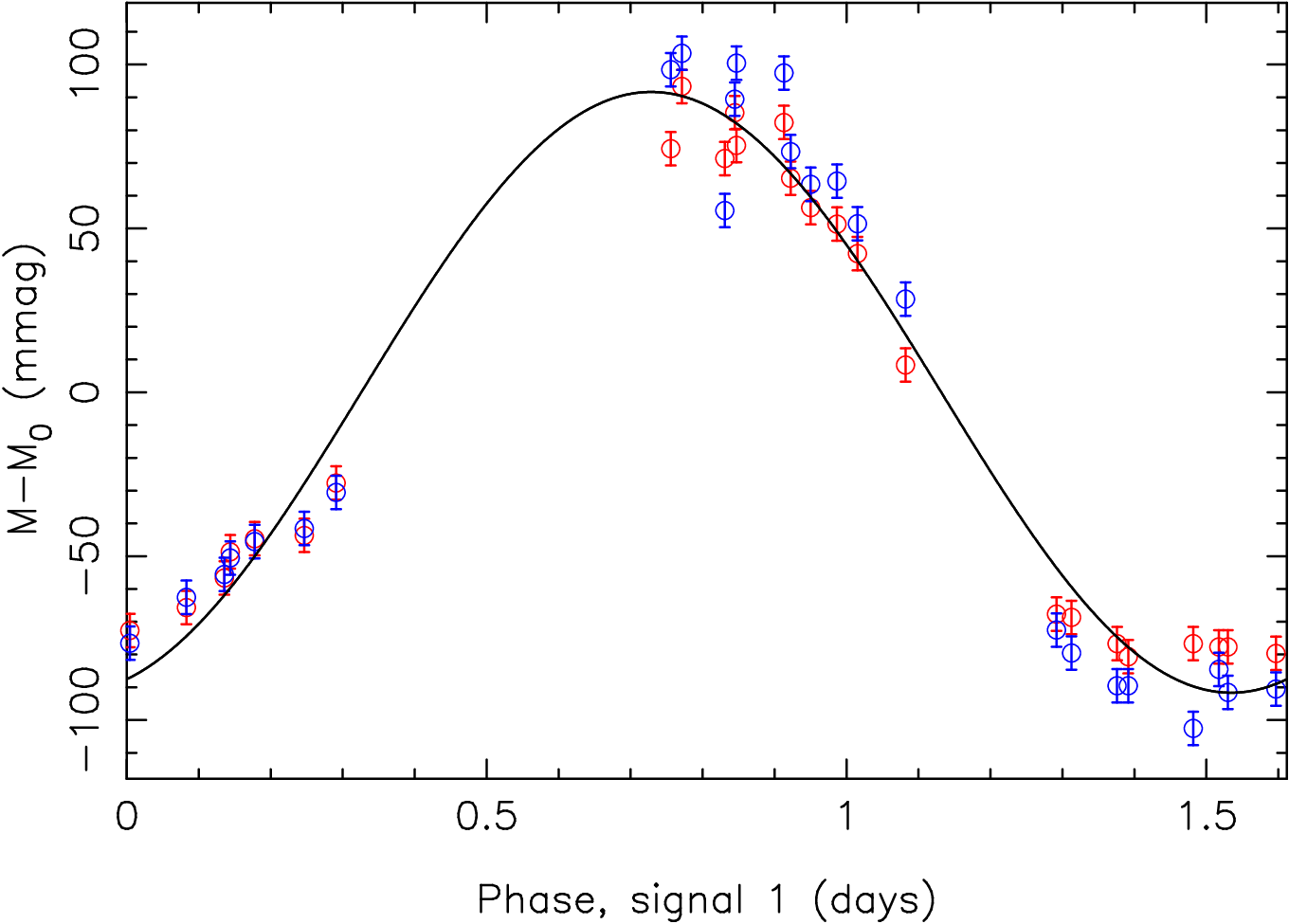}
\caption{Example plots of three data subsets for LQ Hya and 15-day baselines as a function of time (top panels) and signal phase (bottom panels). The red and blue circles represent V- and B-band magnitudes. The solid black curve denotes the fitted sinusoidal function.}
\label{fig:subsets}
\end{figure*}

We have plotted three 15-day subsets of LQ Hya in Fig. \ref{fig:subsets} to visually demonstrate the nature of the identified periodic signals. These three sets have been selected randomly from near the beginning, the middle, and the end of the light curve. The typical numbers of individual photometric observations in such subsets ranged from around 20 for 15-day subsets to roughly 30 for the 20-day subsets. We required that there were at least ten observations in all subsets in order to include them in the analyses.

Although the brightnesses of the two stars vary rather clearly according to some (quasi)periodic activity cycles (Fig. \ref{fig:data}), the statistics of the period and amplitude parameters were not found to depend on the overall brightness. This indicates that although the change in brightness probably corresponds to differences in spot coverage, this has no discernible effects on the statistics of the dominating spot structures.

\subsection{Largest spot statistics}

We assume that there is a long-baseline photometric data set describing the brightness variability of a rapidly rotating star due to co-rotation of star spots on the stellar surface. We further assume that over consequent sub-intervals of the data baseline, the variations in brightness can be approximated with simple sinusoidal functions. It is thus possible to divide the data into $N$ independent subsets and determine the apparent periods ($P$) and amplitudes ($A$) of such sinusoids for each set. Such a determination can be achieved with, e.g., Lomb-Scargle periodograms \citep{lomb1976,scargle1982}. A simple sinusoid represents the first-order series approximation of any periodic behaviour and can be used to approximate the period and amplitude of the variability rather well. Such a method has been used to estimate the rotational shear of nearby stars \citep[e.g.][]{lehtinen2011,lehtinen2016}.

Yet, not all periods and amplitudes are allowed and their distribution therefore contains information on the distribution of the dominating large spots on the stellar surface as well as the orientation of the stellar axis of spin. For instance, depending on the stellar inclination, the spots at different latitudes cause periodicities with different amplitude while the different periods contain information on the differential rotation of the star. But since no information can be inferred based on single periods and their amplitudes, we present a statistical approach as follows.

The period-amplitude space can be divided into an $n \times n$ grid to determine the number of signals in each grid square. This provides $n^{2}$ data points on the shape of the distribution based on the $N$ detected periodicities. This grid is hereafter referred to as the data as we attempt to model the shape of the two-dimensional distribution. We note that we limit our analyses to $n^{2} \leq N$ to avoid artificially increasing the number of degrees of freedom\footnote{Although it is not at all clear that such a limitation is necessary, as most grid squares will end up being without signals, we choose a limited size for our grids to constrain the computational problem.}.

We started by drawing a spot latitude $\psi$ from a Gaussian distribution limited to the interval $[0, \pi/2]$ such that $\psi \sim \mathcal{N}(\mu_{\psi},\sigma_{\psi}^2)$. Half of these draws are randomly assigned a negative sign to place the corresponding latitudes on the opposite hemisphere. Given an inclination $i$, some spots are never visible to the observer, and we thus neglect them in the analysis by drawing new values corresponding to visible locations.

We further drew a value for a signal amplitude parameter $A_{\rm s}$ from a Gaussian distribution $\mathcal{N}(\mu_{\rm A},\sigma^{2}_{\rm A})$ representing the magnitude of the photometric effect caused by the spot. This parameter depends on the combined size and temperature of the spot, but we do not attempt to differentiate between these two factors. We then account for the geometric effects of the position of the spot on the stellar surface. The actual spot-induced amplitude is scaled by a factor of $\mu = \cos i \sin \psi + \sin i \cos \psi$ depending on how close to the centre of the stellar disk the spot moves as the star rotates. This gives a projected amplitude of $A = \mu A_{\rm s}$, and this amplitude is always greater than zero for visible spots.

Next, we defined the angular velocity $\Omega$ as a function of latitude such that it follows some simple model. We can, for instance, choose the family of typical functions used to describe differential rotation \citep[e.g.][]{wohl2010,lamb2017}, and write
\begin{equation}\label{eq:angular_velocity}
  \Omega(\psi) = \sum_{i=0}^{k} a_{i}\sin^{2i}\psi + \epsilon_{\Omega} .
\end{equation}
This is our forward model describing the angular velocity for different latitudes. The Gaussian random variable $\epsilon_{\Omega}$ has a zero mean and variance of $\sigma_{\Omega}^{2}$ and we simply draw a random value from this distribution to obtain a value for $\Omega$ emphasising that the relationship between latitude and angular velocity is only statistical. This can also be translated into a corresponding period. As a result, we have the position of the modelled spot in the period-amplitude grid, and we can repeat this procedure many times to accurately estimate the modelled shape of the distribution based on simple parameters of the model: $i$, $a_{0}, a_{1}, ..., a_{k}$, $\mu_{\rm A}$, $\sigma_{\rm A}$, $\mu_{\psi}$, $\sigma_{\psi}$, and $\sigma_{\Omega}$.

We then constructed a simple Gaussian likelihood function for the grid squares by assuming a standard deviation of $\sigma_{\rm g}$ as yet another free parameter of the model. The log-likelihood function can be written as
\begin{equation}\label{eq:likelihood}
  \log L = - \frac{1}{2} \sum_{i=1}^{n} \sum_{j=1}^{n} \log \bigg(2 \pi \sigma_{\rm g}^{2} \bigg) + \frac{1}{\sigma_{\rm g}^{2}}\Bigg[ f_{\rm d}(A_{i},P_{j}) - f_{\rm m}(A_{i},P_{j} | \theta) \Bigg]^{2} ,
\end{equation}
where $f_{\rm d}(A_{i},P_{j})$ and $f_{\rm m}(A_{i},P_{j}|\theta)$ denote the functions representing the observed and modelled frequencies of $A$ and $P$, respectively, for the $i$th and $j$th grid elements with the latter conditioned on the variable $\theta$ that is a vector containing all the free parameters in the model.

The resulting models are reasonably simple. For instance, for a model with $k=1$, there are nine free parameters of the model characterising the distribution of spots in the period-amplitude space.

\subsection{Statistical methods}

We compared a variety of model choices in order to find the best descriptions for the data and to enable studying the underlying astrophysics as reliably as possible. The data were analysed by assuming a model structure and by applying the adaptive Metropolis posterior sampling algorithm \citep{haario2001} to obtain samples from the parameter posterior distribution. Given uniform prior probability densities in their respective ranges, this sample was used to obtain parameter estimates and to find a maximum-likelihood value given the selected model. This maximum likelihood value was then applied to obtain simple model comparison statistics in terms of Bayesian information criterion values \citep{liddle2007} and the corresponding model probabilities. We also applied the likelihood-ratio tests based on $\chi^{2}$ statistics to compare nested models but relied on the model probabilities due to their higher robustness in practice.

\section{Results}

\subsection{V889 Her}

\begin{figure*}
\center
\includegraphics[angle=270,width=0.24\textwidth,clip]{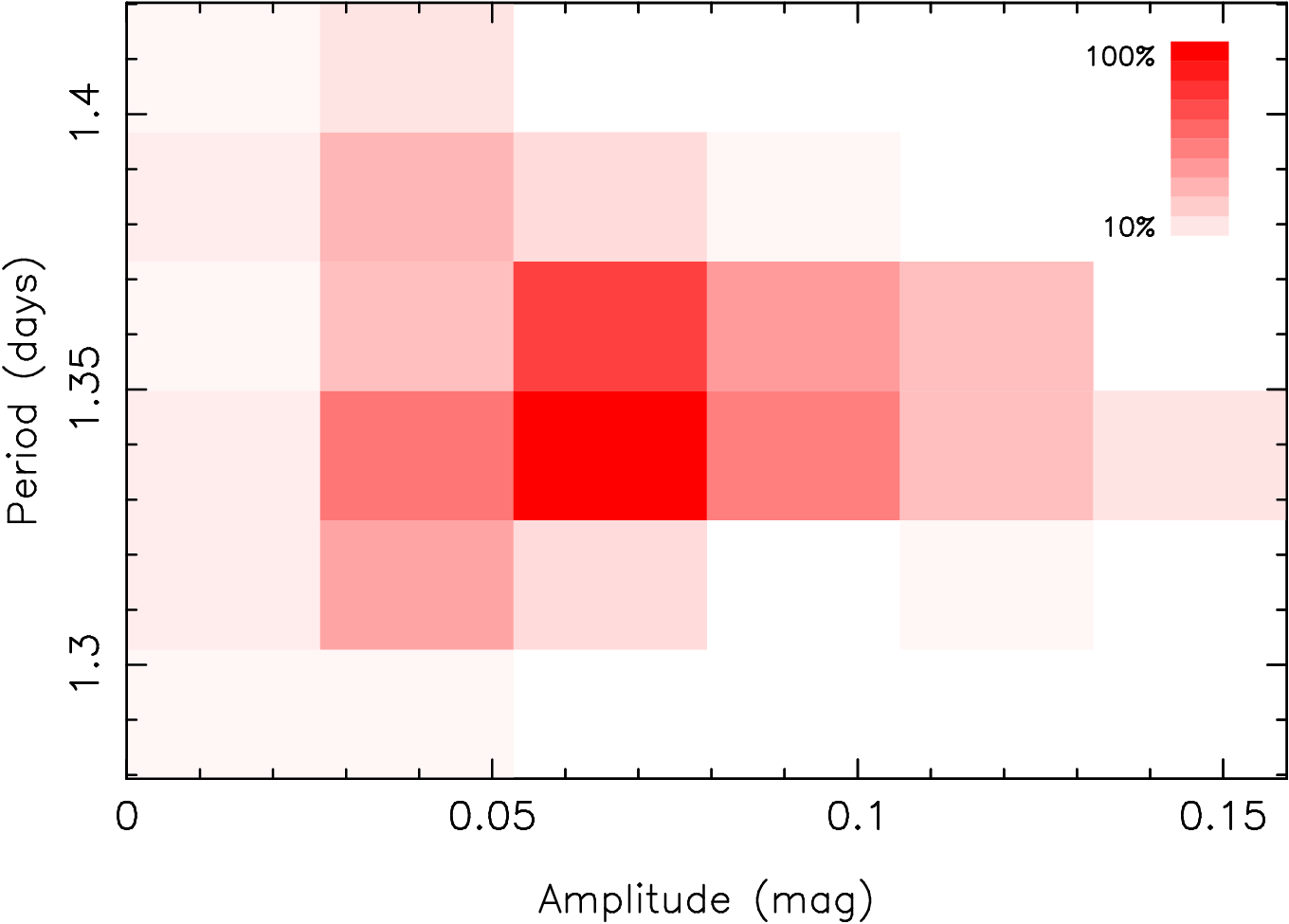}
\includegraphics[angle=270,width=0.24\textwidth,clip]{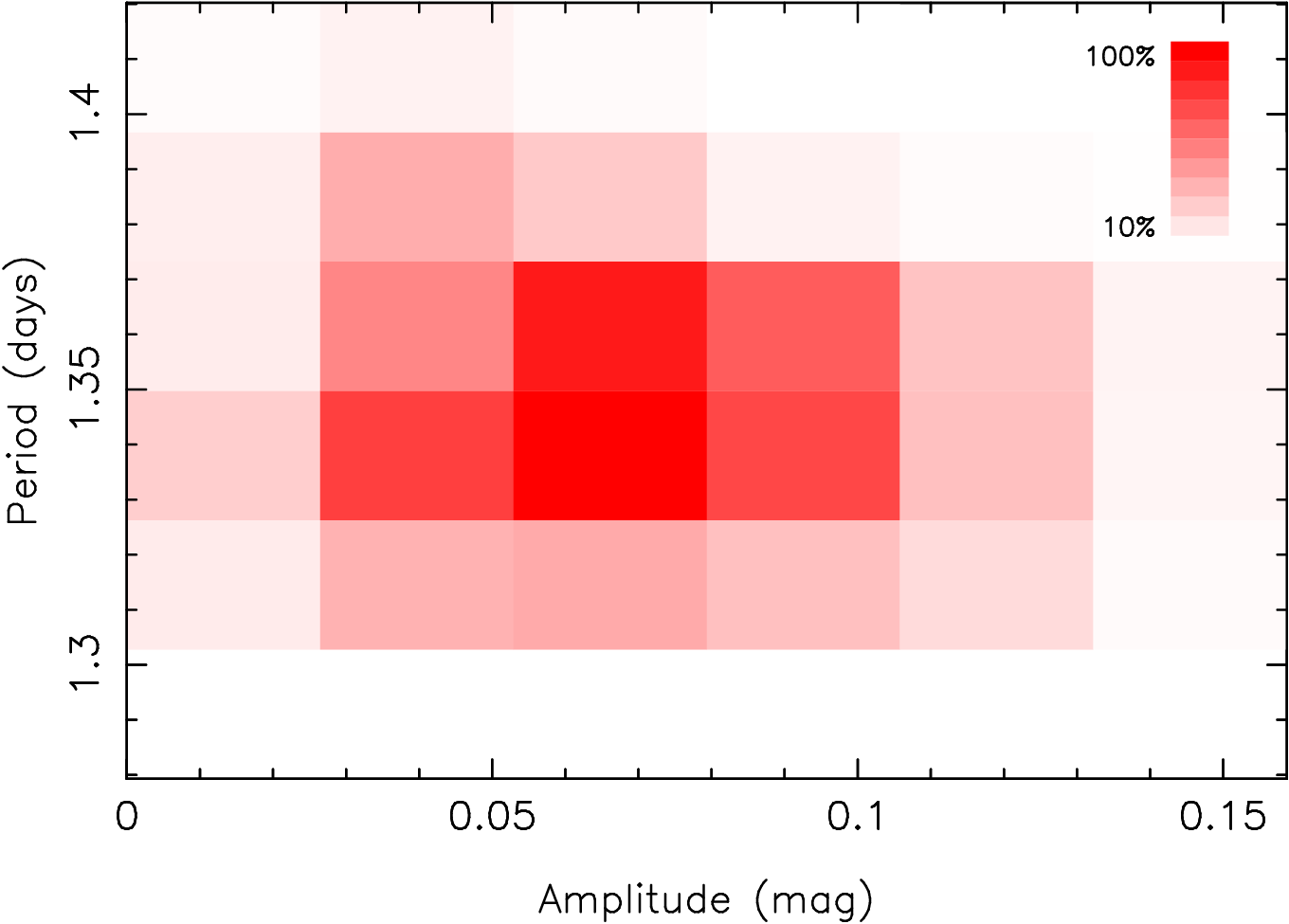}
\includegraphics[angle=270,width=0.24\textwidth,clip]{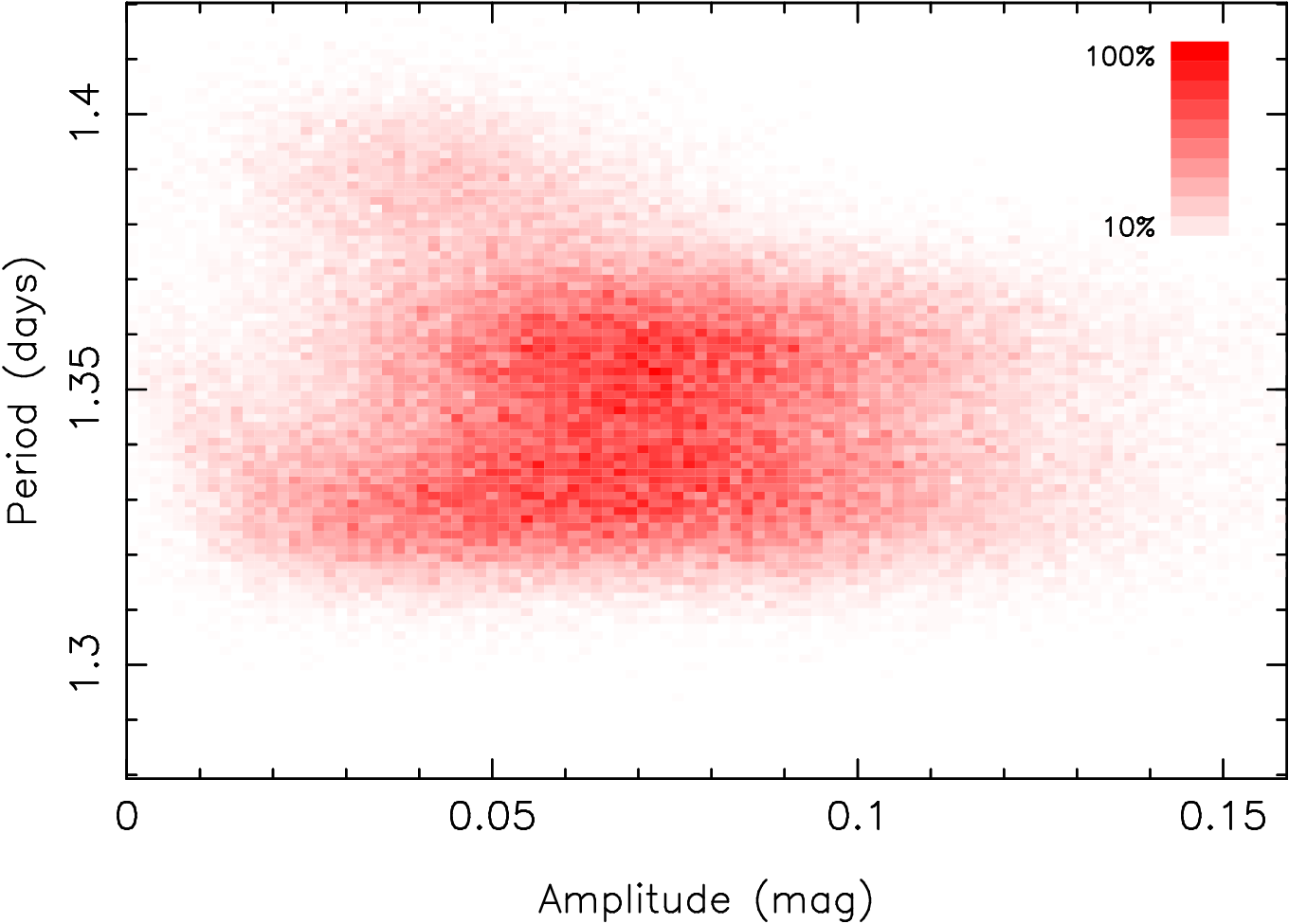}
\includegraphics[angle=270,width=0.24\textwidth,clip]{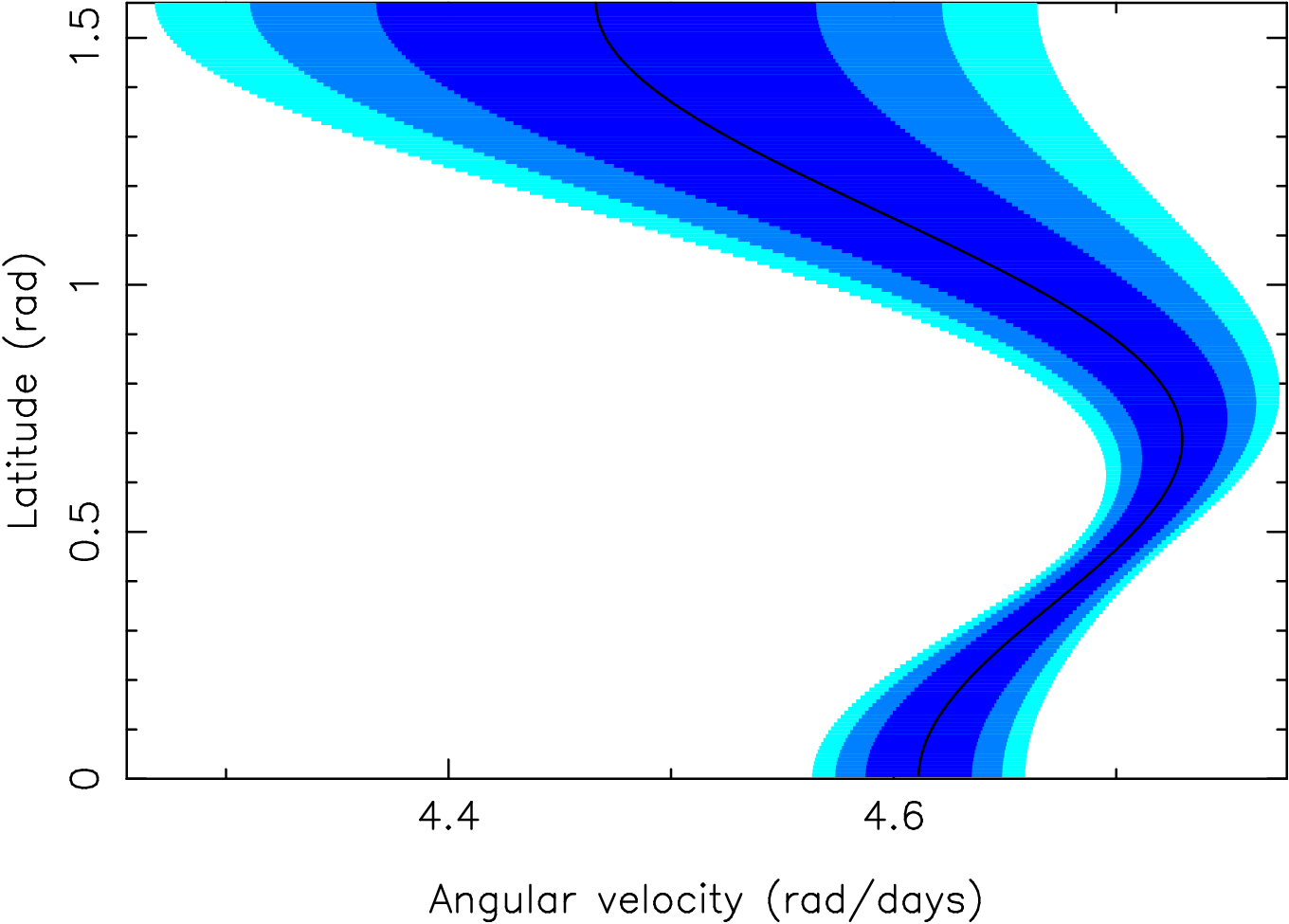}

\includegraphics[angle=270,width=0.24\textwidth,clip]{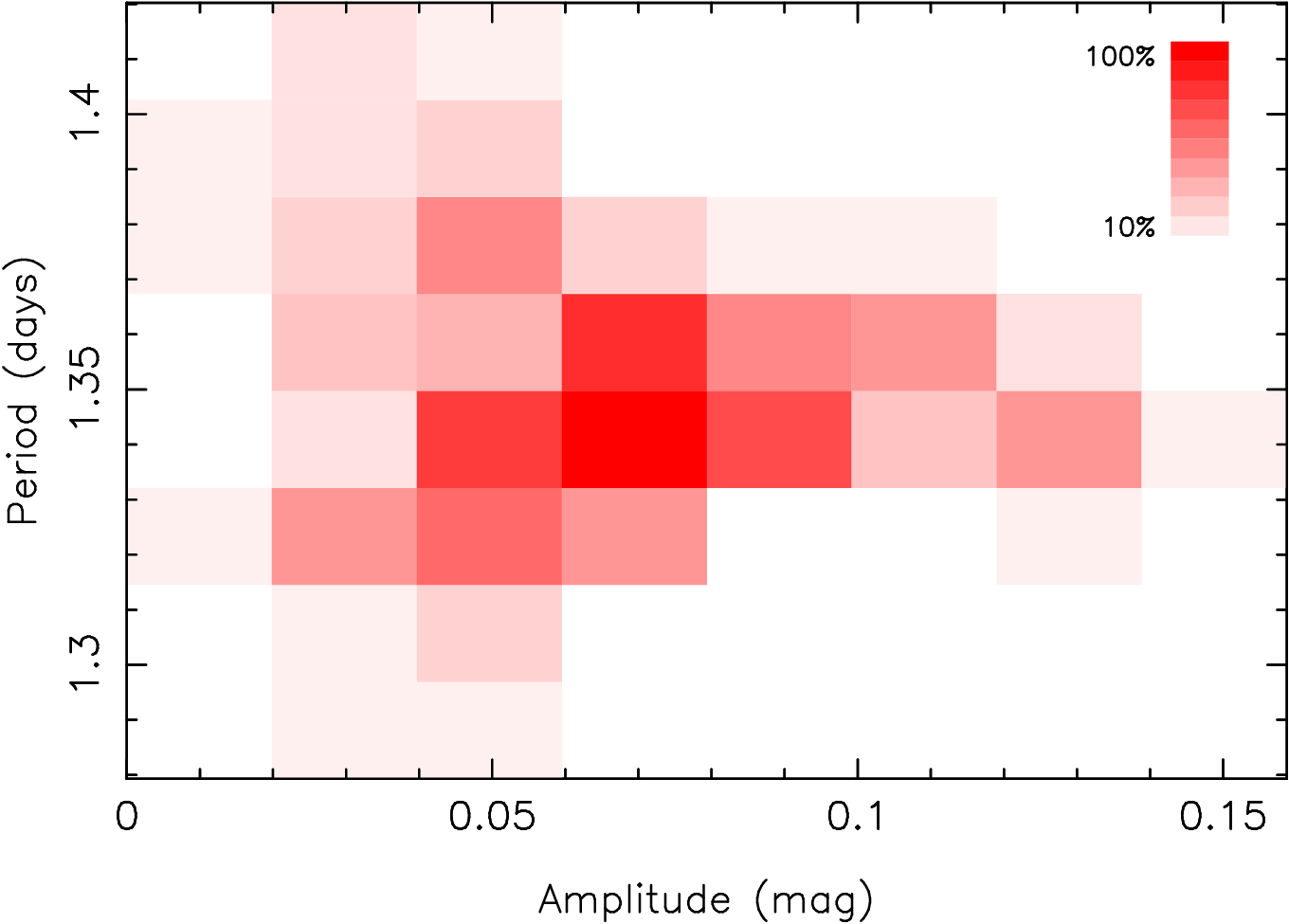}
\includegraphics[angle=270,width=0.24\textwidth,clip]{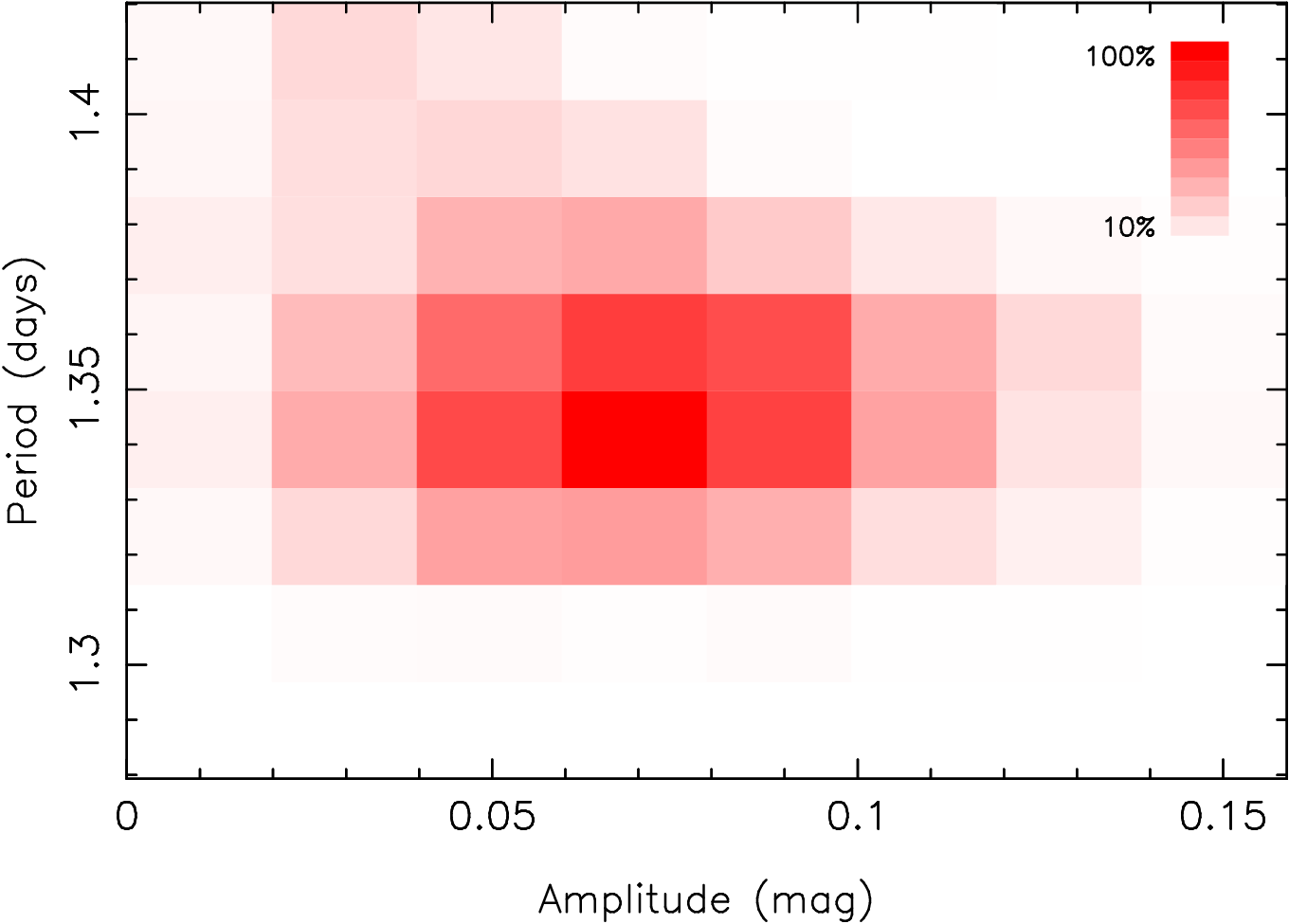}
\includegraphics[angle=270,width=0.24\textwidth,clip]{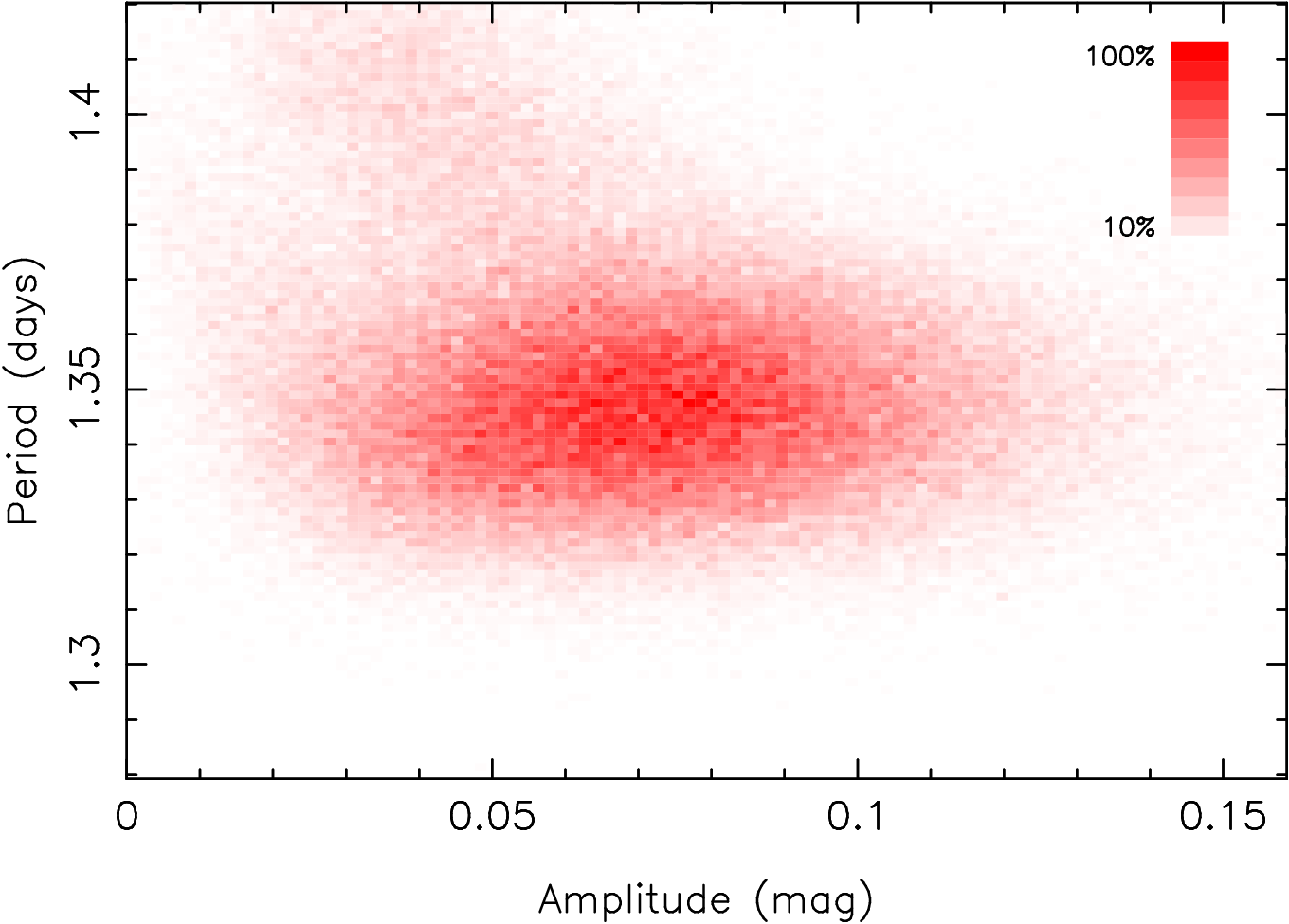}
\includegraphics[angle=270,width=0.24\textwidth,clip]{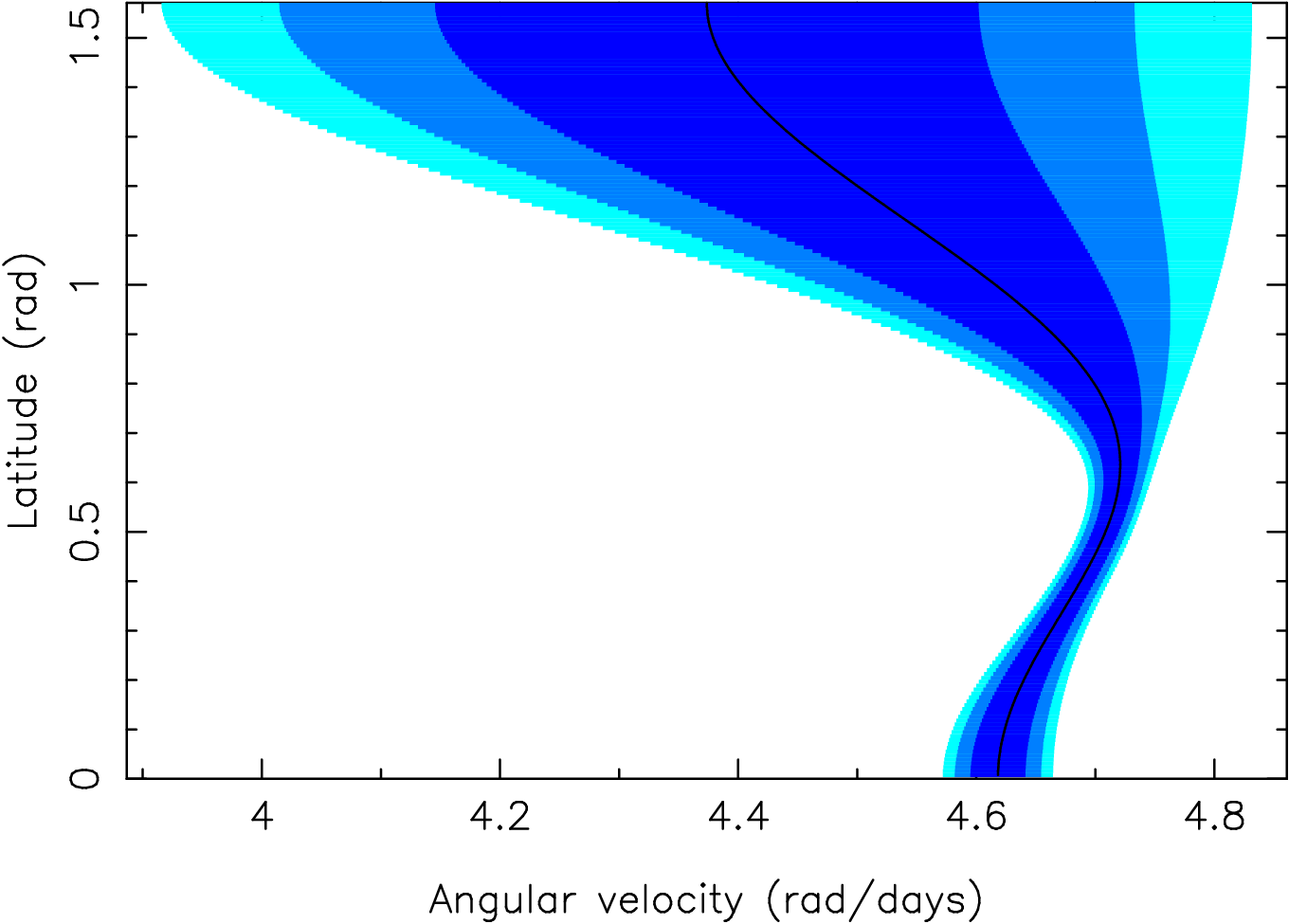}
\caption{Data of V889 Her and maximum \emph{a posteriori} model for $T_{\rm s} = 15$ days and for grids with $n=6$ (top row) and $n=8$ (bottom row). The model has $k=2$ and uniform latitude distribution for spots. The first plot on the left hand side represents data, second one represents the corresponding statistical model, third plot shows a high-precision model in a 100$\times$100 grid, and the last plot visualises the corresponding differential rotation curve with 1, 2, and 3-$\sigma$ uncertainties (dark to light blue). The high-resolution version of the model is added to enable visual inspection of the underlying distribution. The shades of red represent the distribution with respect to its maximum.}\label{fig:plots_15_V889H}
\end{figure*}

\begin{figure*}
\center
\includegraphics[angle=270,width=0.24\textwidth,clip]{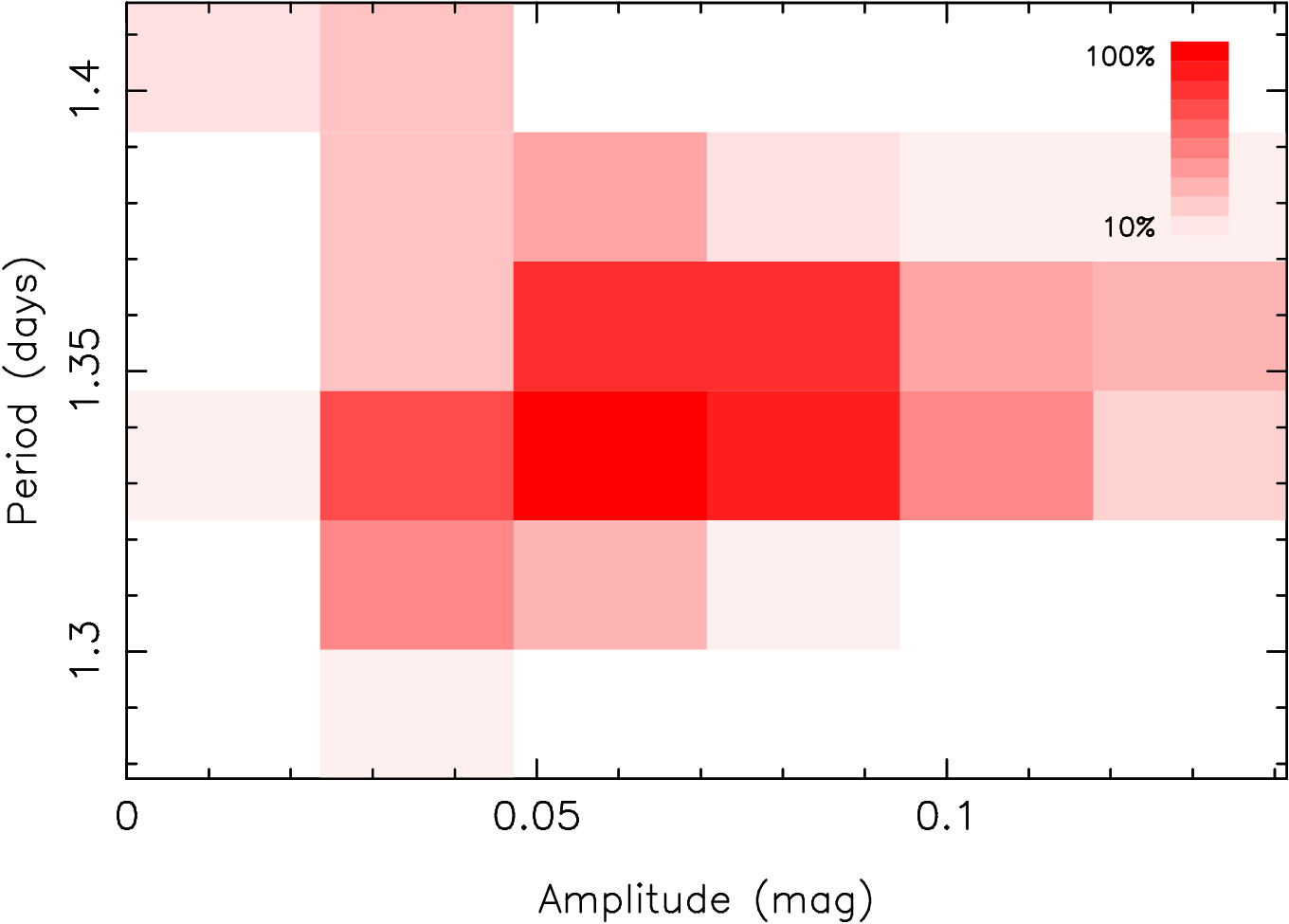}
\includegraphics[angle=270,width=0.24\textwidth,clip]{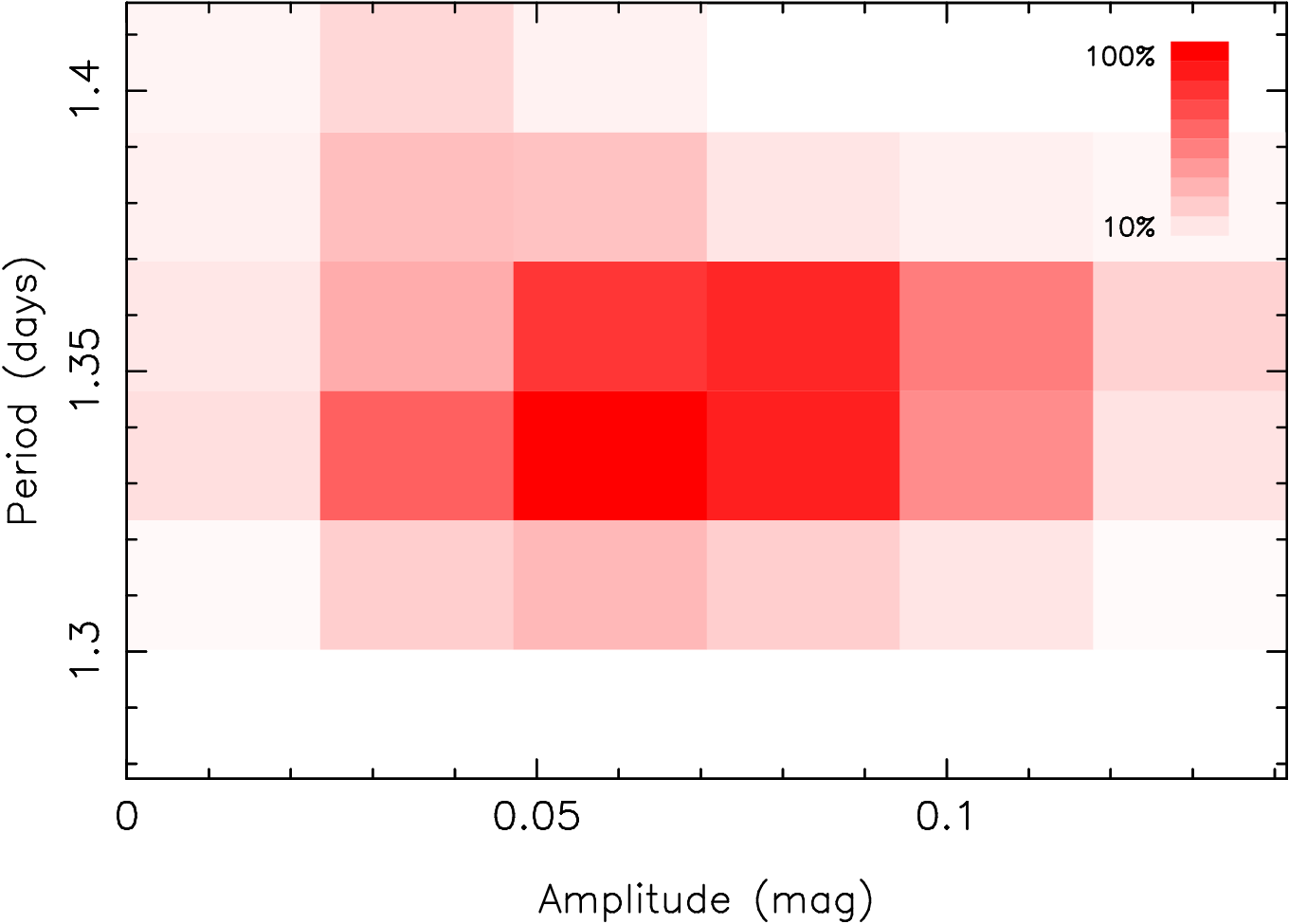}
\includegraphics[angle=270,width=0.24\textwidth,clip]{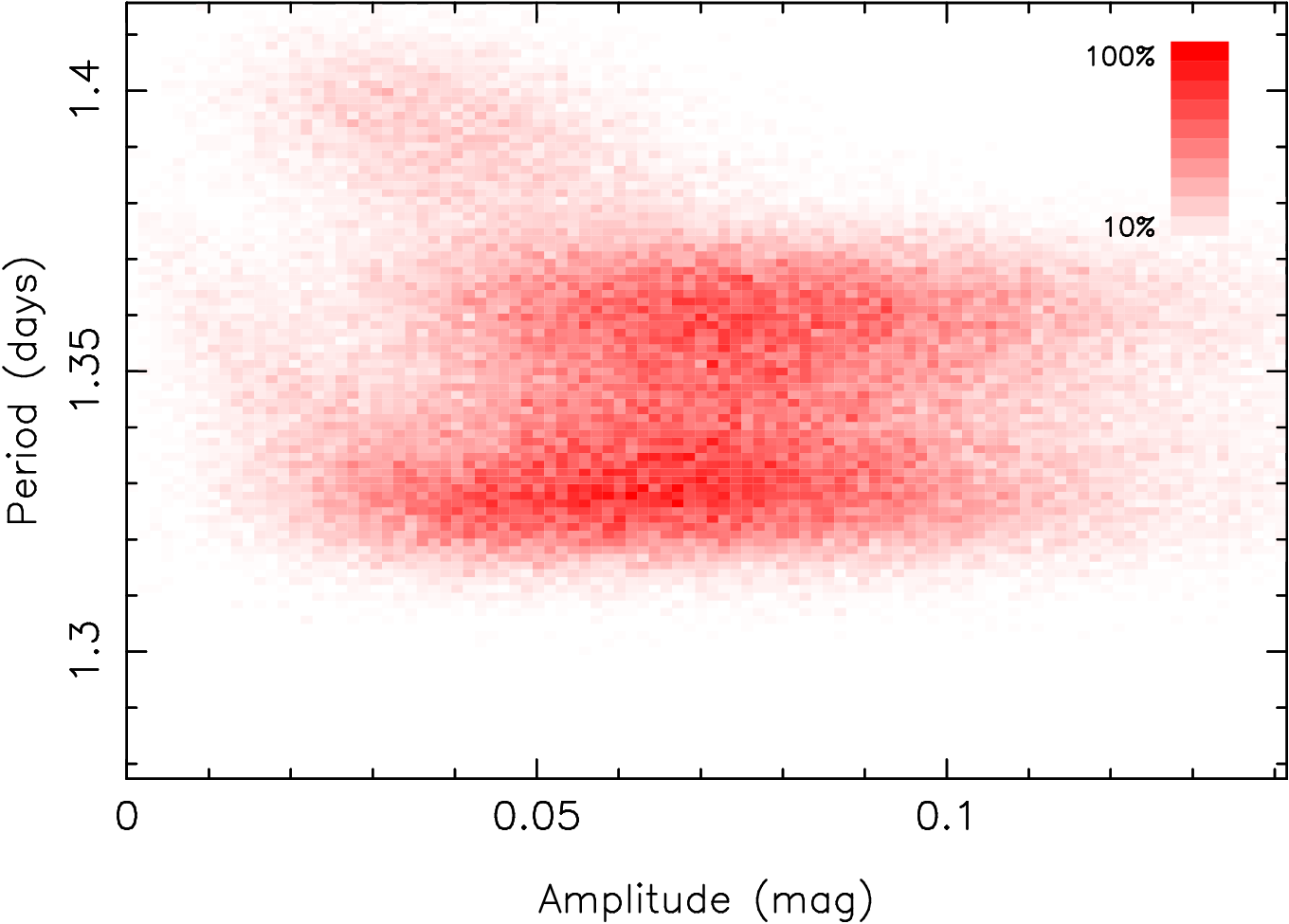}
\includegraphics[angle=270,width=0.24\textwidth,clip]{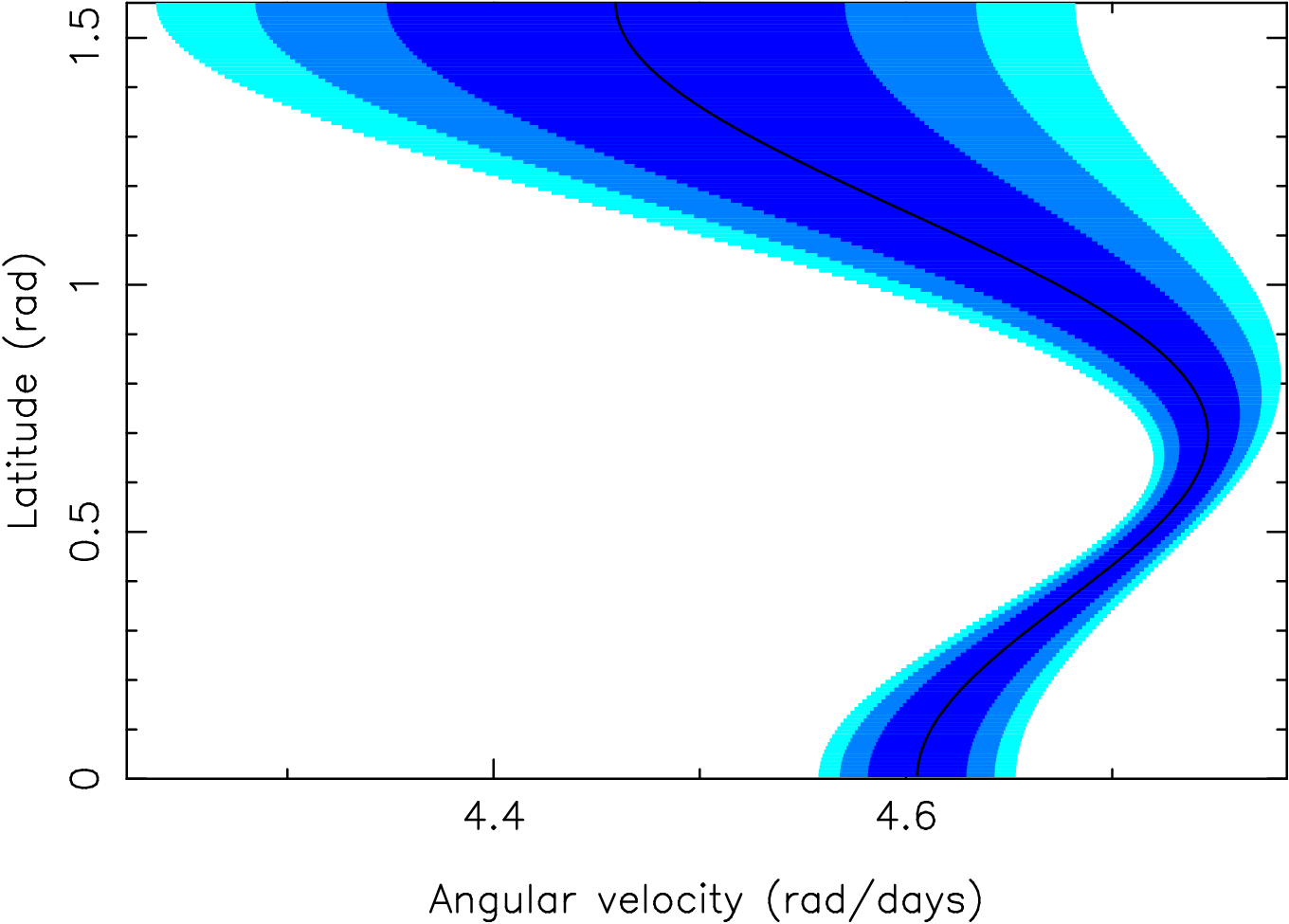}

\includegraphics[angle=270,width=0.24\textwidth,clip]{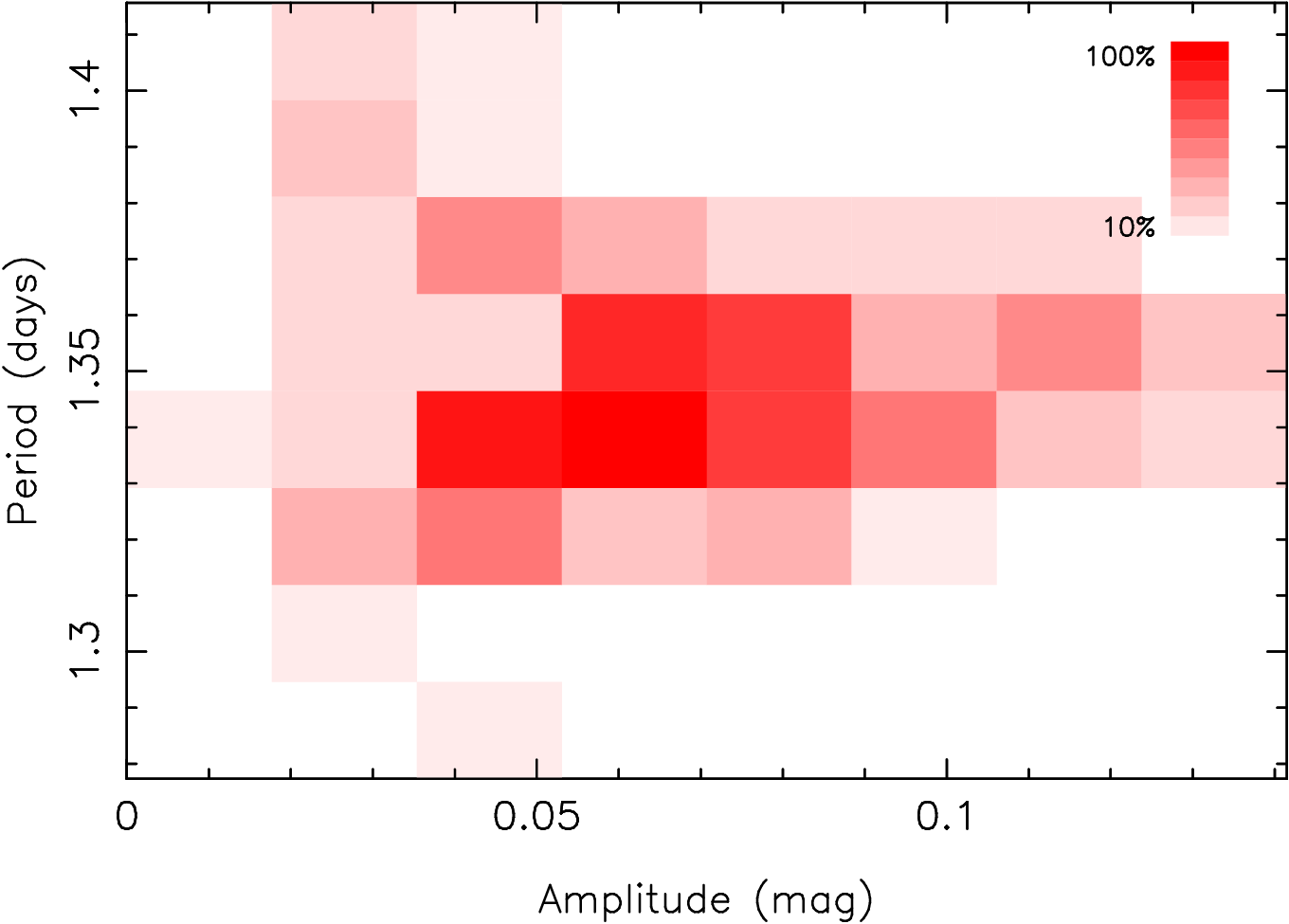}
\includegraphics[angle=270,width=0.24\textwidth,clip]{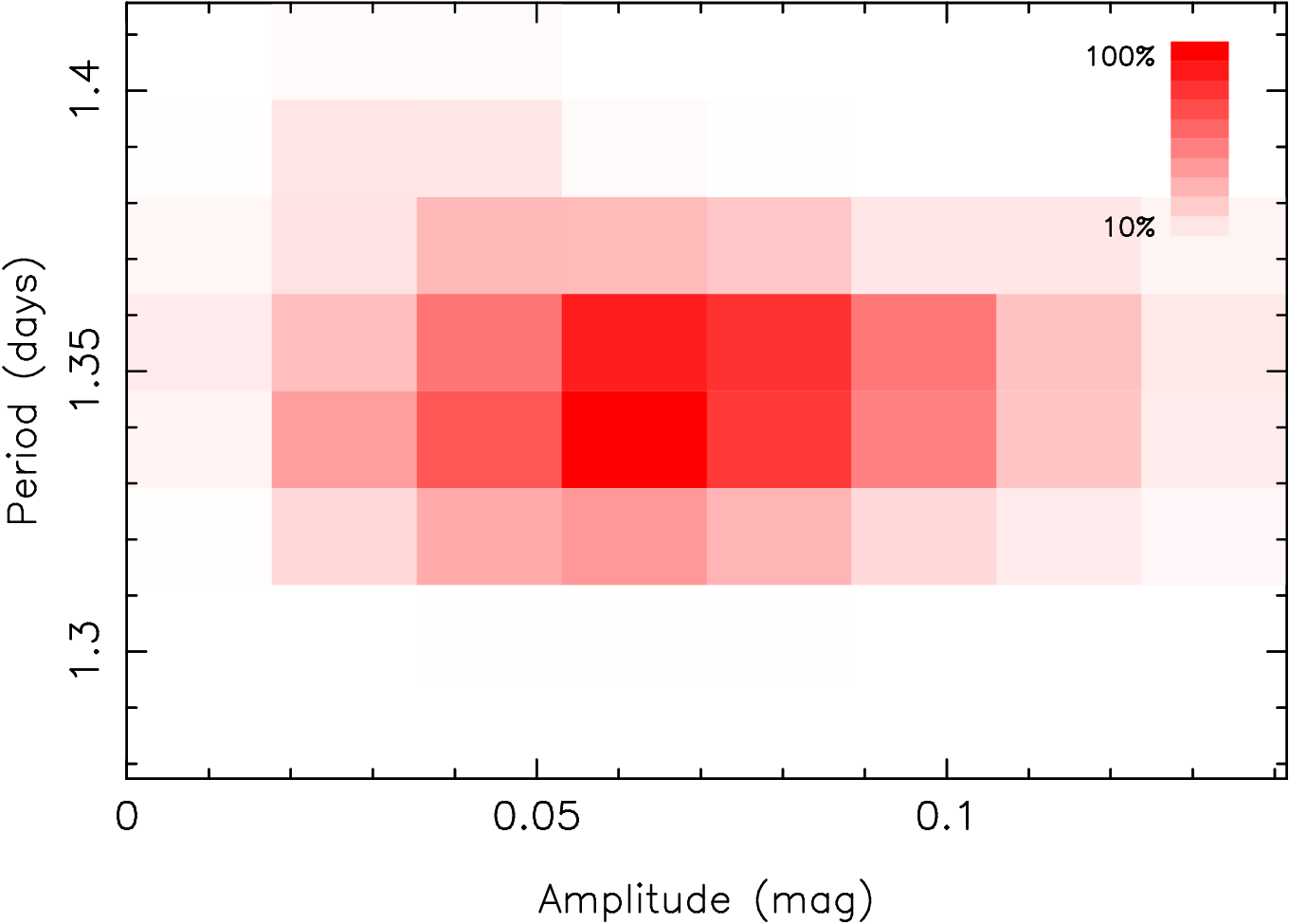}
\includegraphics[angle=270,width=0.24\textwidth,clip]{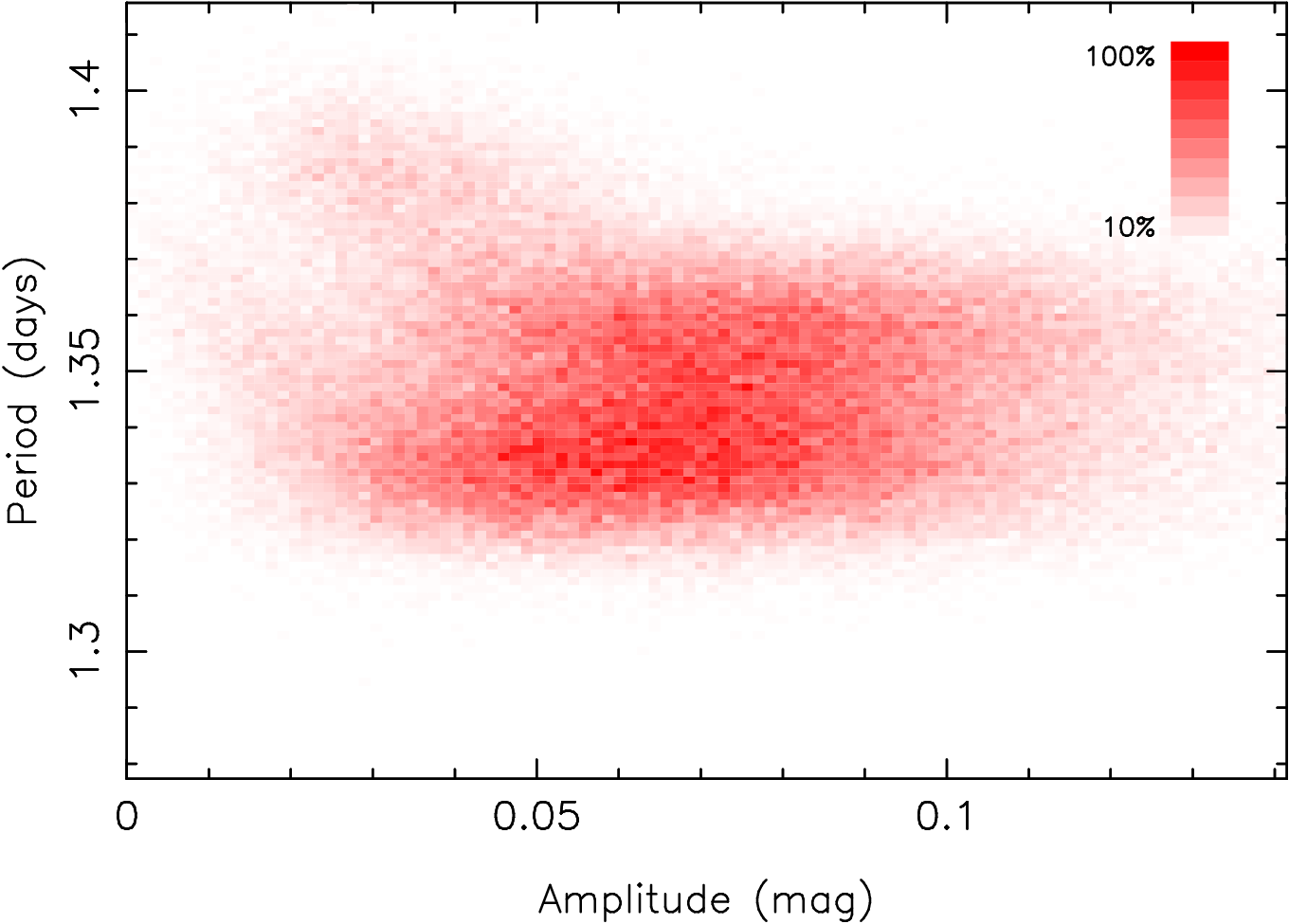}
\includegraphics[angle=270,width=0.24\textwidth,clip]{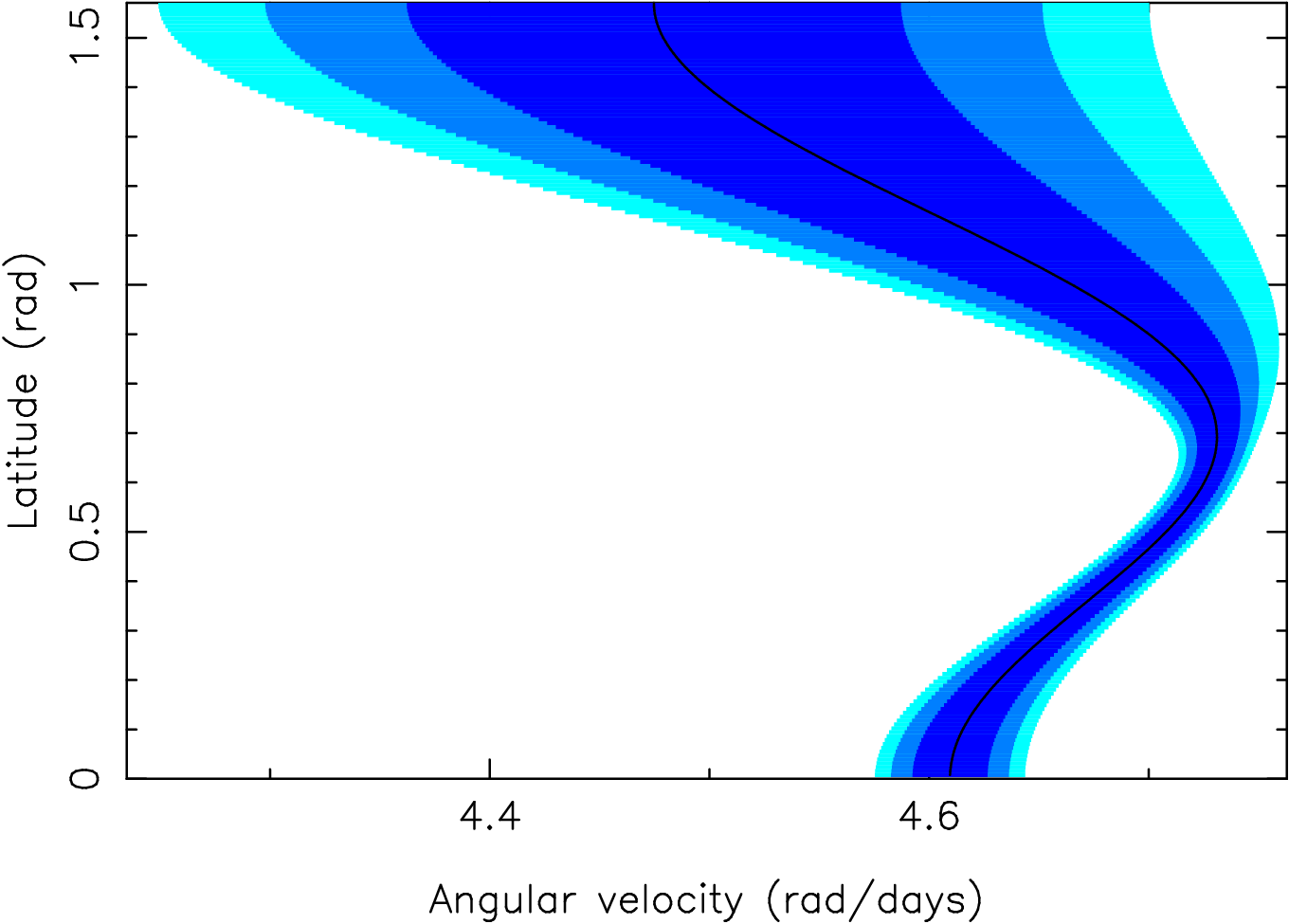}
\caption{Same as in Fig. \ref{fig:plots_15_V889H} but for $T_{\rm s} = 20$ days.}\label{fig:plots_20_V889H}
\end{figure*}

Our model comparison results (Table \ref{tab:model_comparison_V889H}) indicate that a differential rotation curve with $k=2$ is needed to explain the variations in the data. This implies that the differential rotation curve has both a maximum and a minimum that differ from the rotation rate at the equator. We obtain consistent results for a variety of grid sizes and photometric subset baselines, which increases the robustness of our solutions (Figs. \ref{fig:plots_15_V889H} and \ref{fig:plots_20_V889H}). Although a model with monotonic differential rotation and $k=1$ is not preferred, we have presented the corresponding modelled distributions and differential rotation curves in Fig. \ref{fig:plots_1520_V889H_D1} for reference.

As can be seen in Figs. \ref{fig:plots_15_V889H} and \ref{fig:plots_20_V889H}, the modelled distribution has structures that can give rise to the broadening observed in the distribution of periods as a function of decreasing signal amplitudes. Although the grid size introduces limitations, the explanation of this "broadening" as a simple geometric effect arising from differential rotation can be considered viable. It is not possible to produce such effects without differential rotation, or with only the 1st order term in Eq. (\ref{eq:angular_velocity}). Therefore, such broadening indicates that more complex dynamics is required and more complex differential rotation seems to be a plausible choice. Although the data seems to have a more symmetric distribution at low amplitudes, this can be only an apparent effect caused by the fact that there is a limited number of amplitude-period -pairs in each grid cell.

\begin{table}
\caption{Model comparison statistics.}\label{tab:model_comparison_V889H}
\begin{center}
\begin{tabular}{lccccccccc}
\hline \hline 
$k$ & $n$ & $T_{\rm s}$ & Probability & Probability \\
 & & (days) & V889 Her & LQ Hya \\
\hline
0 & 6 & 15 & 2.1$\times 10^{-4}$ & 0.810 \\
1 & 6 & 15 & 1.6$\times 10^{-4}$ & 0.173 \\
2 & 6 & 15 & 1.000 & 0.017 \\
\hline
0 & 8 & 15 & 8.9$\times 10^{-3}$ & 0.841 \\
1 & 8 & 15 & 0.066 & 0.138 \\
2 & 8 & 15 & 0.925 & 0.021 \\
\hline
0 & 10 & 15 & 1.2$\times 10^{-4}$ & 0.288 \\
1 & 10 & 15 & 0.658 & 0.128 \\
2 & 10 & 15 & 0.342 & 0.584 \\
\hline
0 & 6 & 20 & 2.1$\times 10^{-5}$ & 0.849 \\
1 & 6 & 20 & 2.5$\times 10^{-3}$ & 0.140 \\
2 & 6 & 20 & 0.997 & 0.011 \\
\hline
0 & 8 & 20 & 1.1$\times 10^{-4}$ & 0.262 \\
1 & 8 & 20 & 2.5$\times 10^{-3}$ & 0.737 \\
2 & 8 & 20 & 0.997 & 1.1$\times 10^{-3}$ \\
\hline
0 & 10 & 20 & 7.0$\times 10^{-3}$ & 0.874 \\
1 & 10 & 20 & 0.167 & 0.125 \\
2 & 10 & 20 & 0.826 & 9.5$\times 10^{-4}$ \\
\hline \hline
\end{tabular}
\tablefoot{Statistics for models with uniform latitude distribution of spots and different degrees of freedom for differential rotation model $k$, different grid sizes $n$ and different timeseries subset baselines $T_{\rm s}$.}
\end{center}
\end{table}

According to our results, the angular velocity of V889 Her increases as a function of latitude until latitudes of around 0.65--0.70 rad (37$^\circ$--40$^\circ$), where a maximum is reached. At higher latitudes the angular velocity decreases to values below the equatorial angular velocity. This non-monotonic differential rotation curve is the preferred one according to the data and given the simple statistical models. It is qualitatively different from the monotonic curve observed, e.g., for the Sun and commonly assumed for other stars. There is only one combination of $n$ and $T_{\rm s}$ for which the non-monotonic model does not have the highest probability (Table \ref{tab:model_comparison_V889H}). In that case the favoured model is one with monotonic increase in the angular velocity towards the poles. We do not consider this model to represent the overall distribution of signals in the period-amplitude space even though it arises for the largest grid size that is limited by at most 11 signals in each grid cell.

We have tabulated some parameters quantifying the behaviour of the differential rotation curve of V889 Her in Table \ref{tab:parameters}. Parameter $a_{0} = \Omega_{\rm eq}$ represents the equatorial angular velocity, and the other differential rotation coefficients are those in Eq. (\ref{eq:angular_velocity}), describing the change in angular velocity as a function of latitude. We have defined the rotational shear $\Delta \Omega = \Omega_{\rm max} - \Omega_{\rm min}$ as the difference between minimum and maximum angular velocities, respectively, as given by the non-monotonic curve. The parameter $\alpha$ is defined by dividing the rotational shear by equatorial angular velocity such that $\alpha = \frac{\Delta \Omega}{\Omega_{\rm eq}}$. We denote the latitude of the maximum angular velocity as $\psi_{\rm max}$.

\begin{table*}
\caption{Estimates for parameters.}\label{tab:parameters}
\begin{center}
\begin{tabular}{lccccccccc}
\hline \hline 
$n$ & $T_{\rm s}$ & $i$ & $a_{0}$ & $a_{1}$ & $a_{2}$ & $\psi_{\rm max}$ & $\Delta \Omega$ & $\alpha$ \\
 & (days) & (rad) & (rad d$^{-1}$) & (rad d$^{-1}$) & (rad d$^{-1}$) & (rad) & (rad d$^{-1}$) & (--) \\
\hline
6  & 15 & 1.27$\pm$0.11 & 4.608$\pm$0.013 & 0.655$\pm$0.106 & -0.868$\pm$0.247 & 0.673$\pm$0.048 & 0.338$\pm$0.160 & 0.073$\pm$0.035 \\
8  & 15 & 1.34$\pm$0.12 & 4.617$\pm$0.013 & 0.598$\pm$0.113 & -0.862$\pm$0.248 & 0.640$\pm$0.051 & 0.369$\pm$0.157 & 0.080$\pm$0.034 \\
10 & 15 & 1.24$\pm$0.06 & 4.614$\pm$0.008 & 0.579$\pm$0.045 & -0.707$\pm$0.060 & 0.696$\pm$0.020 & 0.246$\pm$0.032 & 0.053$\pm$0.007 \\
6  & 20 & 1.25$\pm$0.07 & 4.617$\pm$0.013 & 0.598$\pm$0.113 & -0.862$\pm$0.248 & 0.640$\pm$0.051 & 0.369$\pm$0.157 & 0.080$\pm$0.034 \\
8  & 20 & 1.33$\pm$0.08 & 4.609$\pm$0.010 & 0.596$\pm$0.063 & -0.731$\pm$0.114 & 0.696$\pm$0.033 & 0.257$\pm$0.068 & 0.056$\pm$0.015 \\
10 & 20 & 1.32$\pm$0.11 & 4.612$\pm$0.011 & 0.612$\pm$0.078 & -0.798$\pm$0.207 & 0.679$\pm$0.052 & 0.306$\pm$0.138 & 0.066$\pm$0.030 \\
\hline \hline
\end{tabular}
\tablefoot{Parameter estimates of the differential rotation model of V889 Her given the $k=2$ model. The uncertainties correspond to standard errors.}
\end{center}
\end{table*}

We obtain consistent estimates for the rotational shear for different choices of $n$ and $T_{\rm s}$ ranging from 0.246 to 0.369 rad d$^{-1}$ (Table \ref{tab:parameters}). We conservatively adopt the median value and the maximal uncertainty and obtain an estimate of 0.322$\pm$0.160 rad d$^{-1}$ for the rotational shear. This estimate corresponds to a differential rotation coefficient, or the relative rotational shear $\alpha$, of 0.070$\pm$0.035. In a similar manner, we estimate that the turnover of the rotational curve occurs at the latitude of 0.676$\pm$0.052 rad (38.7$\pm$3.0$^\circ$).

\subsection{LQ Hya}

\begin{figure*}
\center
\includegraphics[angle=270,width=0.24\textwidth,clip]{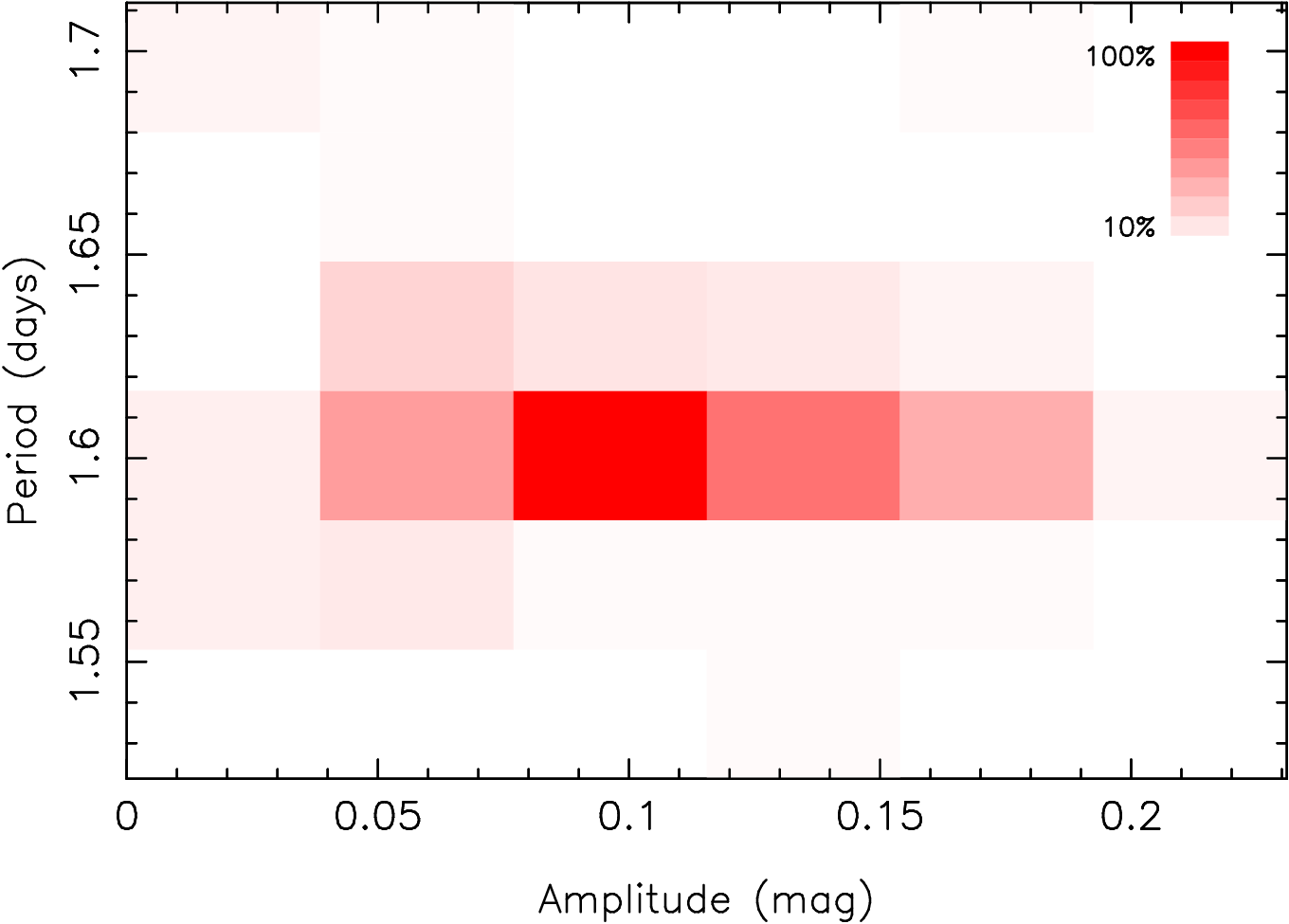}
\includegraphics[angle=270,width=0.24\textwidth,clip]{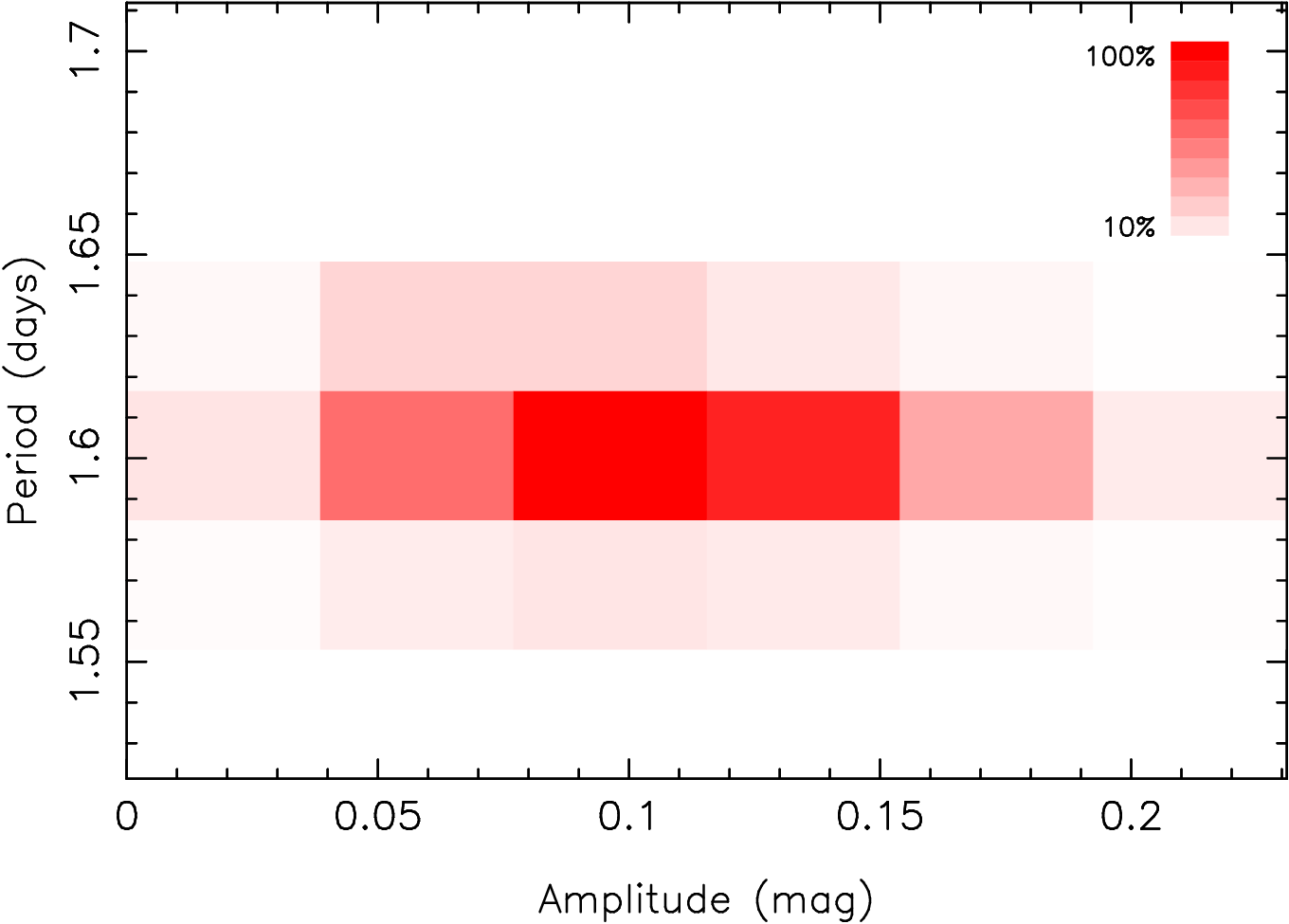}
\includegraphics[angle=270,width=0.24\textwidth,clip]{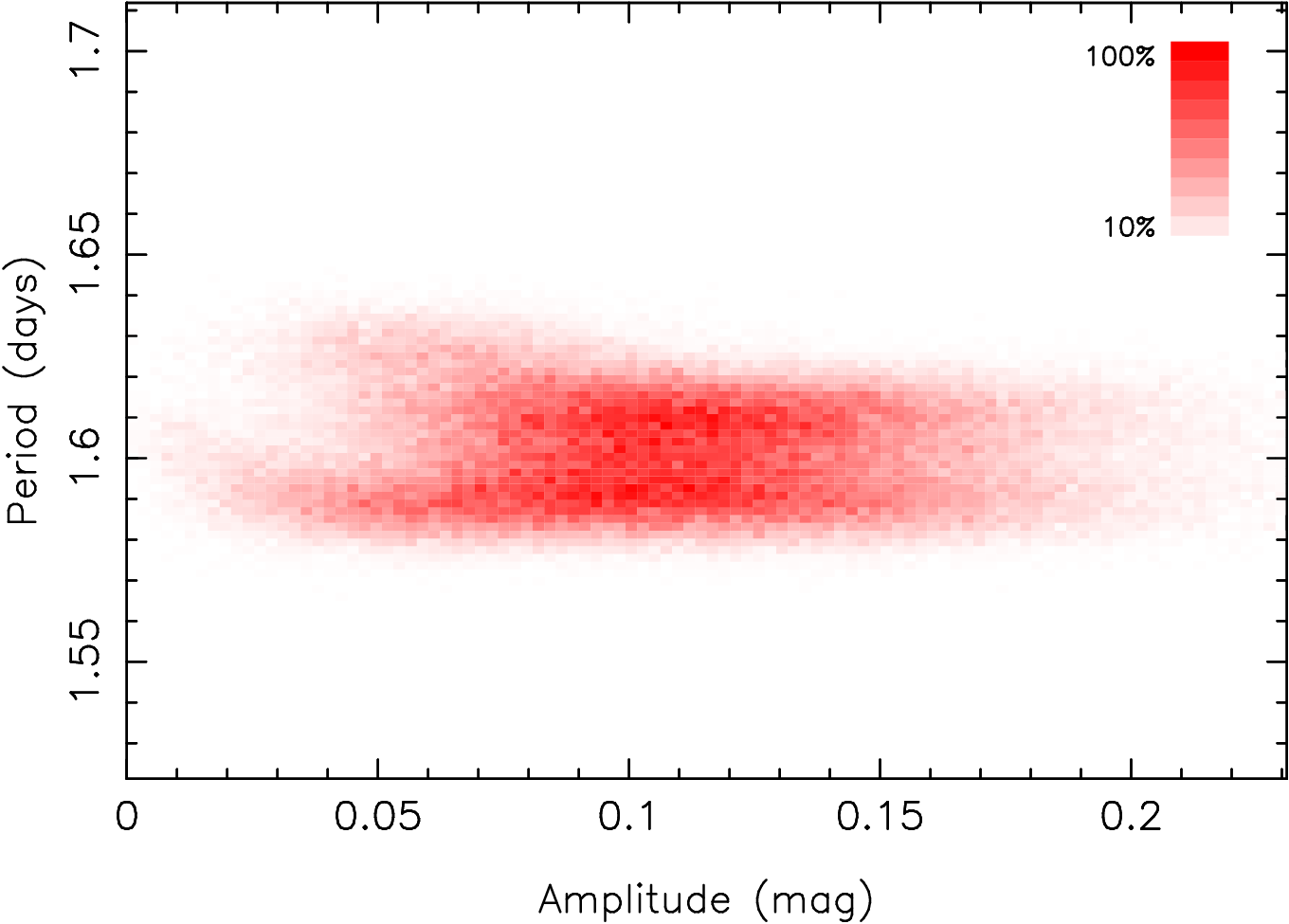}
\includegraphics[angle=270,width=0.24\textwidth,clip]{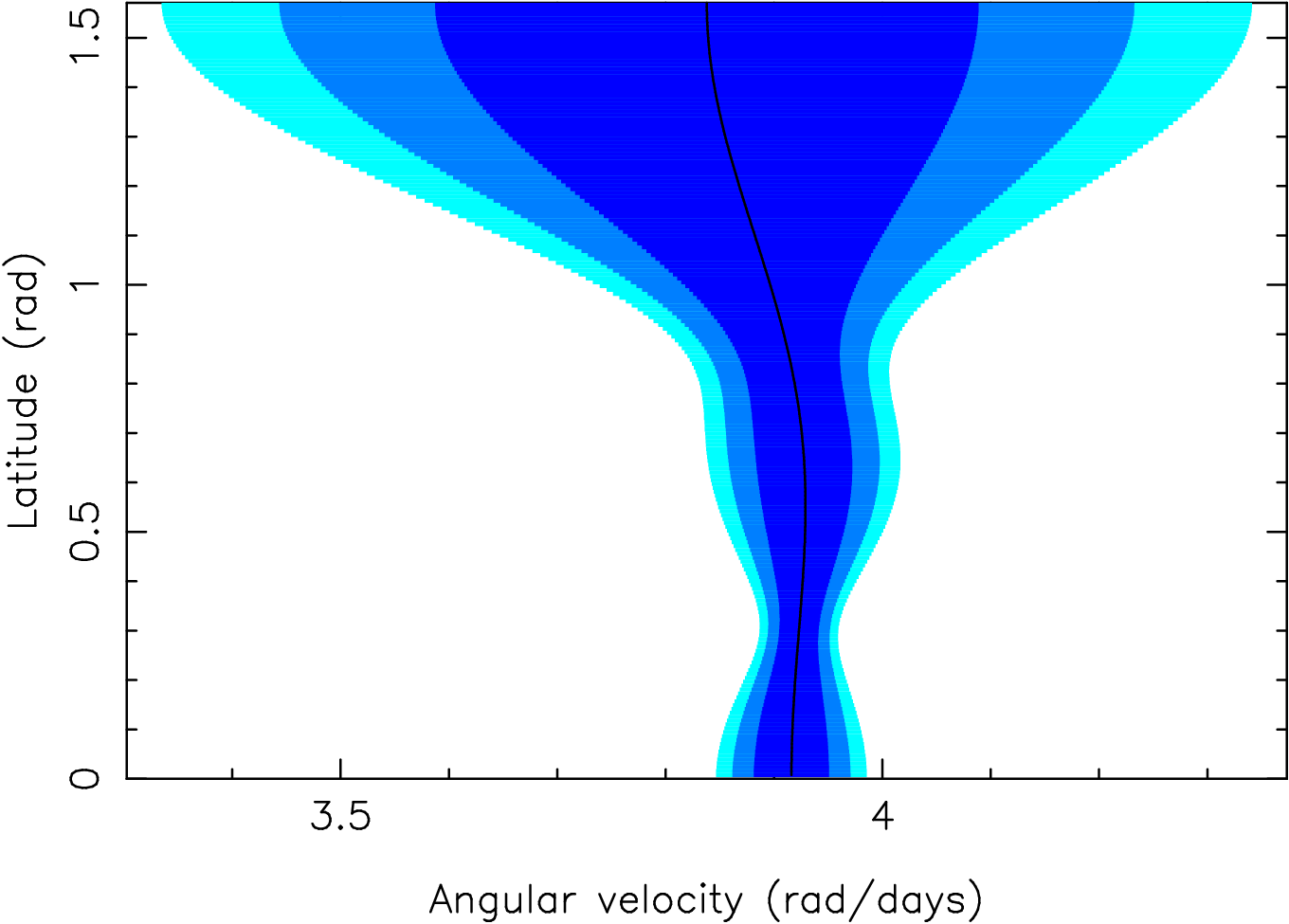}

\includegraphics[angle=270,width=0.24\textwidth,clip]{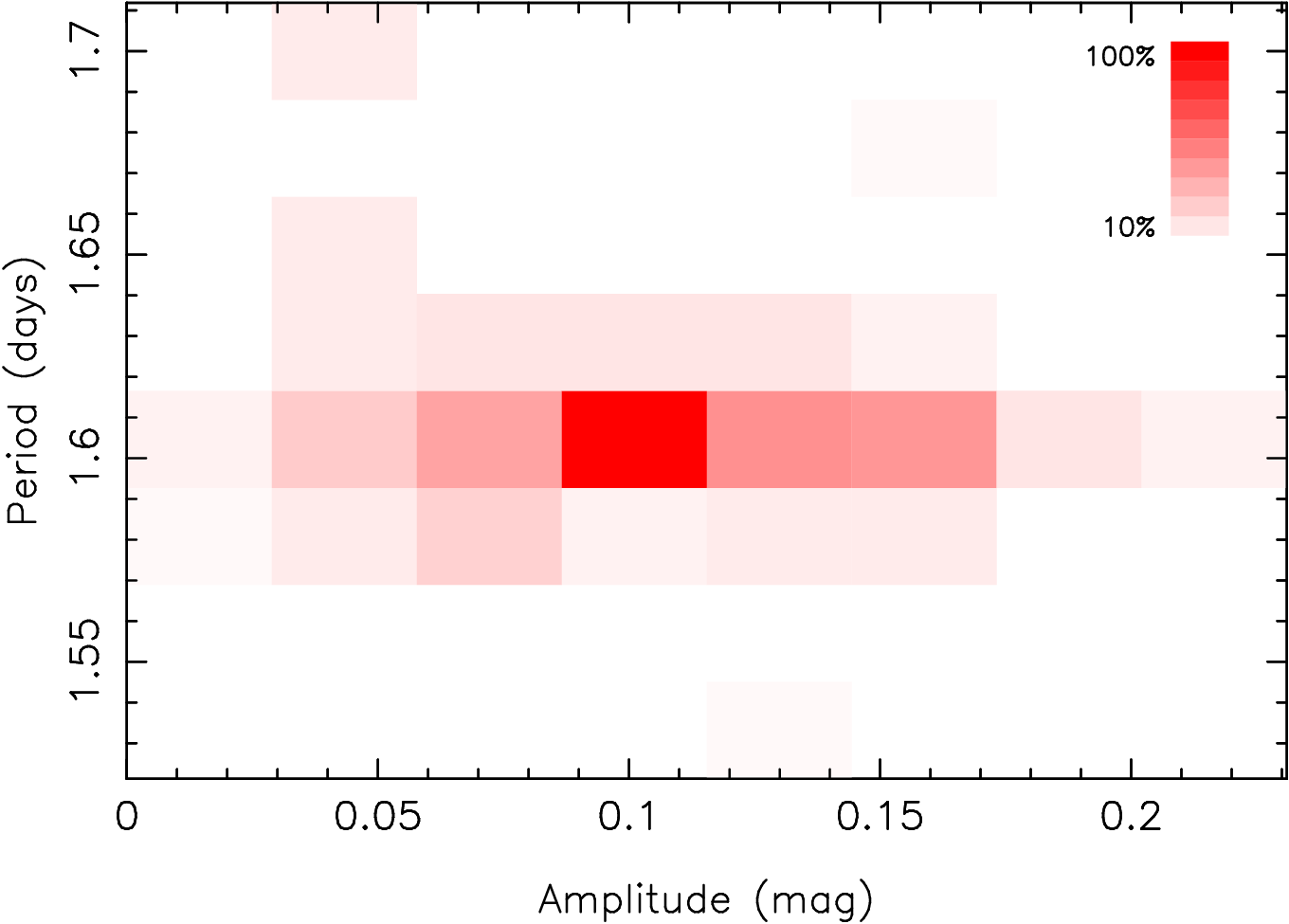}
\includegraphics[angle=270,width=0.24\textwidth,clip]{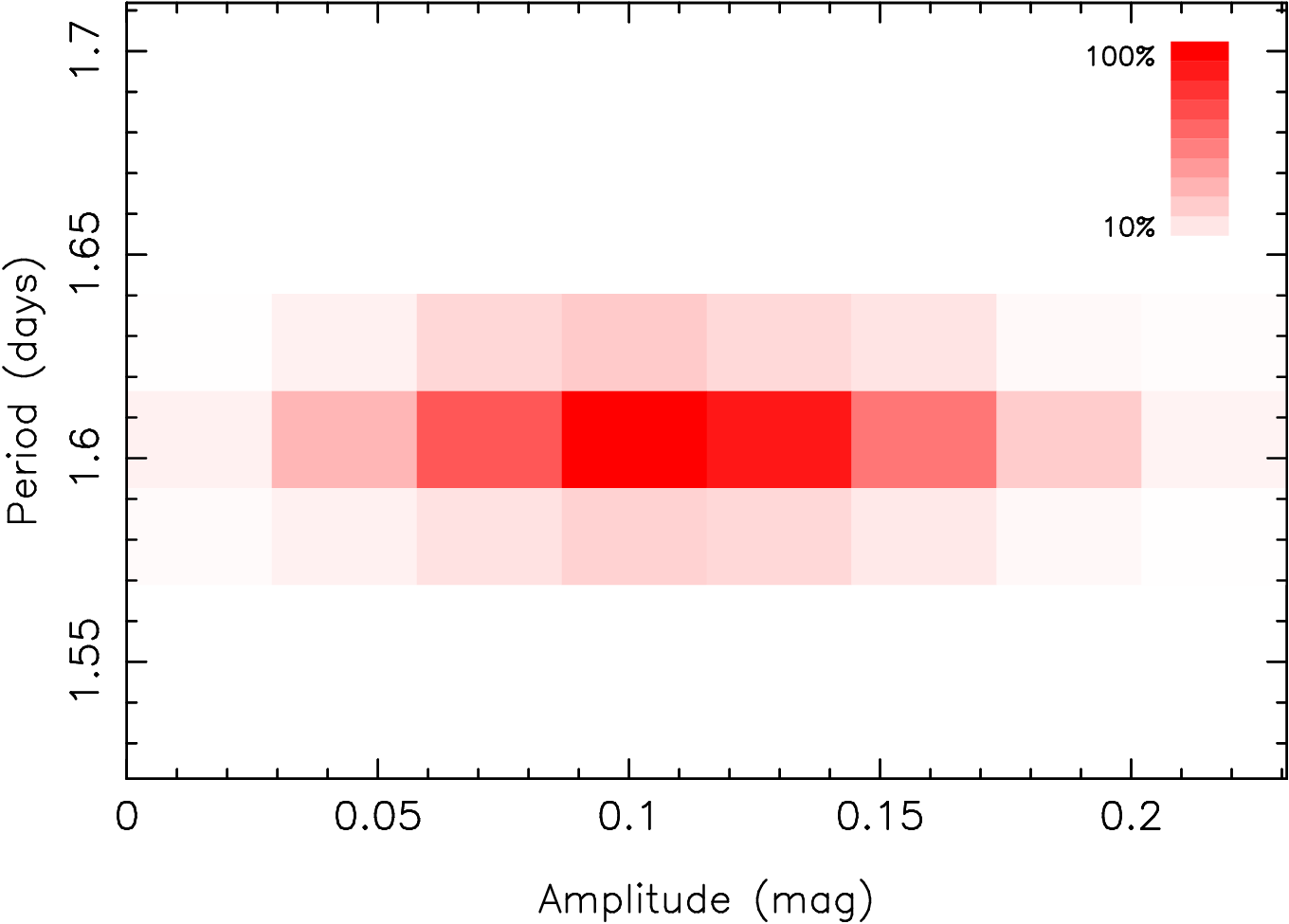}
\includegraphics[angle=270,width=0.24\textwidth,clip]{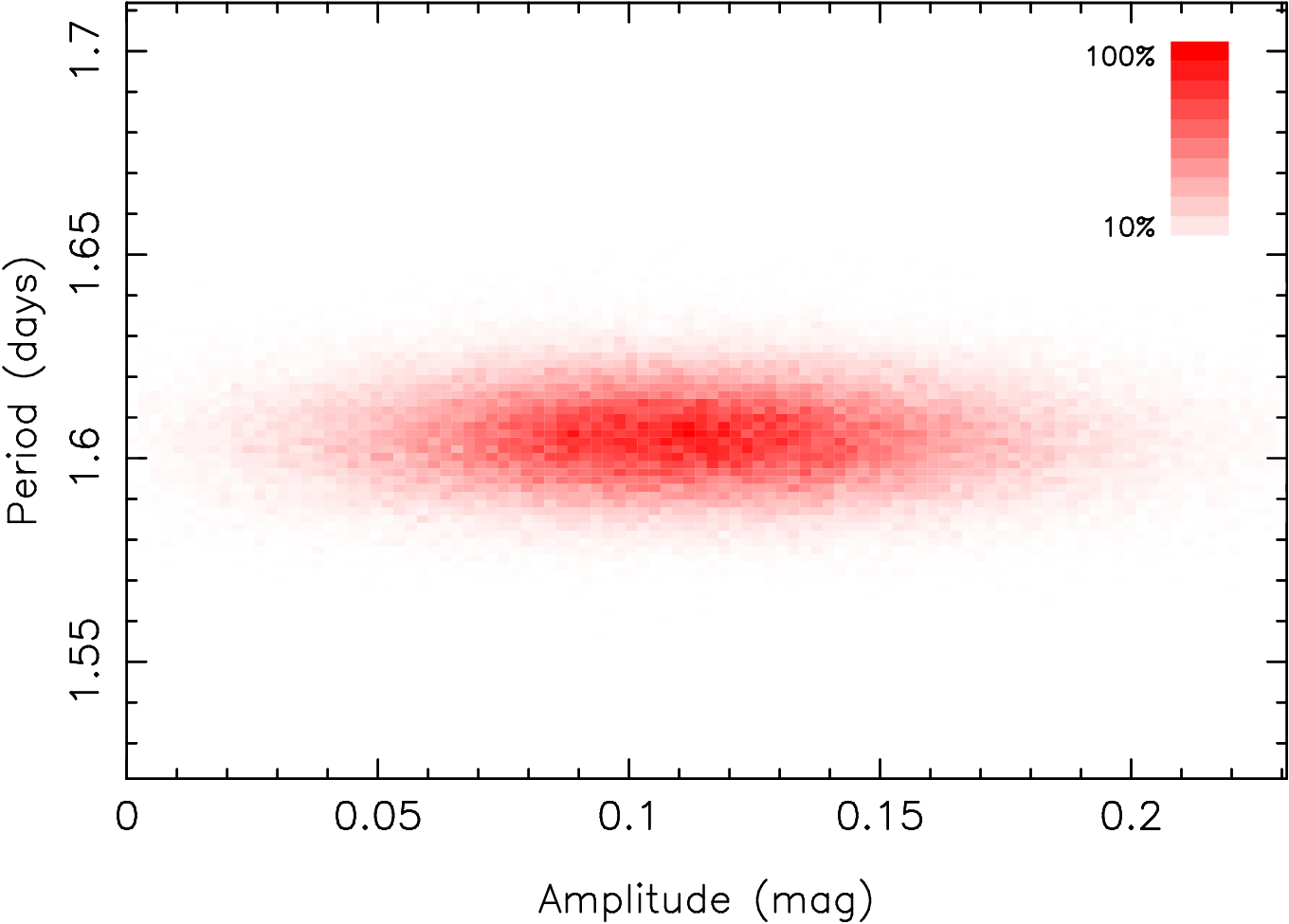}
\includegraphics[angle=270,width=0.24\textwidth,clip]{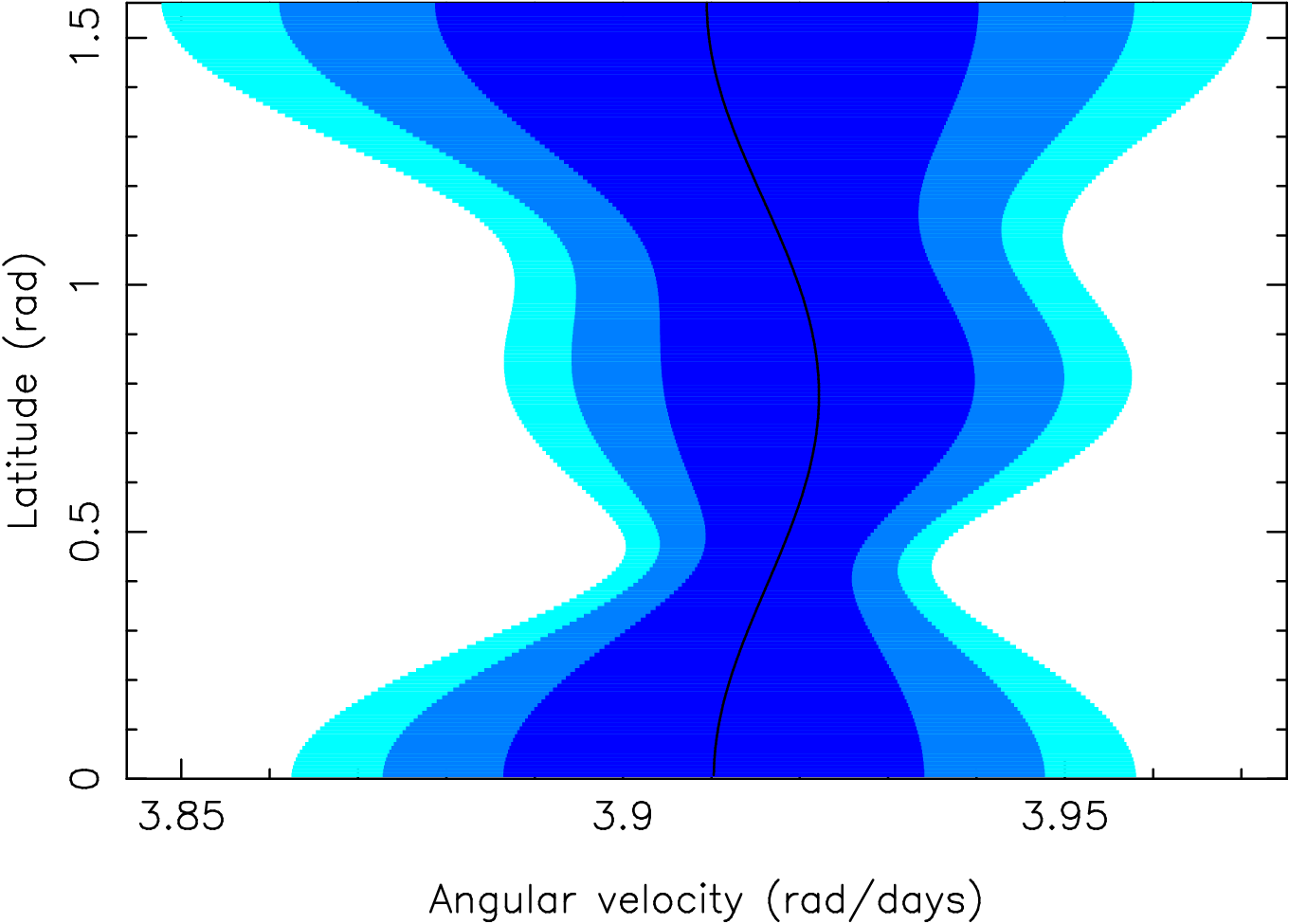}
\caption{Same as in Fig. \ref{fig:plots_15_V889H} but for LQ Hya. The applied differential rotation model is that with $k=2$.}\label{fig:plots_15_LQHYA}
\end{figure*}

\begin{figure*}
\center
\includegraphics[angle=270,width=0.24\textwidth,clip]{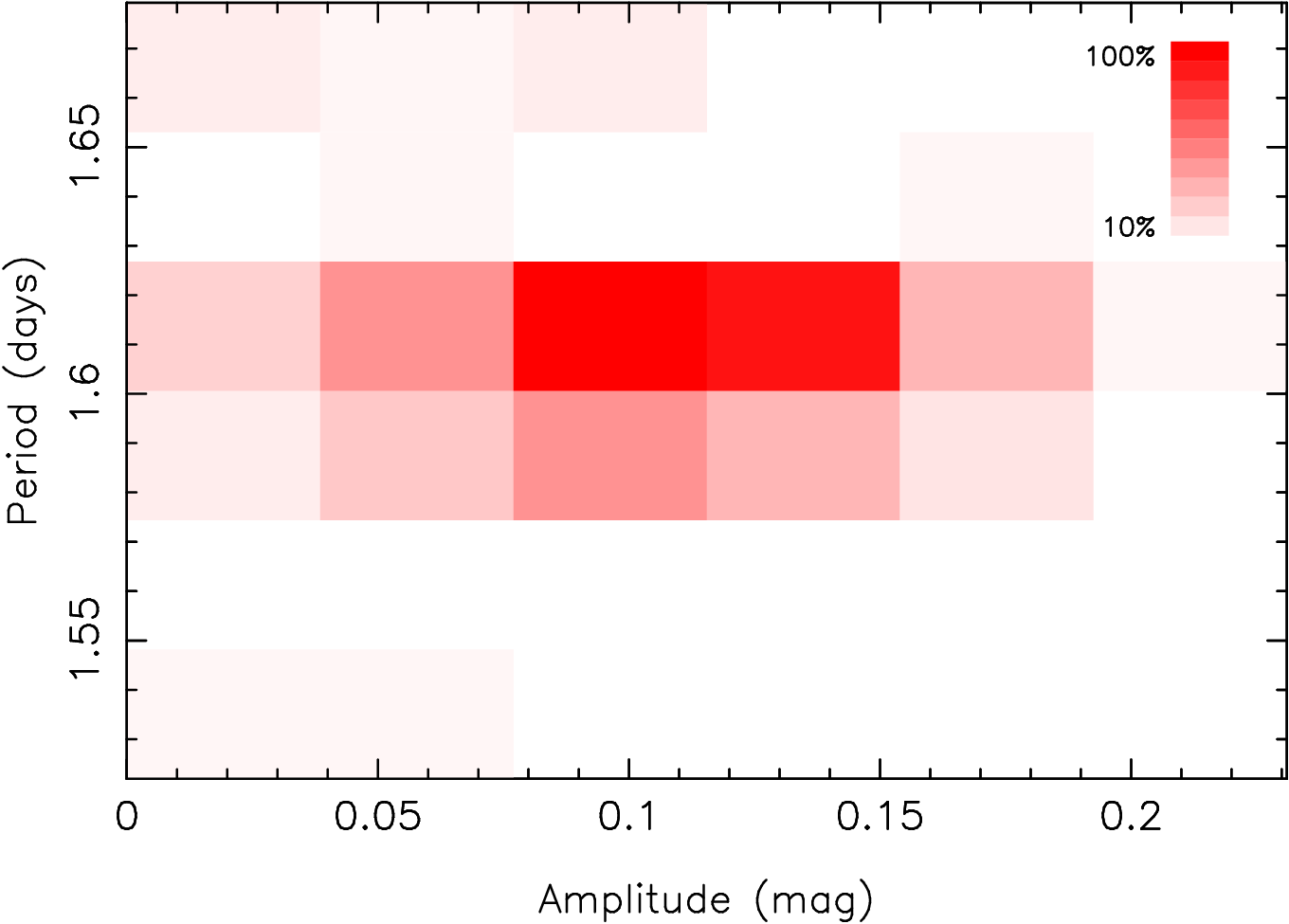}
\includegraphics[angle=270,width=0.24\textwidth,clip]{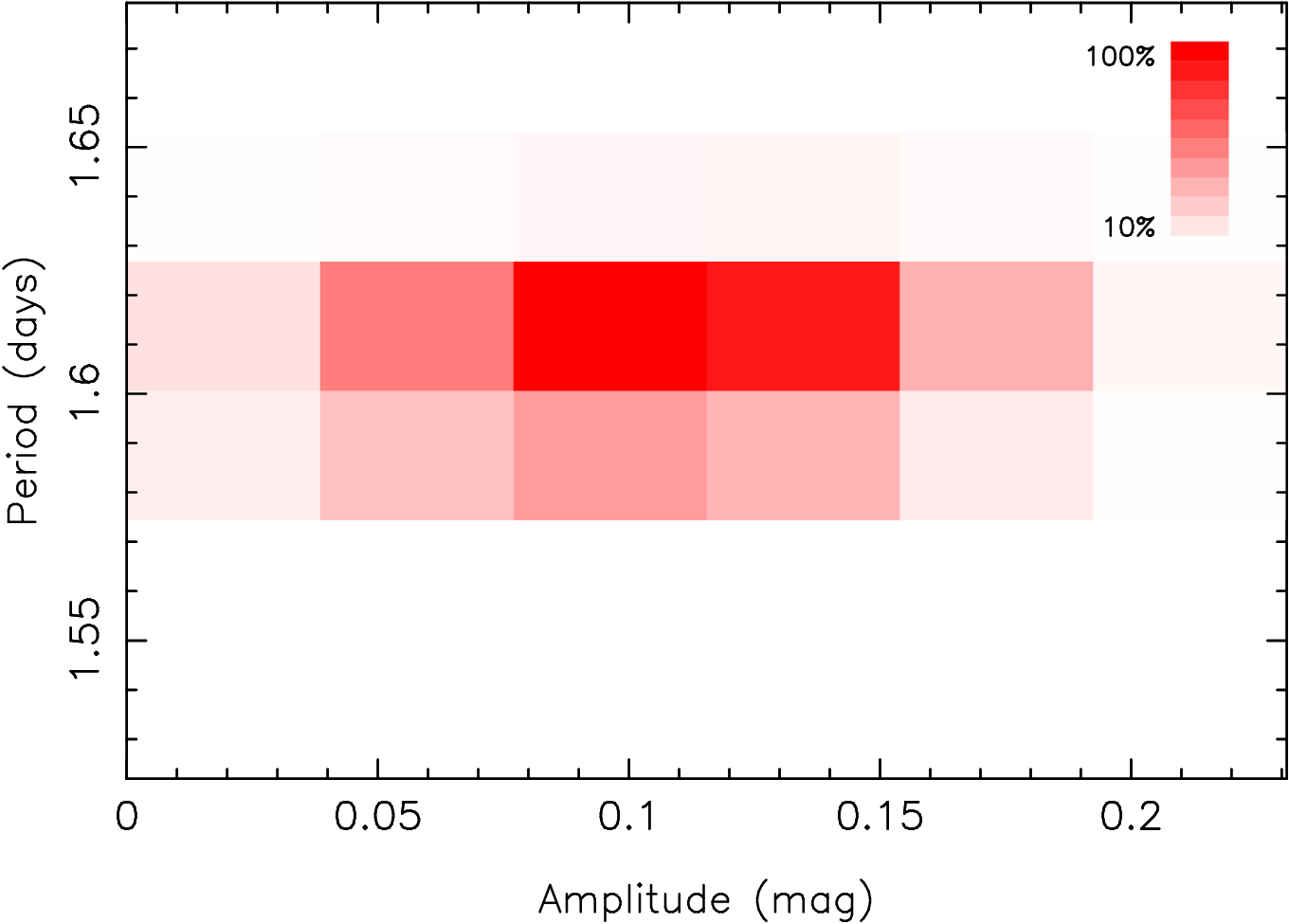}
\includegraphics[angle=270,width=0.24\textwidth,clip]{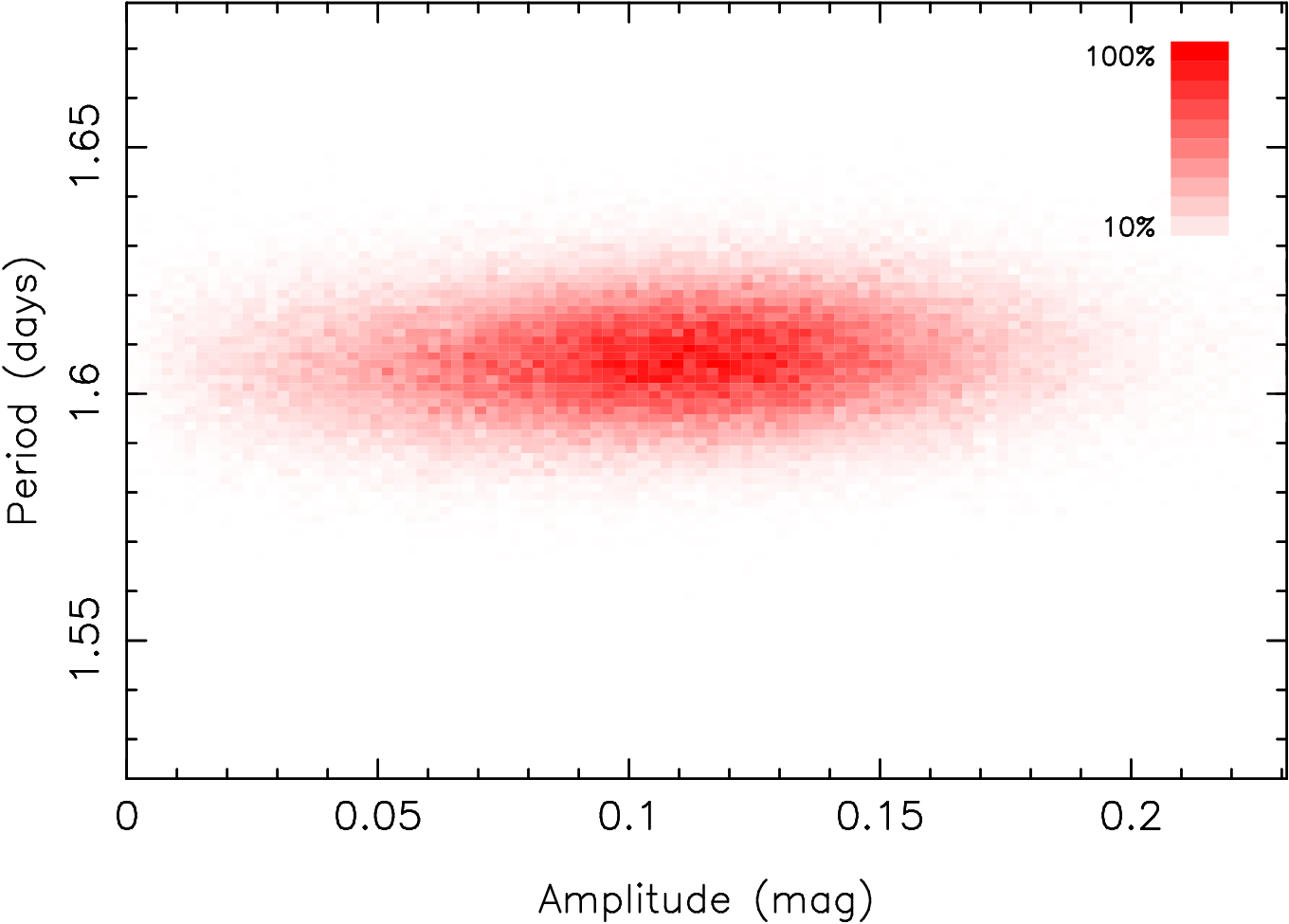}
\includegraphics[angle=270,width=0.24\textwidth,clip]{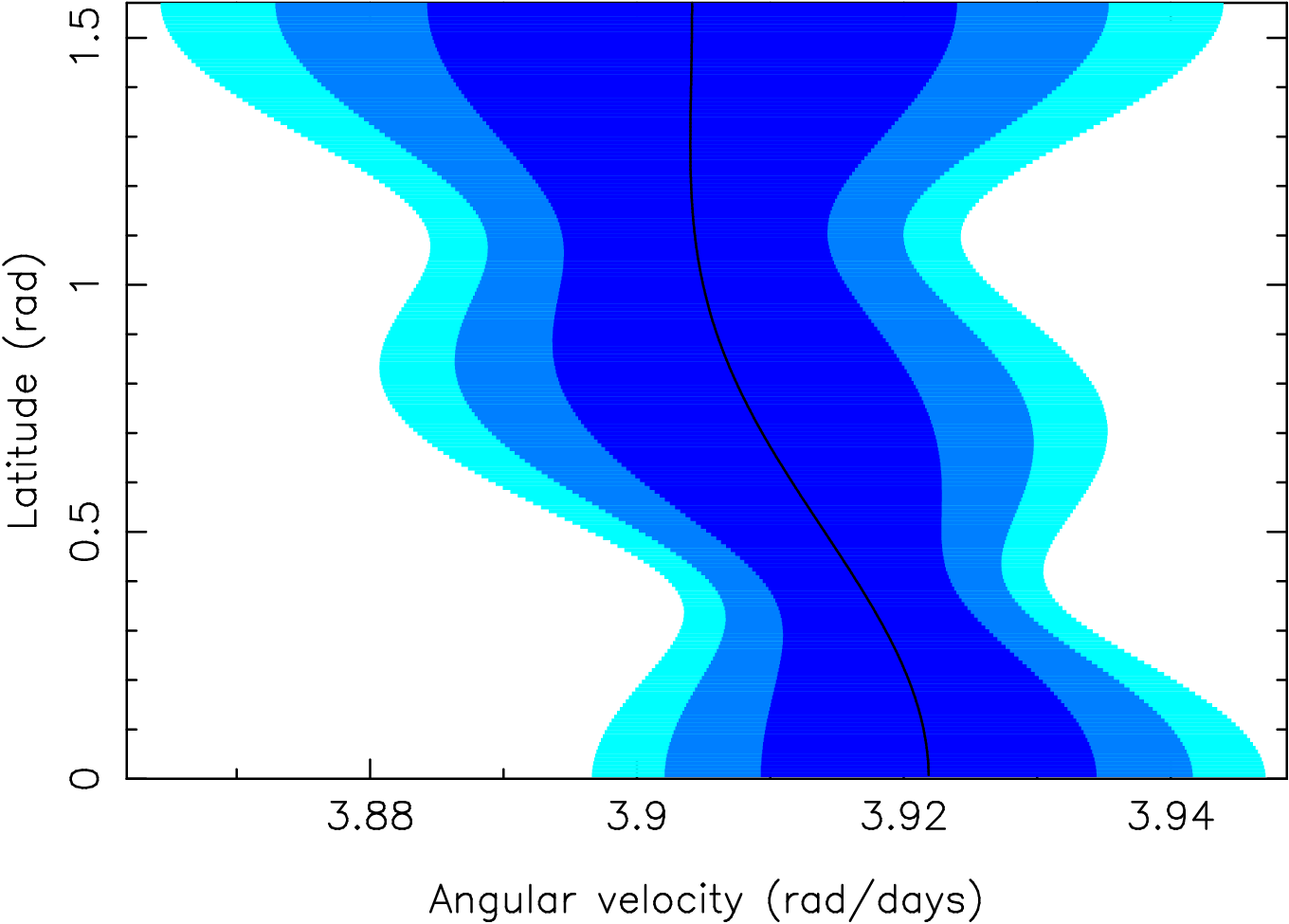}

\includegraphics[angle=270,width=0.24\textwidth,clip]{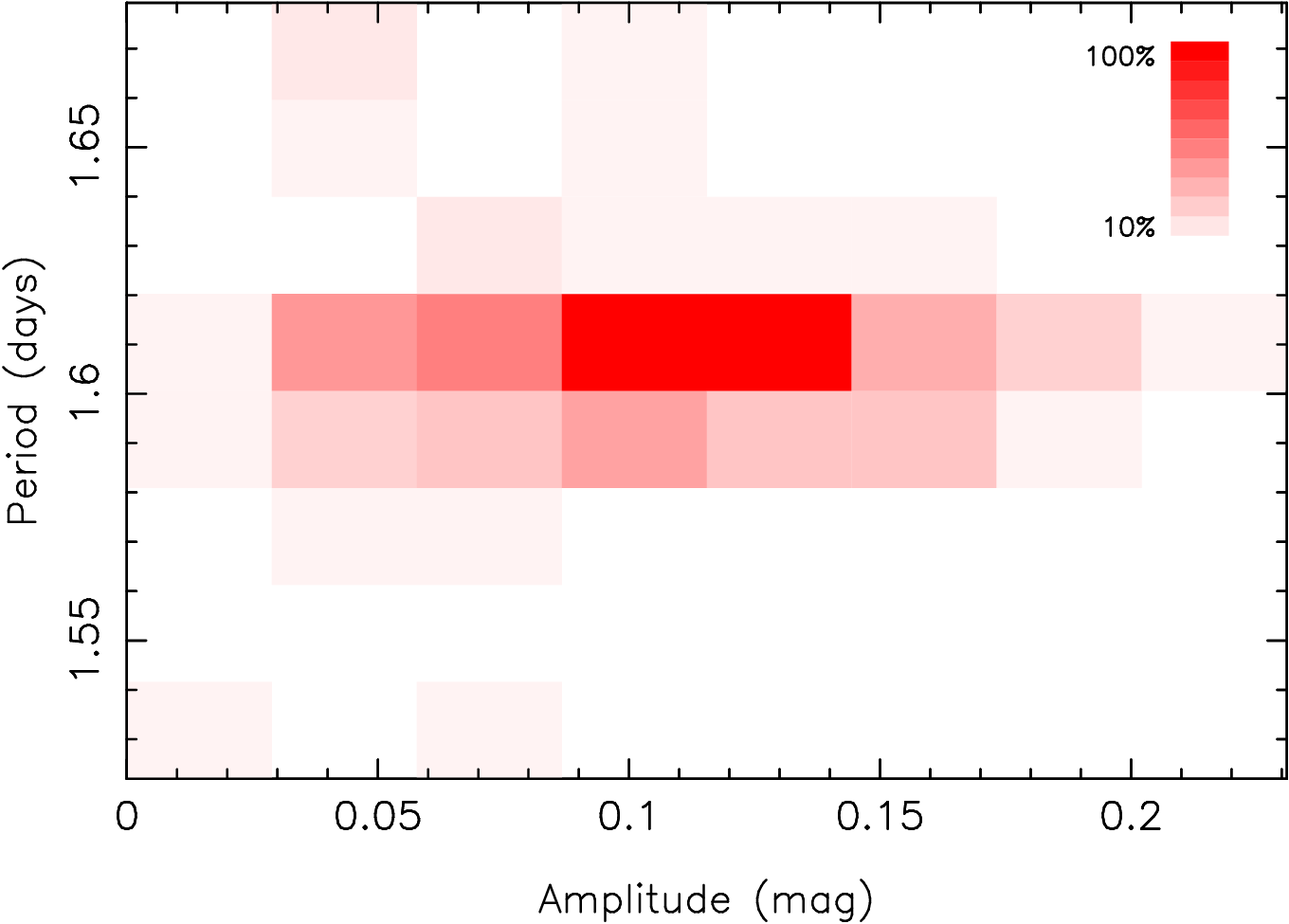}
\includegraphics[angle=270,width=0.24\textwidth,clip]{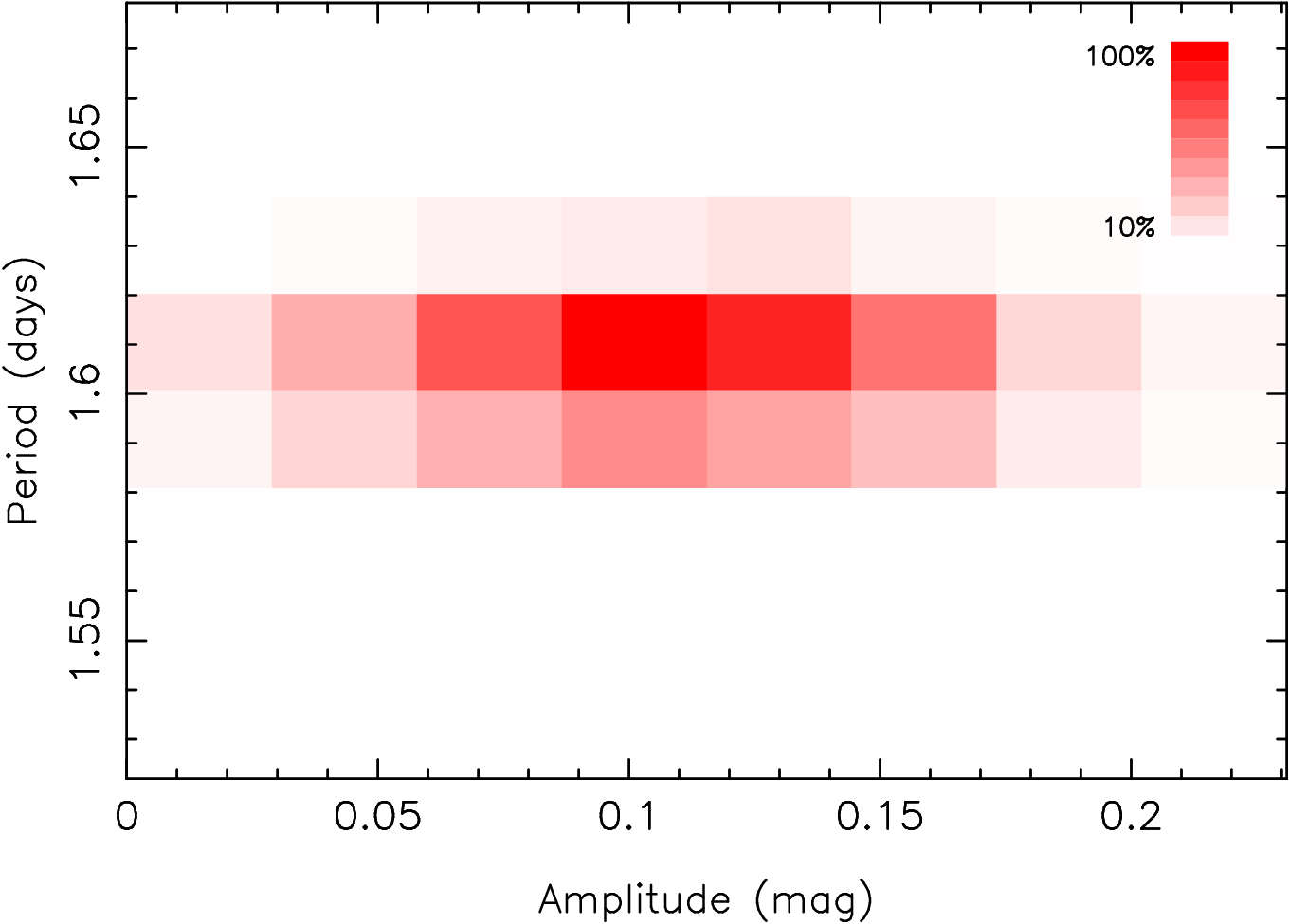}
\includegraphics[angle=270,width=0.24\textwidth,clip]{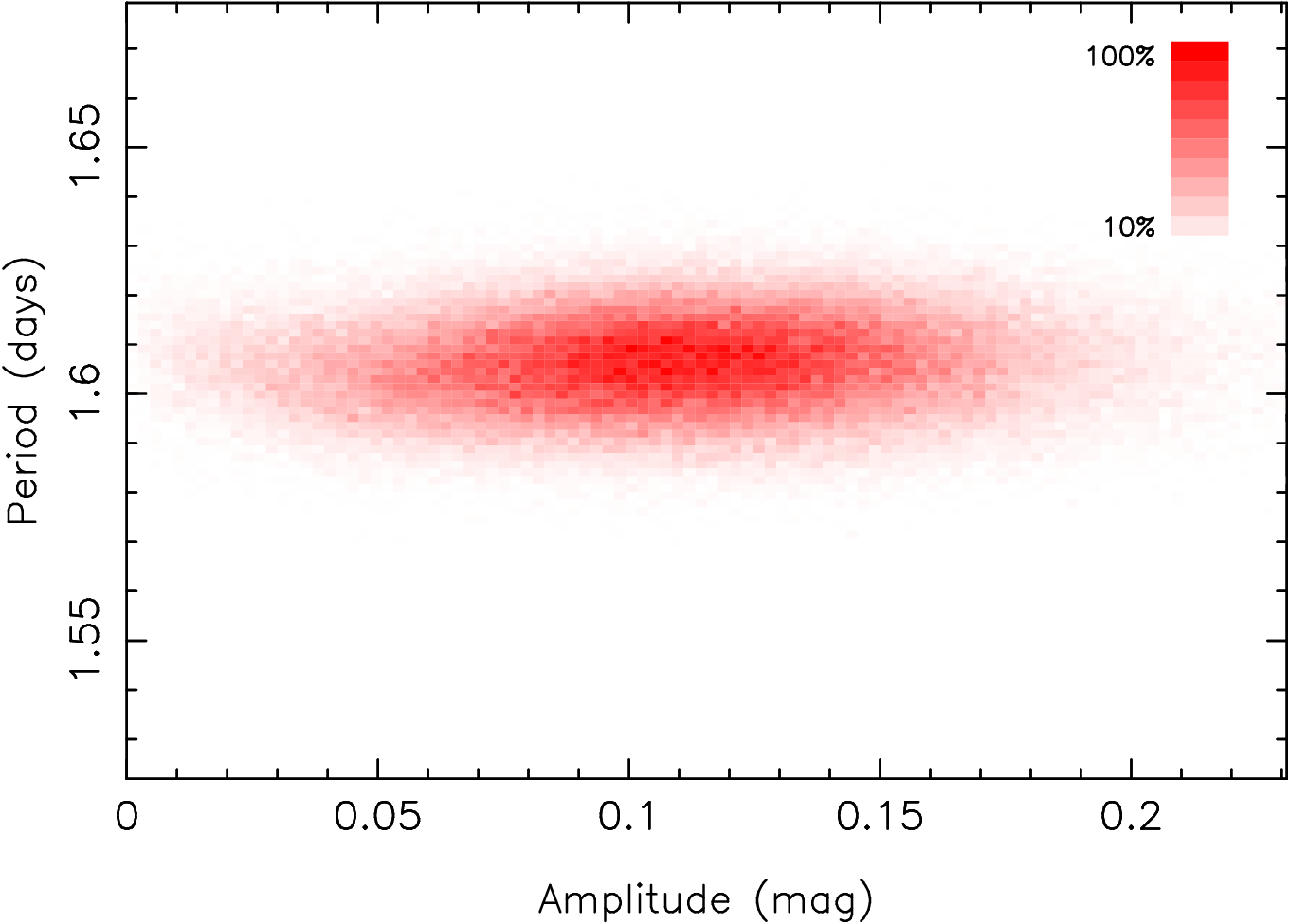}
\includegraphics[angle=270,width=0.24\textwidth,clip]{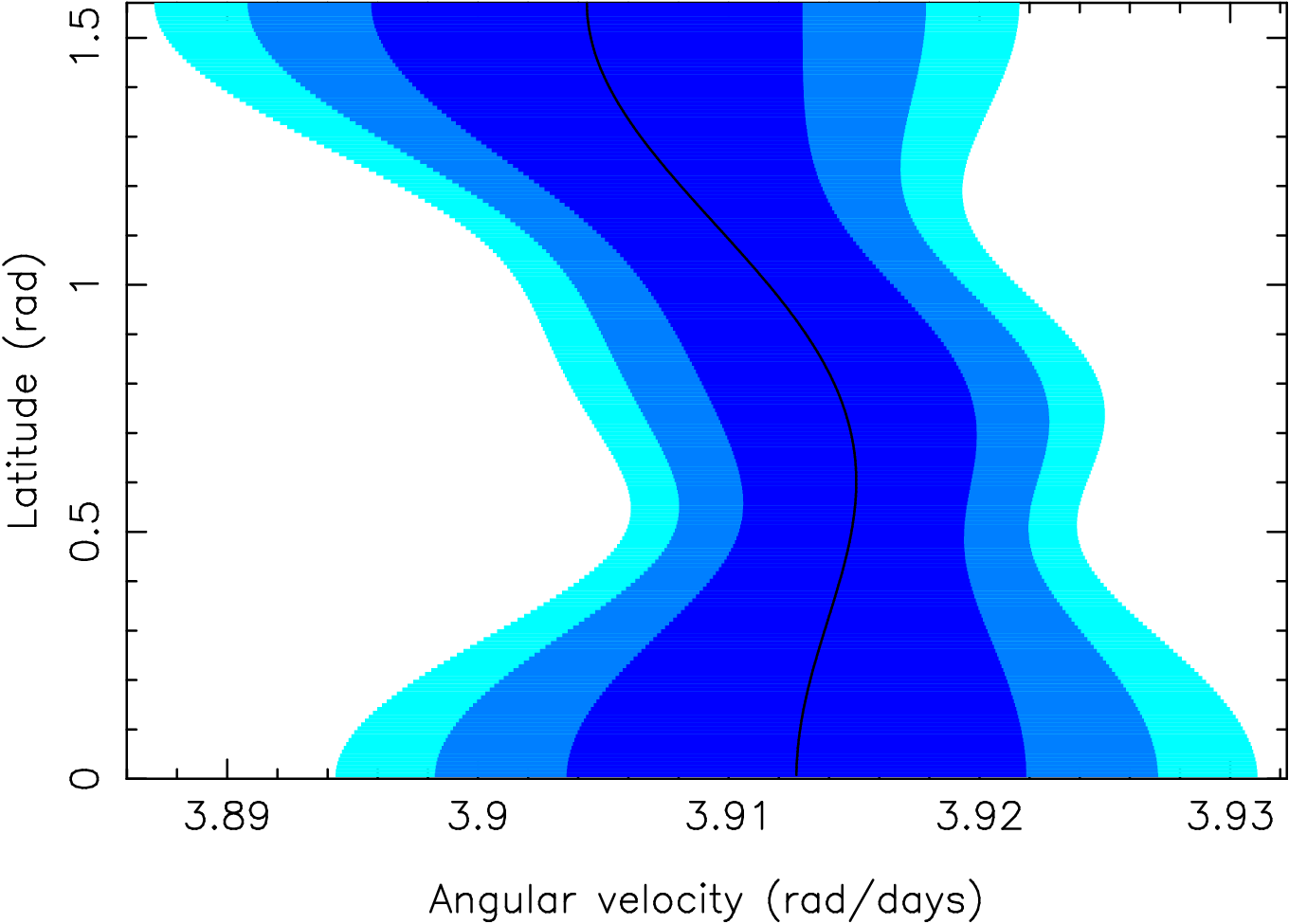}
\caption{Same as in Fig. \ref{fig:plots_15_V889H} but for LQ Hya and $T_{\rm s} = 20$ days.}\label{fig:plots_20_LQHYA}
\end{figure*}

According to our results, LQ Hya behaves qualitatively differently from V889 Her. LQ Hya shows no statistically significant evidence for differential rotation, and estimated angular velocity curves are fully consistent with rigid body rotation (Figs. \ref{fig:plots_15_LQHYA} and \ref{fig:plots_20_LQHYA}). We resort to the principle of parsimony in the interpretation of the results in Table \ref{tab:model_comparison_V889H} -- if the simpler model is not ruled out with a reasonable probability, we consider it to represent the most viable description of the data. The $k=0$ models have consistently significant probabilities for LQ Hya, and it is thus not possible to claim that there is evidence for deviations from rigid body rotation. That is, independent distributions of period and amplitude are sufficient in describing the data and postulating more complex behaviour is not warranted. We note that in Figs. \ref{fig:plots_15_LQHYA} and \ref{fig:plots_20_LQHYA} we have plotted the differential rotation for models with $k=2$ to visually demonstrate that the curves are rather uniform even for a model that does not implicitly assume such uniformity.

We cannot calculate estimates for the rotational shear of LQ Hya in the same way that we did for V889 Her, as only upper limits are available. Instead, we only report the rotation period to be 1.6048$\pm$0.0036 d. Although there are deviations in the observed periodic signals from 1.53~d to 1.70~d for the observed individual periodicities (Fig. \ref{fig:data}), these outliers do not allow concluding that differential rotation plays a role but suggest that other statistical processes are at play, such as the fact that multiple spots with similar observable photometric amplitudes can cause outliers in the sample of observed periodicities. We obtain the maximal and minimal rotation values from our modelling with $k=2$ and adopt the mean of the resulting rotational shear estimates. As a result, we obtain an upper limit of 0.026 rad d$^{-1}$ for the rotational shear corresponding to an estimate for parameter $\alpha$ of 0.0067. These numbers can be considered, at most, only suggestive values given that they are based on a model that is overparameterised and not supported by the data.

It is noteworthy that the apparent structure in the high-precision model in Fig. \ref{fig:plots_15_LQHYA} (top row), which is visible as two separate horizontal clusters of modelled signals at two distinct periods is simply an artefact of the small grid size. The structure is not significantly present in the data.

\section{Conclusions and discussion}

Even in the era of important space observatories capable of precision photometry such as TESS \citep{ricker2014}, it is worth emphasising that dedicated ground-based photometric surveys with multi-decadal baselines can be highly valuable in obtaining new astrophysical information on nearby stars. We have studied the statistical properties of the photometric rotation two nearby Solar-type stars, V889 Her and LQ Hya, with a new approach modelling the largest-spot statistics. This has enabled us to determine the nature of rotation for the two stars: V889 Her demonstrates non-monotonic differential rotation such that the angular velocity first increases to reach a maximum between latitudes of 37-40 deg and then decreases towards a minimum that is lower at the pole than at the equator. On the other hand, the rotation of LQ Hya is indistinguishable from that of a rigid body.

Our estimate for the rotational shear of V889 Her of 0.257$\pm$0.068 rad d$^{-1}$ is consistent with the estimate of 0.319$\pm$0.114 rad d$^{-1}$ reported by \citet{brown2024}. The estimate of \citet{marsden2006} of 0.402$\pm$0.044 rad d$^{-1}$ is somewhat higher, but broadly consistent remembering that the errors are 1-$\sigma$ values. This implies that our method provides reasonable means for determining the rotational shear, being consistent with independent methods.

Interestingly, the qualitative behaviour of the solution for differential rotation is unlike the monotonic behaviour observed for the Sun, where poles rotate more slowly than equatorial regions \citep{howard1970}. Anti-solar behaviour, such that poles rotate faster than equator, has also been observed \citep{brun2017,benomar2018}. However, qualitatively non-monotonic differential rotation profiles have not been previously reported based on observations. Although such more complex rotation profiles are commonly produced in magnetohydrodynamic simulations of solar-type stars \citep{viviani2018,viviani2021,kapyla2023}, these simulations do not readily produce situations where there is a maximum in angular velocity at mid-latitudes. This could be due to the fact that V889 Her has an anomalously large rotational shear and LQ Hya could instead be more representative of the population of young solar-type stars. Yet, the appearance of these non-monotonic rotation profiles now both in observations and numerical models gives reason to consider them physical and not merely numerical artefacts.

We estimate the stellar inclination of V889 Her to be 1.332$\pm$0.077 rad or 76.3$\pm$4.5 deg, with a 99\% credibility interval of [1.15, 1.57] rad. This estimate is broadly consistent with the estimate of 60$\pm$10$^{\circ}$ (or 1.05$\pm$0.18 rad) based on the star's rotation velocity \cite{marsden2006}.

Our results for LQ Hya are also consistent with results reported in the literature. For instance, \citet{lehtinen2022} reported a relative differential rotation of only $\alpha =$ 0.017, which they concluded to be consistent with solid-body rotation. Several consistent results have been reported in the earlier literature \citep{berdyugina2002,you2007,lehtinen2016}.

Although we could not obtain evidence for differential rotation of LQ Hya in our analysis, we still obtained constraints for stellar inclination. The observed distribution of photometric periods enables ruling out nearly pole-on inclinations and qualitative constraints can be given. We estimate the inclination to be 1.03$\pm$0.16 rad or 59$\pm$9$^\circ$, and the 99\% credibility interval is estimated as [0.55, 1.57] rad. This implies that only close to pole-on orientations of the star can be ruled out.

We also considered drawing spot latitude values from non-uniform distributions, such that the distribution of spots was characterised by a Gaussian distribution with additional two free parameters. However, as increasing model complexity in this manner did not result in statistically significantly better models, we resorted to uniform distribution.

The obtained solutions were slightly affected by the chosen grid size. This choice was essentially about balancing between a larger grid and the corresponding greater resolution on the underlying distribution, but limitations regarding the low number of signals in each grid cell became problematic for higher-resolution grids. However, given that the results did not change statistically significantly depending on the selected grid indicates that they are reasonably robust. Likewise, there were no qualitative or otherwise significant differences with respect to the chosen photometry subset baseline either. The results for both stars were essentially consistent for all tested combinations of $n$, and $T_{\rm s}$ (Table \ref{tab:model_comparison_V889H}). This indicates that the results are probably trustworthy for both targets.

Spot evolution can cause problems when determining their properties in accordance with our approach as is often the case for studies of stellar photometry \citep{basri2020}. However, although we rely on detections of simple periodicities indicative of rotational modulation of the light curve due to starspots, we also reject all subsets that show evidence for a superposition of more than one periodicity. Such situations can arise from existence of two spots at different longitudes or from significant spot evolution that causes spots to disappear and/or appear on the stellar surface. Moreover, it has been demonstrated that the light curves of the two stars are stable on time-scales of 30 and 50 days for V889 Her \citep{lehtinen2012} and LQ Hya \citep{lehtinen2016}.

It is also noteworthy that typical large spots on the surfaces of similar targets are rather long-lived. Spot lifetime estimates range 10-350 days \citep{namekata2019,basri2022}, which implies that most large spots probably live longer than our data subset baselines. Moreover, larger spots tend to have longer lifetimes in general \citep{giles2017}, which suggests that the amplitude-period pairs underlying our largest-spot statistics are probably rather robust.

\begin{acknowledgements}
The authors acknowledge Research Council of Finland project SOLSTICE (decision No. 324161). This work has made use of the University of Hertfordshire's high-performance computing facility. GWH acknowledges long-term support for the robotic telescopes from NASA, NSF, and Tennessee State University.
\end{acknowledgements}

\appendix

\section{Solutions for monotonic differential rotation of V889 Her}

\begin{figure*}
\center
\includegraphics[angle=270,width=0.24\textwidth,clip]{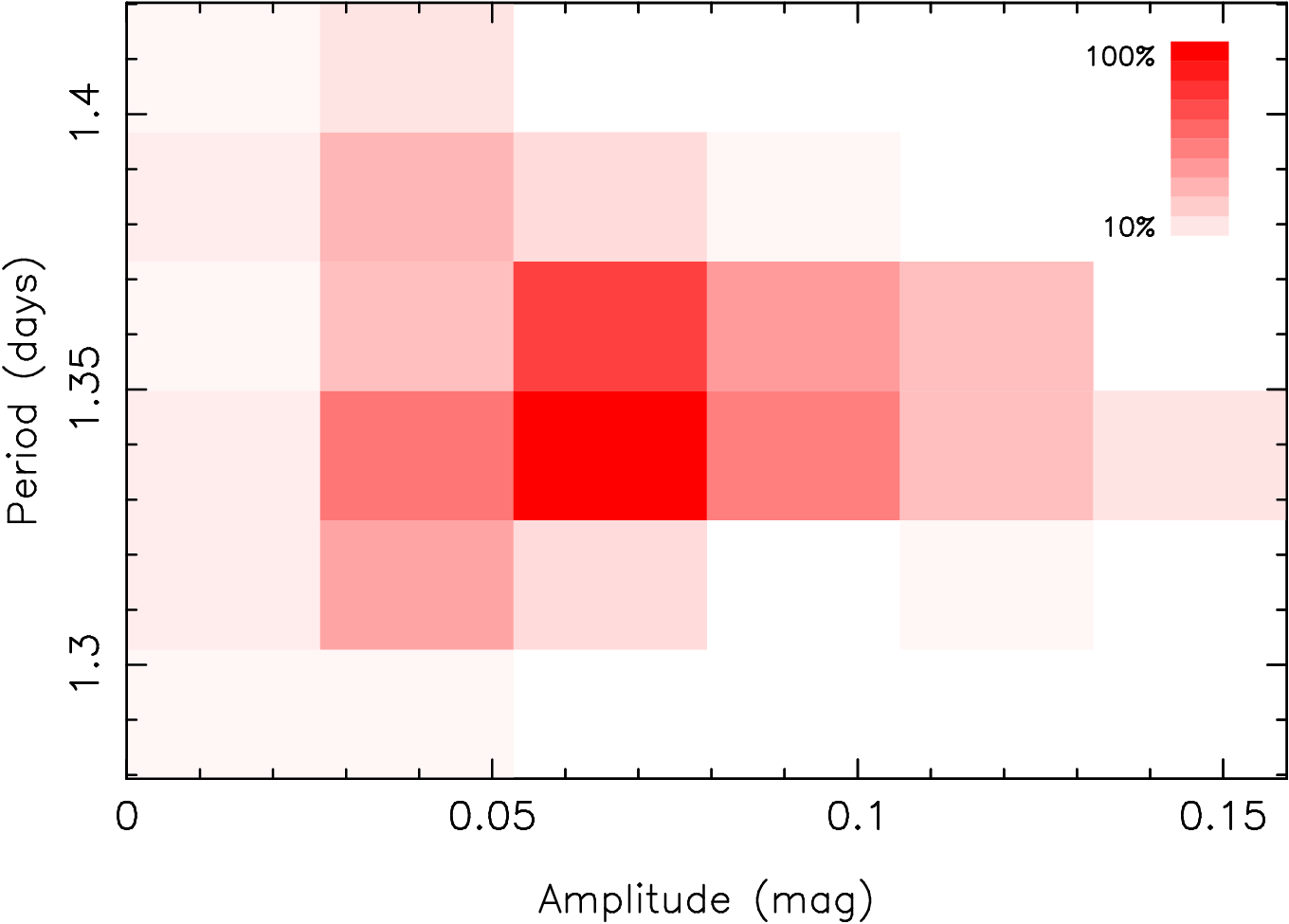}
\includegraphics[angle=270,width=0.24\textwidth,clip]{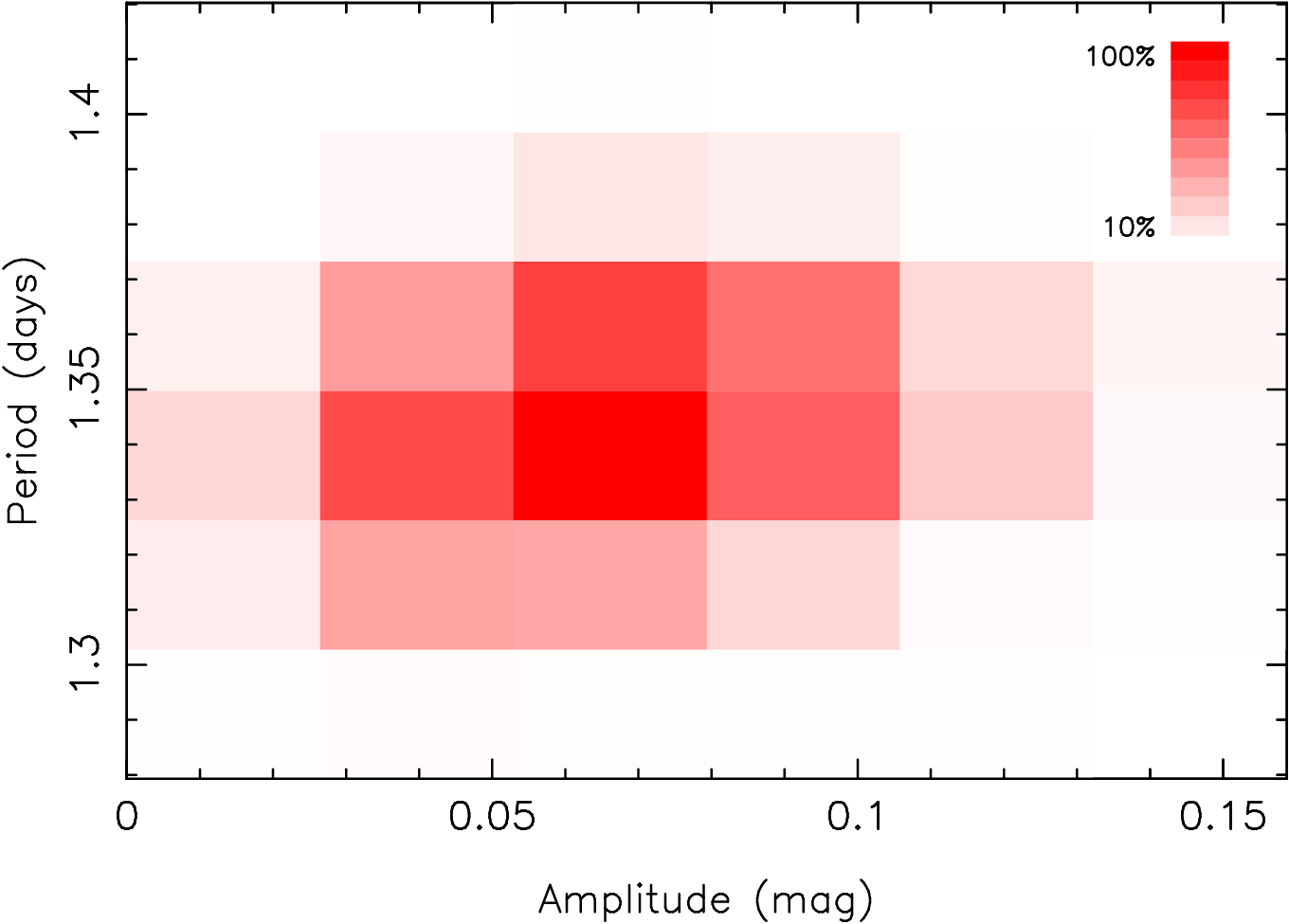}
\includegraphics[angle=270,width=0.24\textwidth,clip]{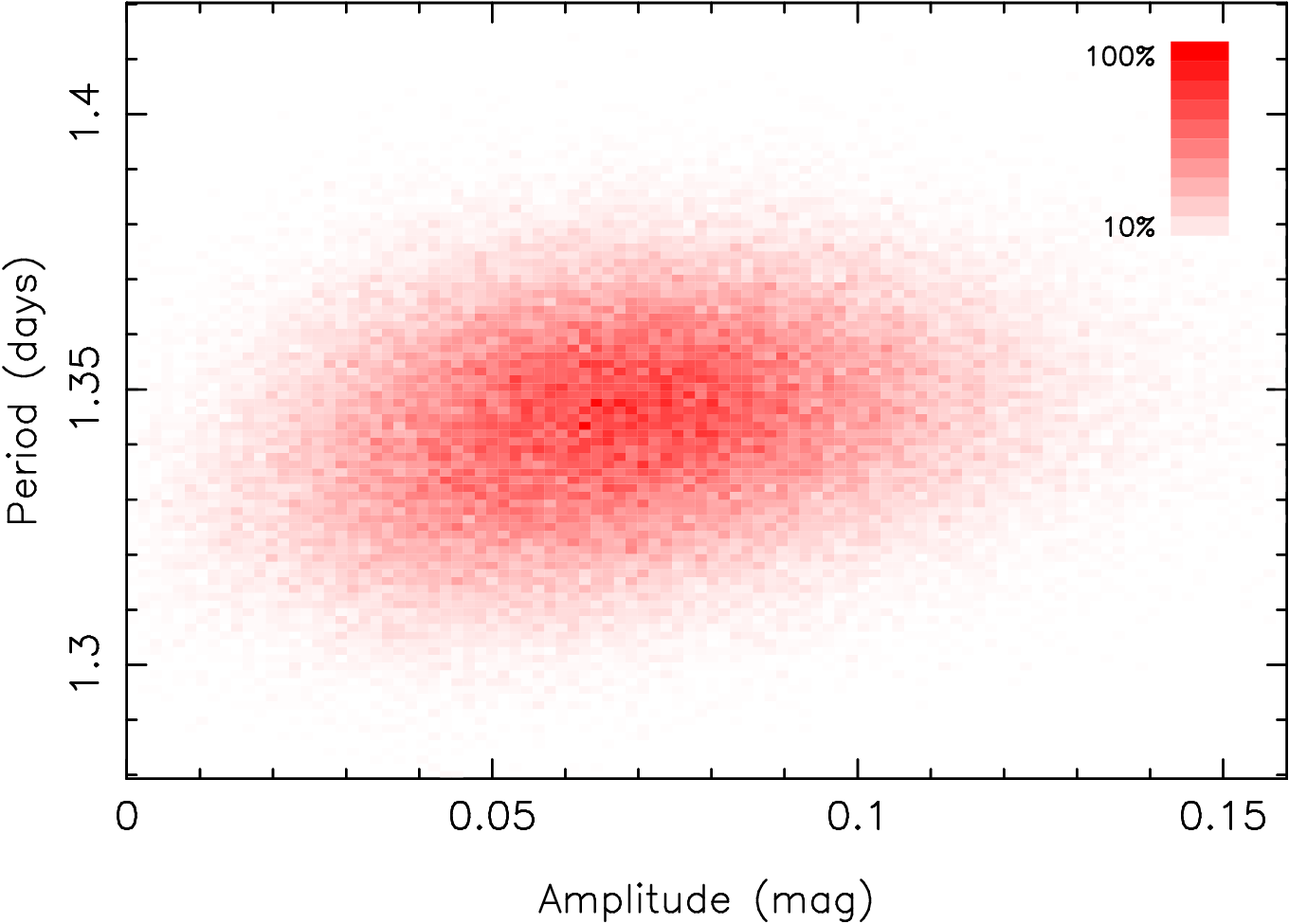}
\includegraphics[angle=270,width=0.24\textwidth,clip]{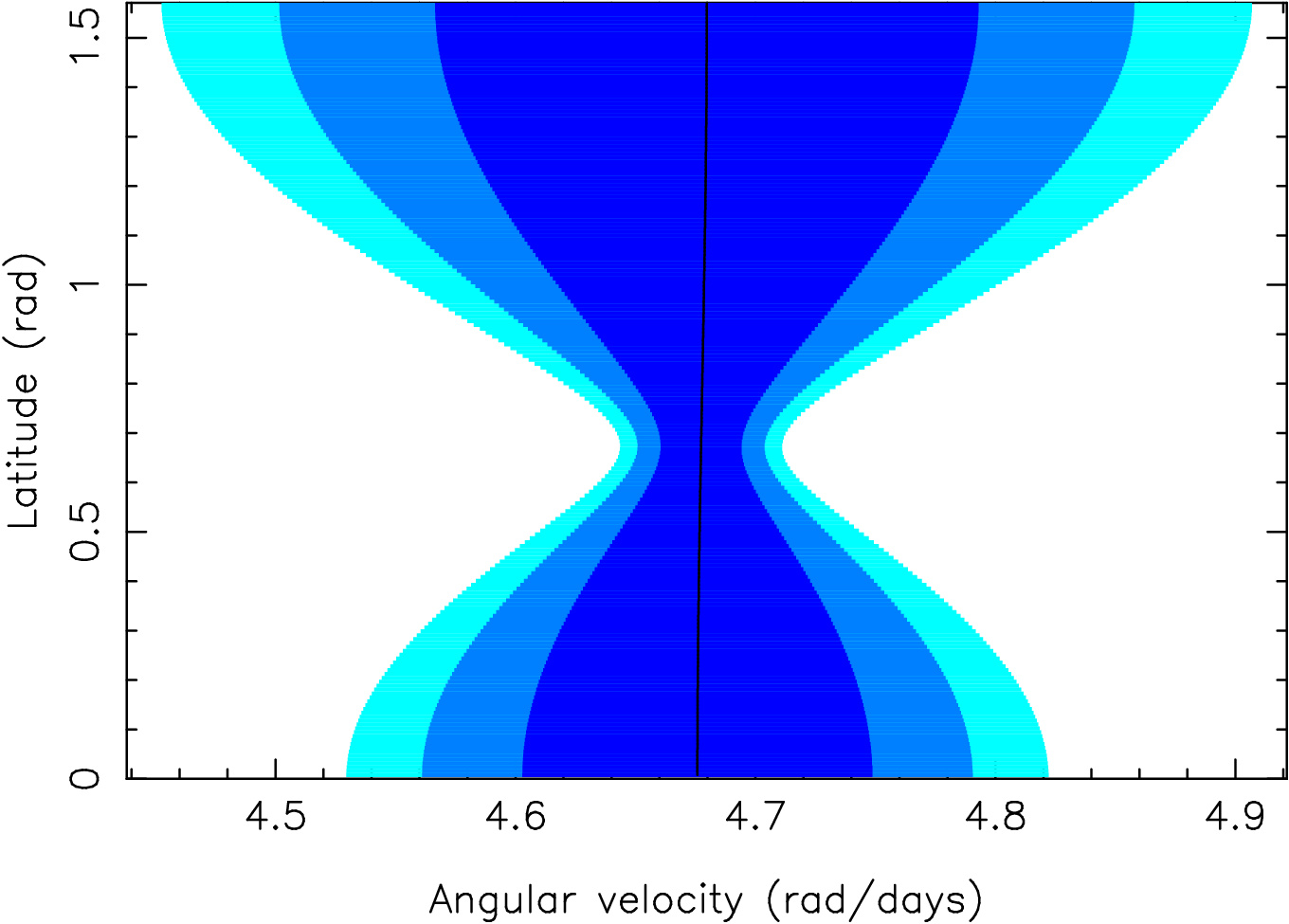}

\includegraphics[angle=270,width=0.24\textwidth,clip]{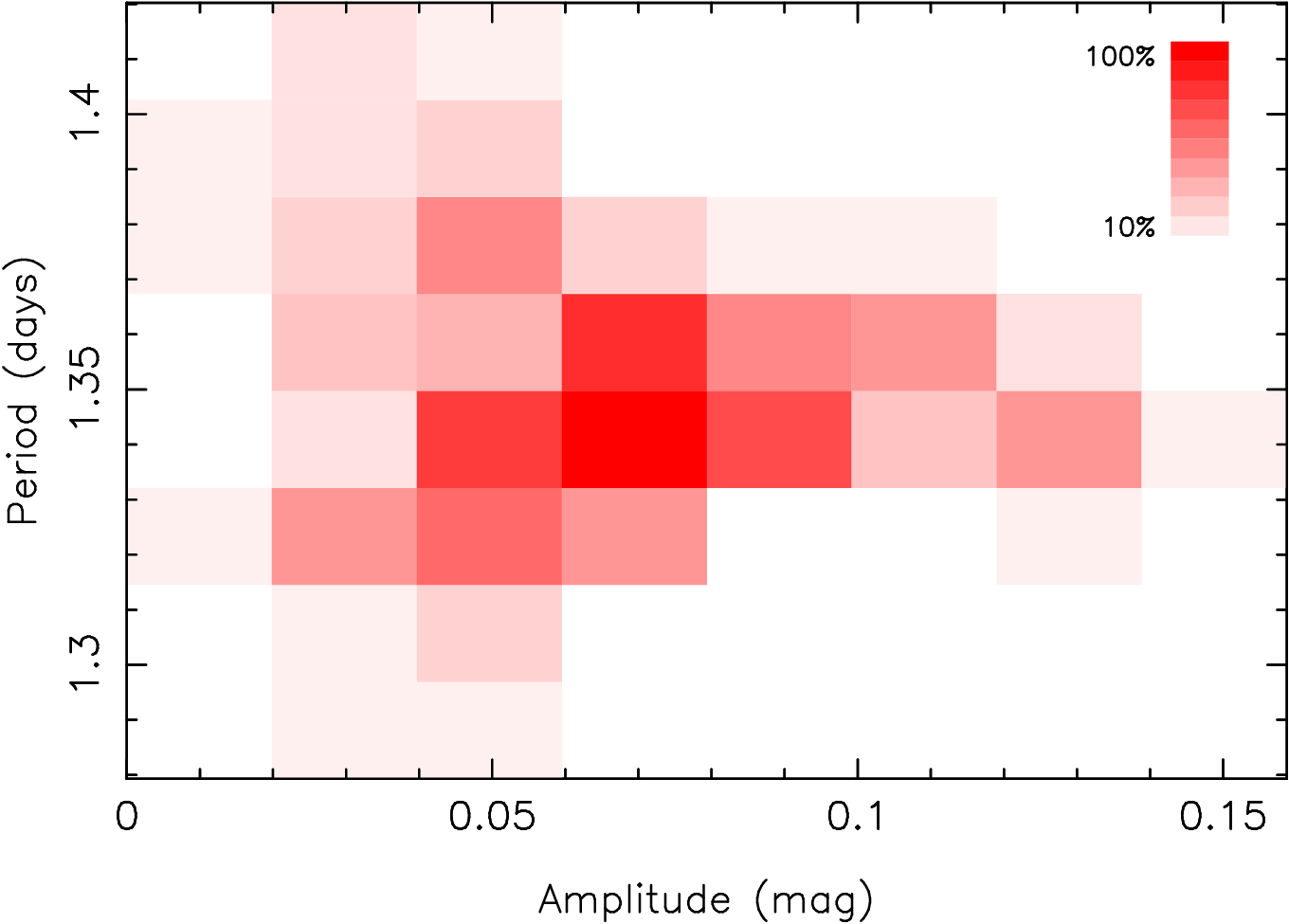}
\includegraphics[angle=270,width=0.24\textwidth,clip]{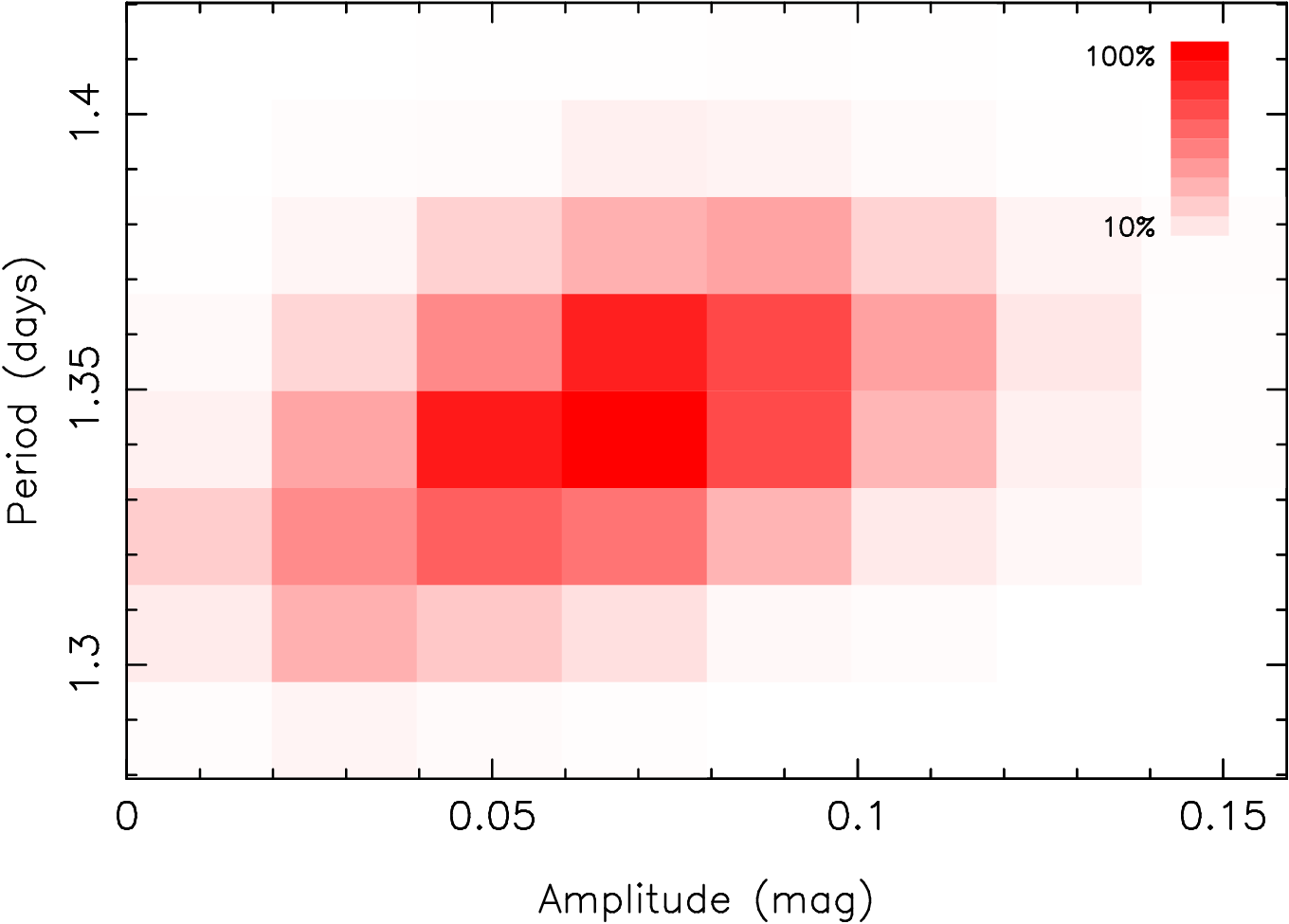}
\includegraphics[angle=270,width=0.24\textwidth,clip]{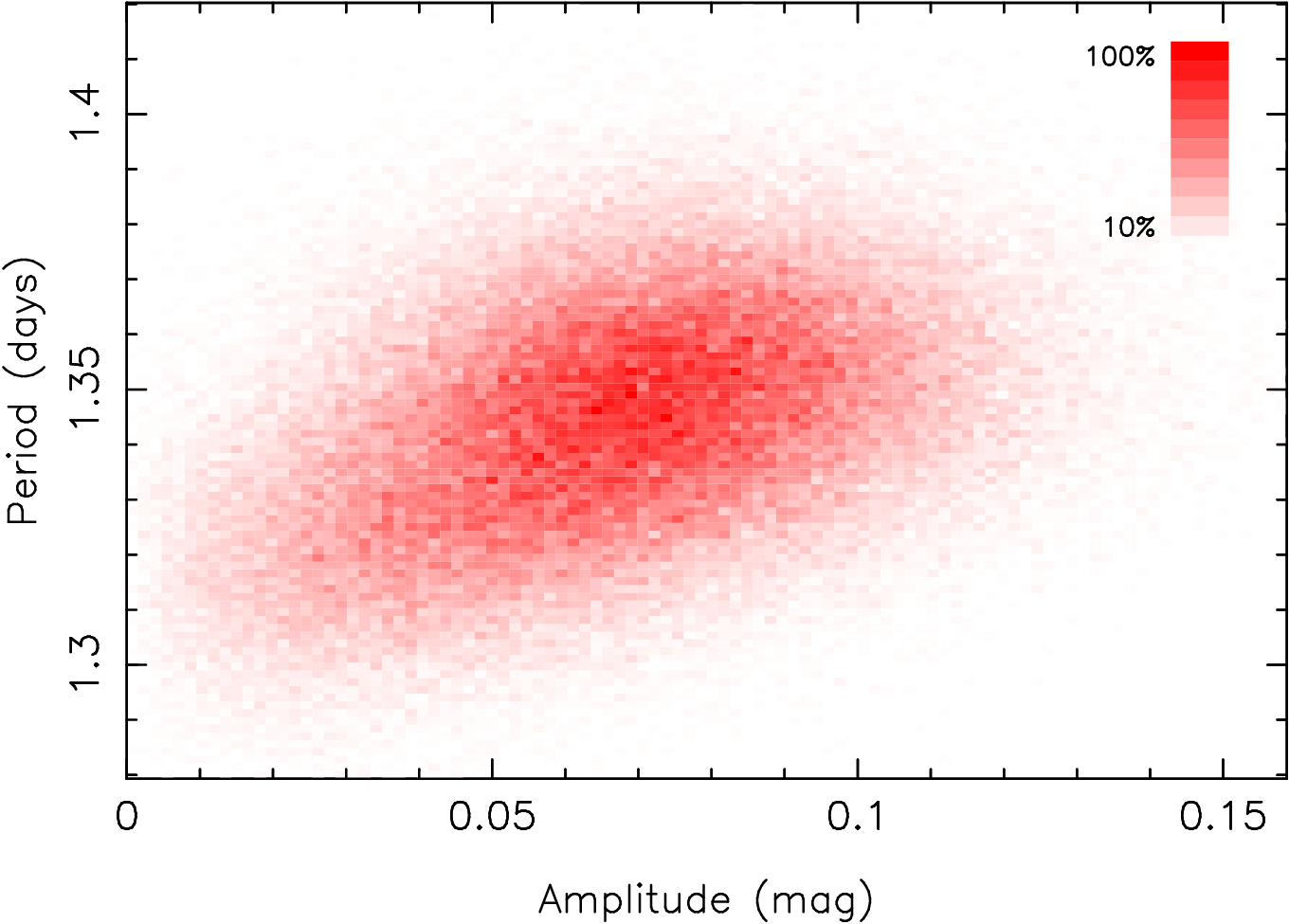}
\includegraphics[angle=270,width=0.24\textwidth,clip]{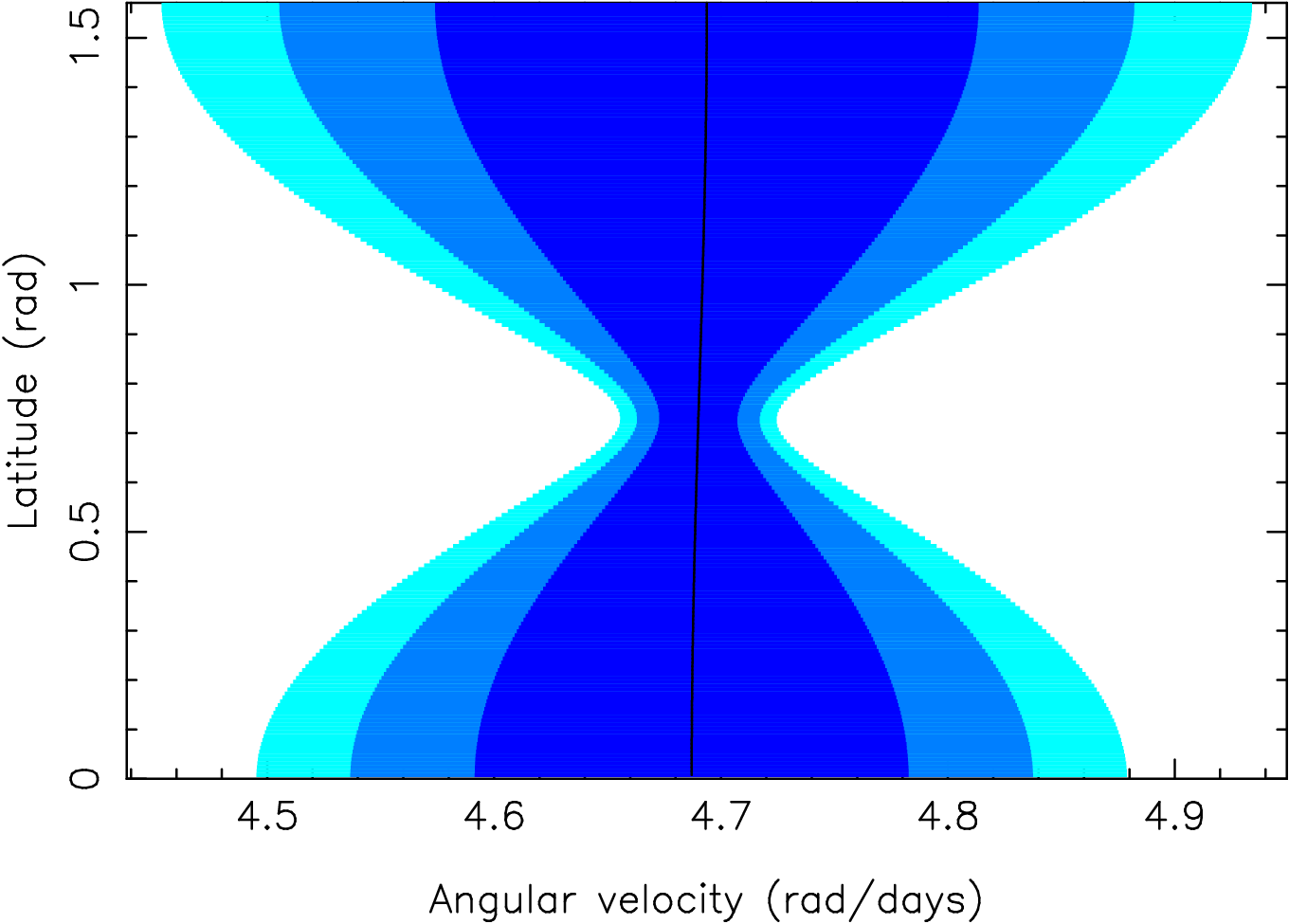}

\includegraphics[angle=270,width=0.24\textwidth,clip]{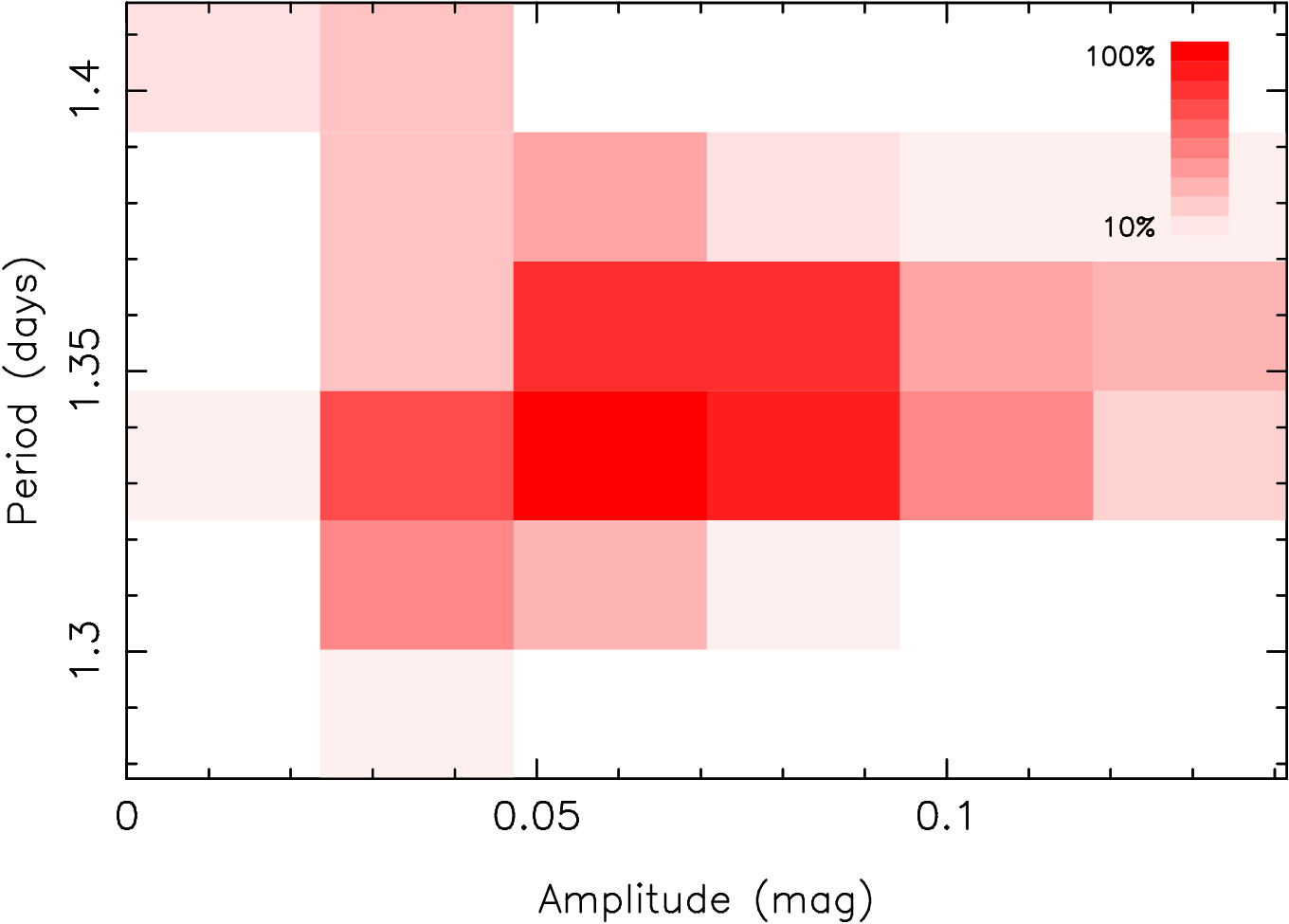}
\includegraphics[angle=270,width=0.24\textwidth,clip]{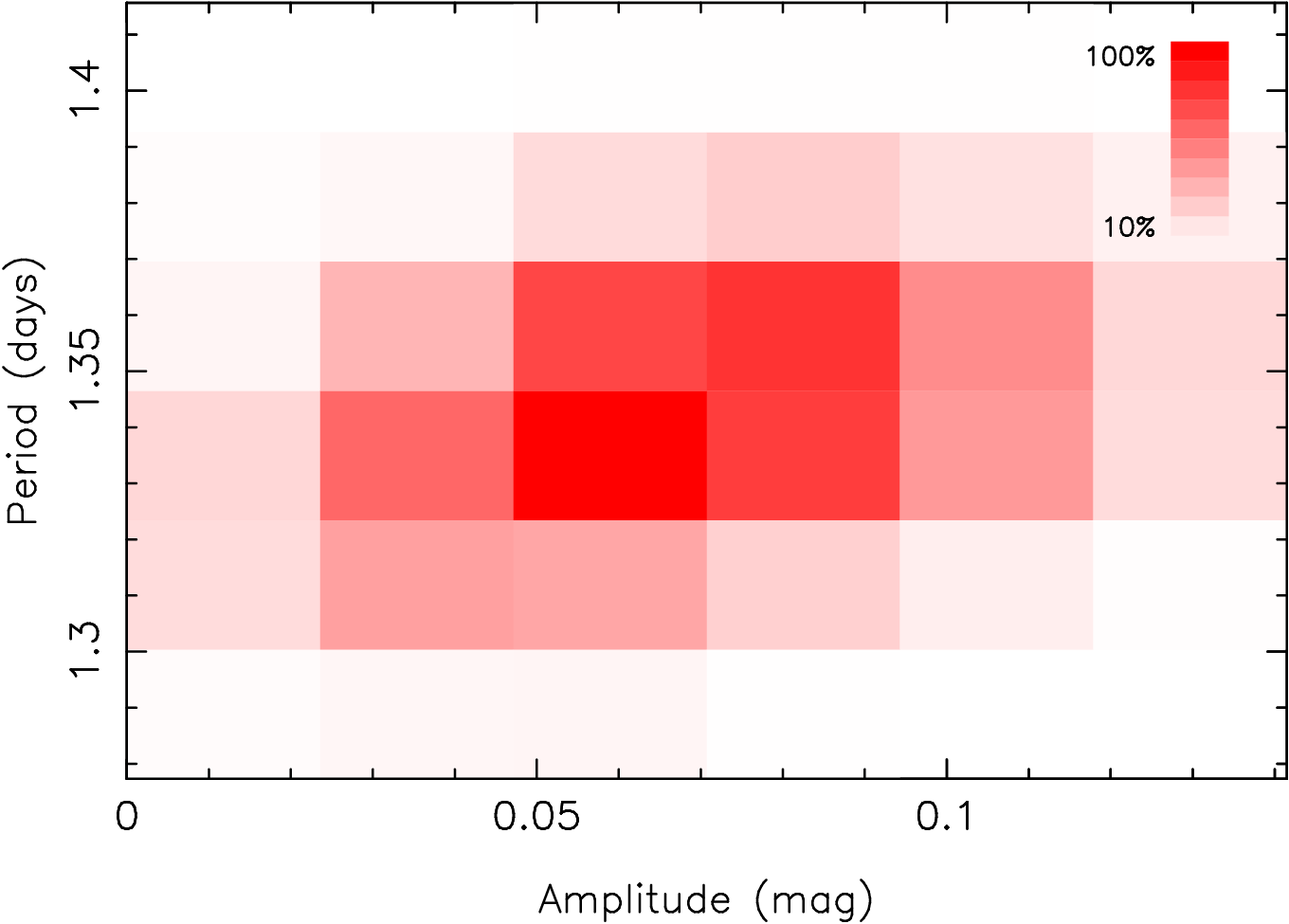}
\includegraphics[angle=270,width=0.24\textwidth,clip]{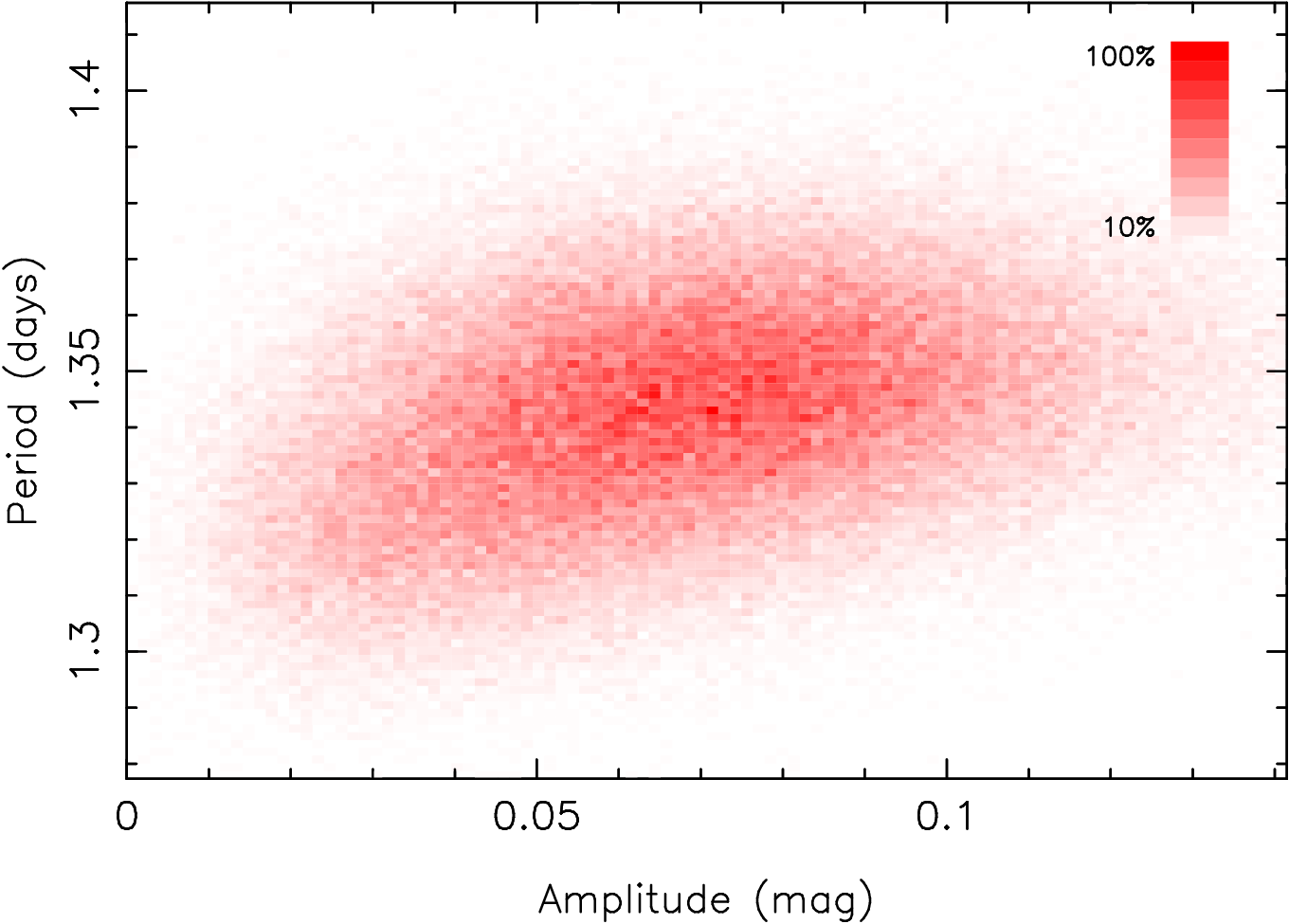}
\includegraphics[angle=270,width=0.24\textwidth,clip]{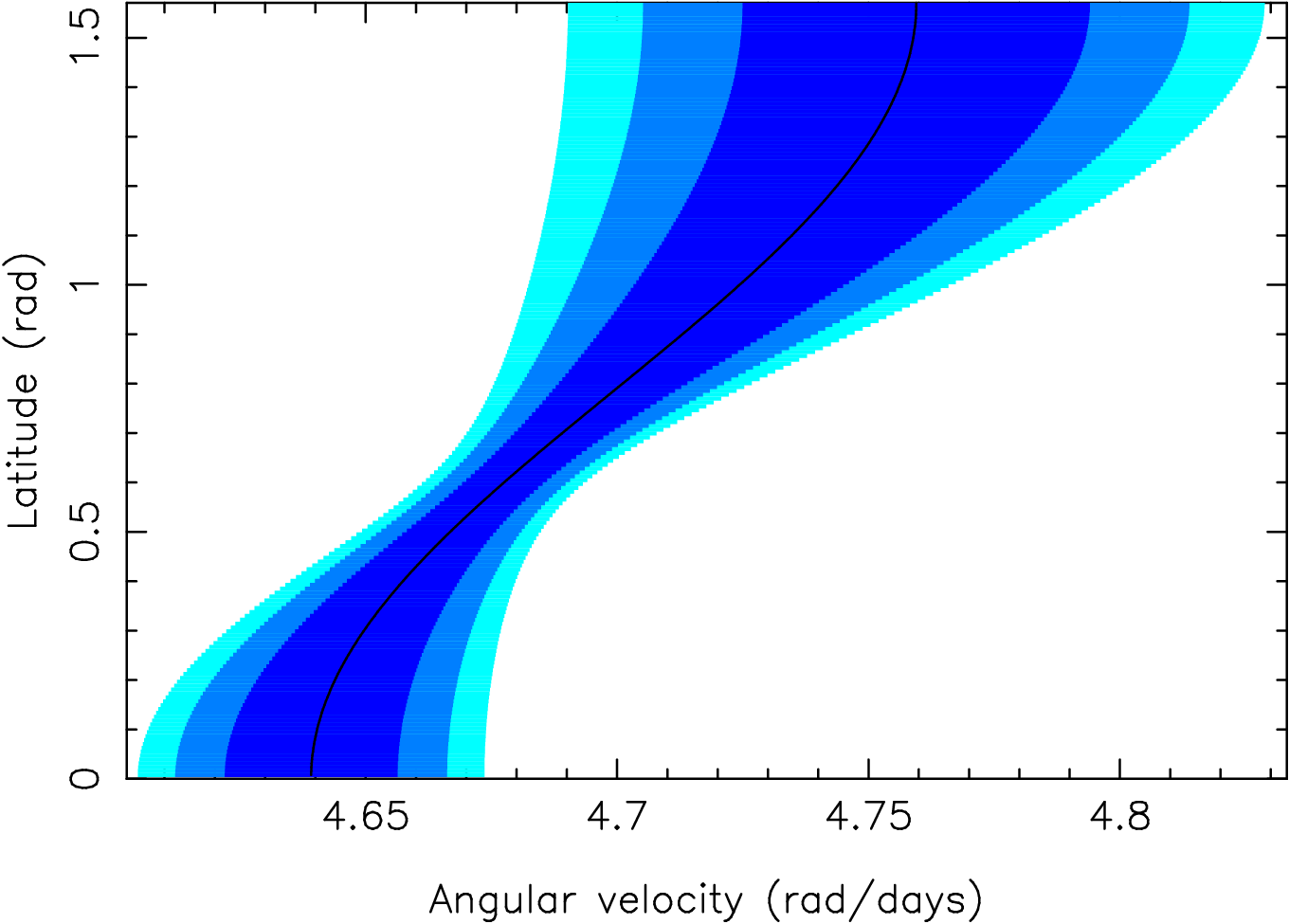}

\includegraphics[angle=270,width=0.24\textwidth,clip]{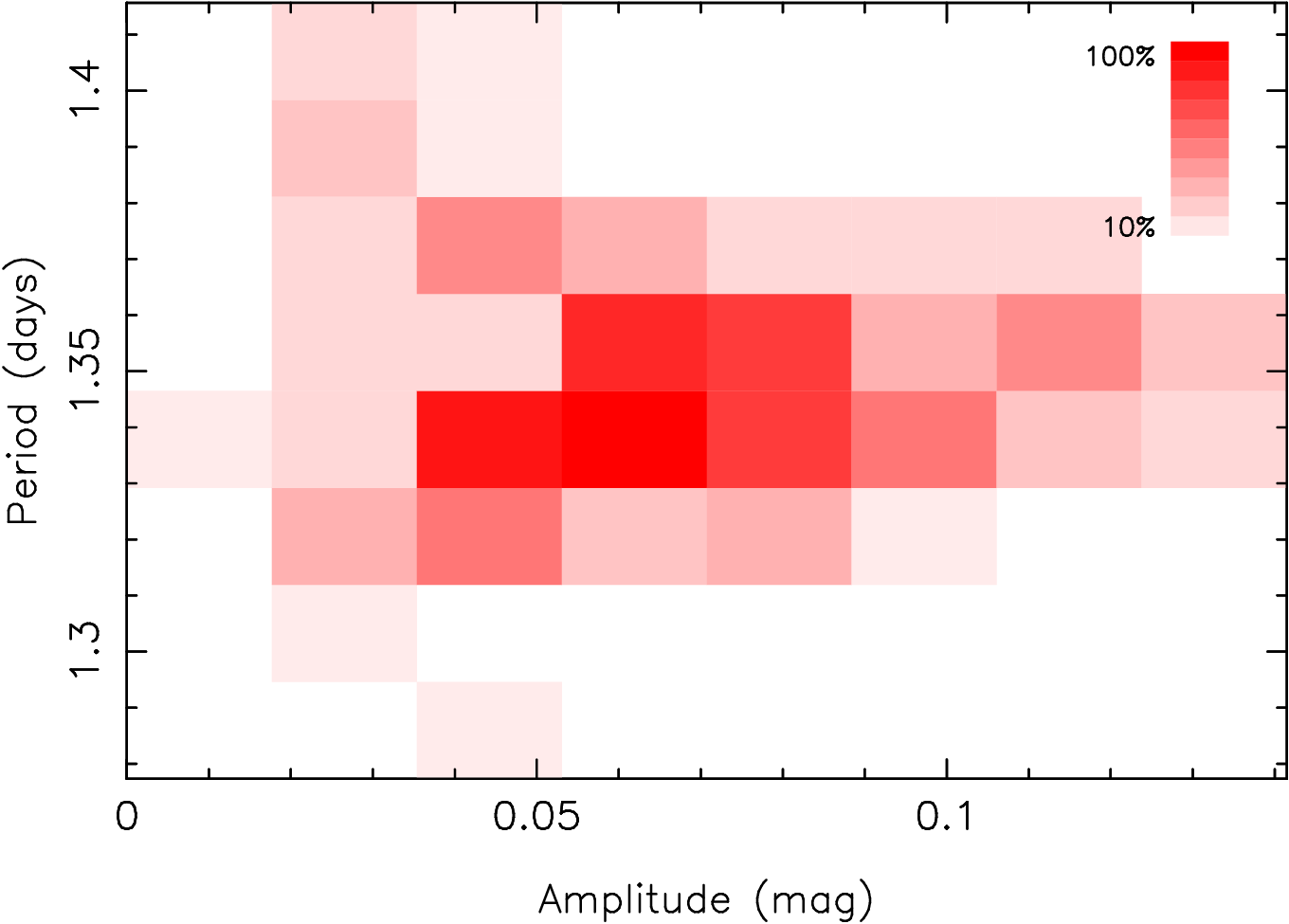}
\includegraphics[angle=270,width=0.24\textwidth,clip]{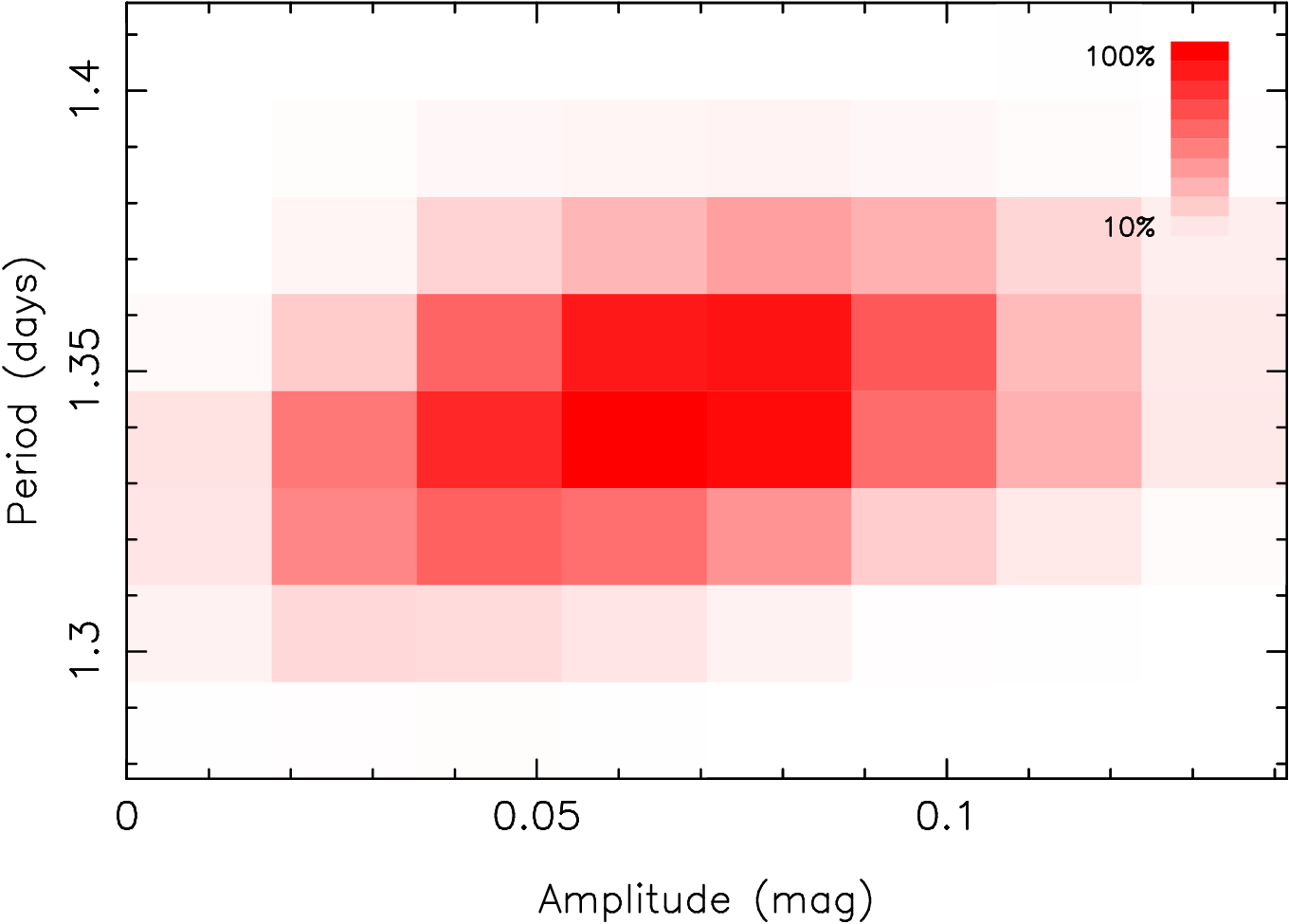}
\includegraphics[angle=270,width=0.24\textwidth,clip]{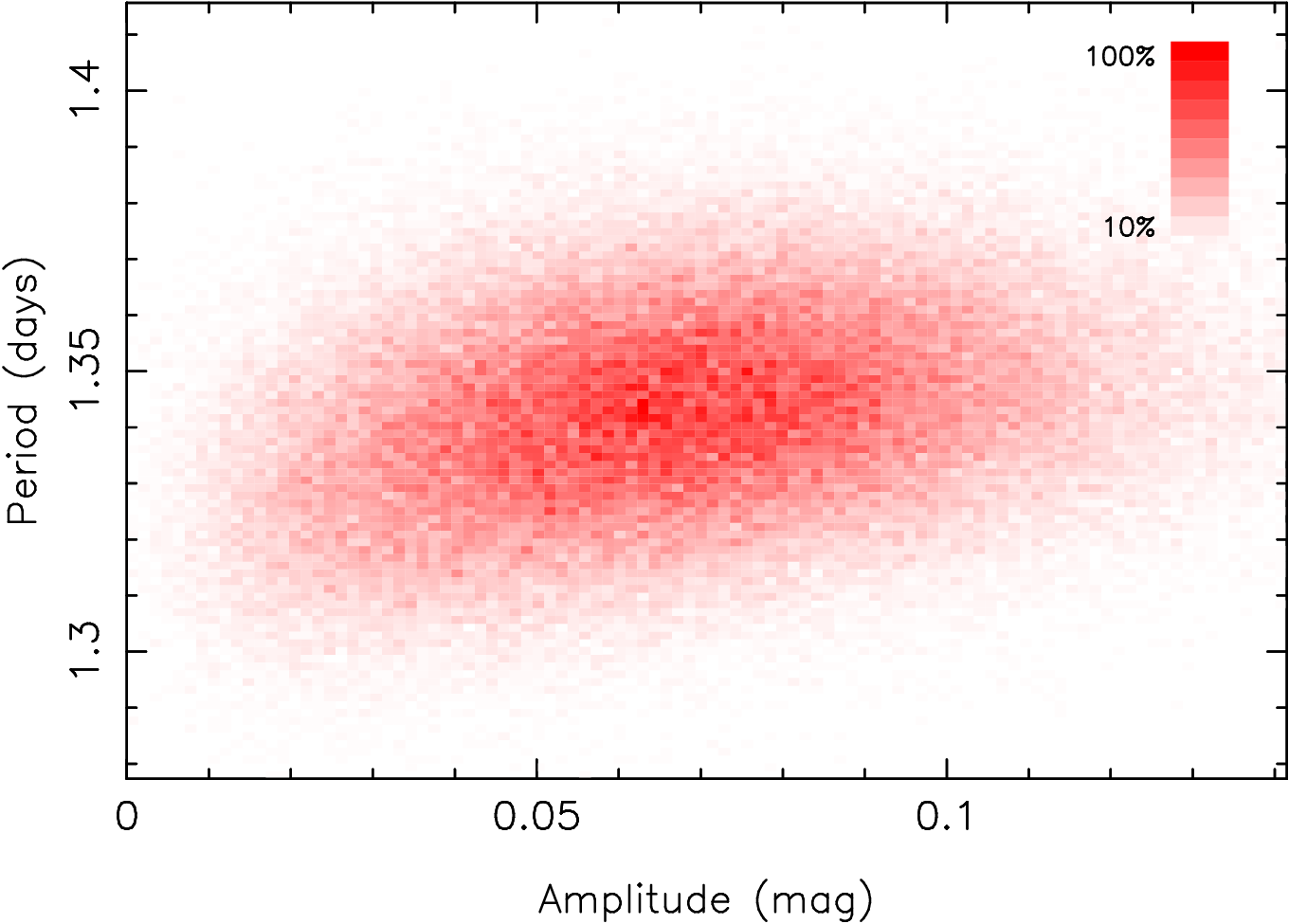}
\includegraphics[angle=270,width=0.24\textwidth,clip]{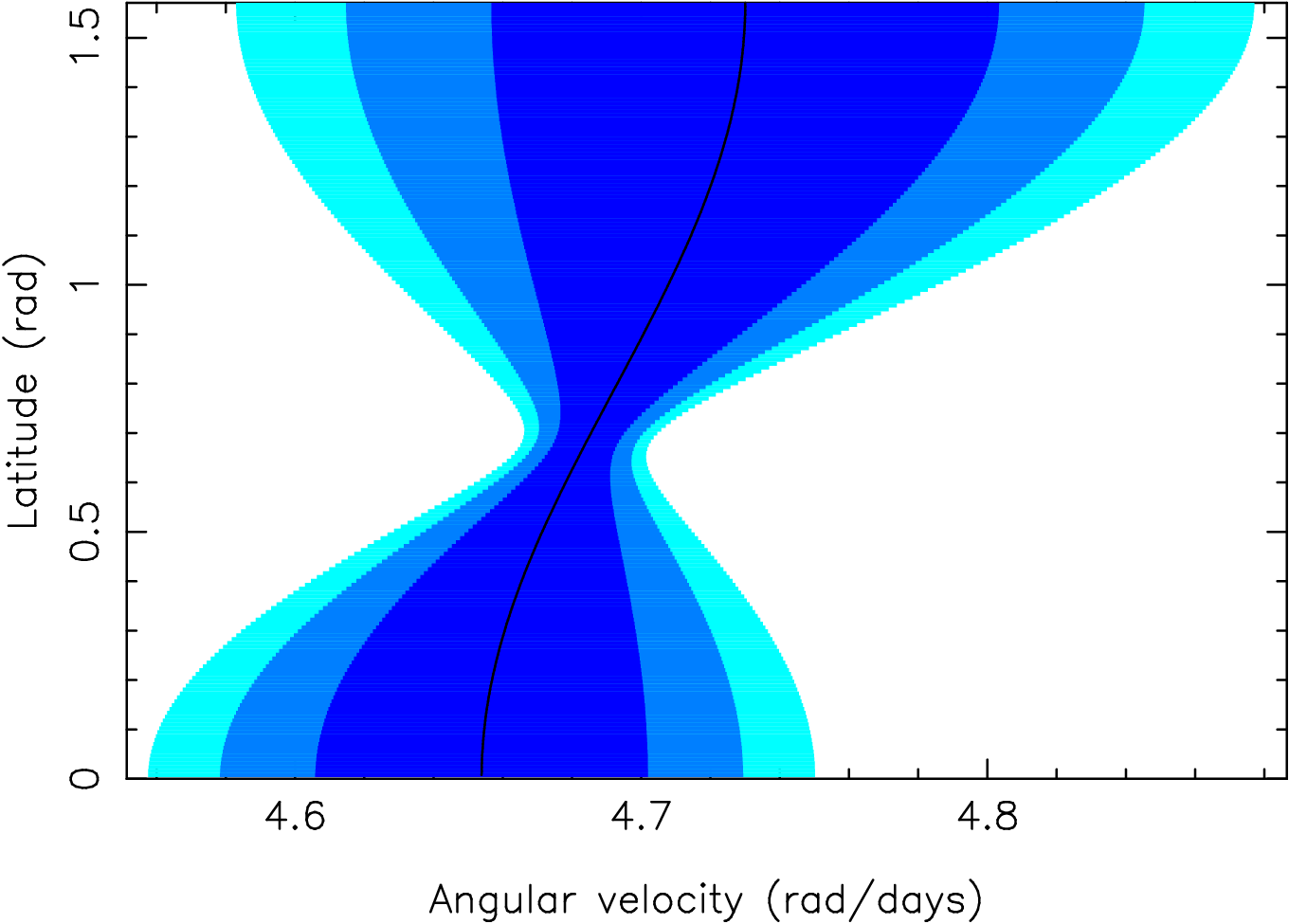}
\caption{As in Fig. \ref{fig:plots_15_V889H} but for monotonic differential rotation model with $k=1$. Rows from top to bottom: $T_{\rm s} = 15$ days and $n=6$; $T_{\rm s} = 15$ days and $n=8$; $T_{\rm s} = 20$ days and $n=6$; $T_{\rm s} = 20$ days and $n=8$. The target star is V889 Her.}\label{fig:plots_1520_V889H_D1}
\end{figure*}

\end{document}